\documentclass{JHEP3}

\input epsf
\usepackage{epsfig}
\usepackage{amssymb}
\usepackage{amsmath}

\newcommand{\bea}{\begin{eqnarray}}
\newcommand{\eea}{\end{eqnarray}}
\newcommand{\bean}{\begin{eqnarray*}}
\newcommand{\eean}{\end{eqnarray*}}

\def\O #1{\overline{#1}}

\def\W #1{\widetilde{#1}}
\def\WH #1{\widehat{#1}}

\def\braket#1{\left\langle #1 \right\rangle}

\def\ket#1{\left| #1\right\rangle}
\def\bket#1{\left| #1\right]}
\def\gb #1{ \left\langle #1 \right]}

\def\a{{\alpha}}

\def\b{{\beta}}

\def\la{\lambda}
\def\eps{\epsilon}

\def\vev{\braket}

\def\Label#1{\label{#1}}%
\title{Unitarity Method with Spurious Pole}
\author{Bo Feng$^{\natural}$, Gang Yang$^{\sharp}$\\
~~~~\\
$^\natural$Center of Mathematical Science, Zhejiang University,
Hangzhou, China\\
E-mail: {\tt b.feng@cms.zju.edu.cn}\\ \\
$^\sharp$Institute of Theoretical Physics, Chinese Academy of Sciences \\
P. O. Box 2735, Beijing 100190, China \\
E-mail: {\tt yangg@itp.ac.cn} }

\abstract{In unitarity cut method, compact input of on-shell tree
level amplitudes is crucial to simplify calculations. Although BCFW
on-shell recursion relation gives very compact tree level
amplitudes, they usually contain spurious poles. In this paper, we
present a method to deal with this issue and provide explicit simple
algebraic functions for various coefficients in the presence of
spurious poles. As an application, we present analytic result (not
just rational term) for one-loop five gluon $A(-++++)$ with scalar
propagator for the first time. }
\begin{document}
%%%%%%%%%%%%%%%%%%%%%%%%%%%%%%%%%%%%%%%%%%%%%%%%%%%%%%%%%%%%%%%%%%

%%%%%%%%%%%%%%%%%%%%%%%%%%%%%%%%%%%%%%%%%%%%%%%%%%%%%%%%%%%%%%%%%
\section{Introduction}
%%%%%%%%%%%%%%%%%%%%%%%%%%%%%%%%%%%%%%%%%%%%%%%%%%%%%%%%%%%%%%%%%

The computation of scattering amplitudes at the next-to-leading
order (NLO) and even beyond is not only a pure theoretical interest,
but also a definite requirement of  experiments, such as the current
Tevatron collider and the forthcoming Large Hadron Collider (LHC)
experiments. In last few years, there have been spectacular
progresses for this subject, which are extensively reviewed in the
report \cite{Report08}.

The art of the NLO calculation is based on the knowledge that, any
one-loop amplitudes can be expanded into a set of master integrals
as \cite{{Passarino-Veltman},IntegralRecursion,MasterIntegrals}
\bea {\cal A} = c_{5,i}(\eps) I_5^{D,(i)}[1] + c_{4,i}(\eps)
I_4^{D,(i)}[1] + c_{3,i}(\eps) I_3^{D,(i)}[1] + c_{2,i}(\eps)
I_2^{D,(i)}[1] + c_{1,i}(\eps) I_1^{D,(i)}[1], \quad
\Label{BasisExp-I} \eea
where the master integral (i.e. scalar integral) is defined as%
\footnote{Here we consider the general massive one loop amplitude,
i.e. the propagator is massive. The massless case can be obtained by
taking $M_1=M_2=m_i=0$. Notice that in the massless case the set of
master integrals is simpler, since there are no one-point functions
(tadpoles) or two-point functions with massless external momenta
(massless bubbles).}
\bea   I_n^D[1]  \equiv -i  (4\pi)^{D/2}  \int {d^D p \over
(2\pi)^D}{1\over (p^2-M_1^2) ((p-K)^2-M_2^2) \prod_{i=1}^{n-2}
((p-K_i)^2-m_{i}^2)}. \quad \Label{n-scalar} \eea
The coefficients $c_{n,i}(\eps)$ are rational functions of external
momenta and polarization vectors. They are also rational functions
of $\eps$, where $D=4-2\eps$ by using dimensional regularization.
Since these master integrals are relatively well understood
\cite{IntegralRecursion, MasterIntegrals}, the problem is reduced to
the computation of these coefficients.

There is another equivalent expression for (\ref{BasisExp-I}) that
is often used in practical calculations as
\bea {\cal A} &=& c_{5,i}(\eps=0) I_5^{D,(i)}[1] + c_{4,i}(\eps=0)
I_4^{D,(i)}[1] + c_{3,i}(\eps=0) I_3^{D,(i)}[1] \nonumber\\ && +~
c_{2,i}(\eps=0) I_2^{D,(i)}[1] + c_{1,i}(\eps=0) I_1^{D,(i)}[1] +
(\textrm{rational part}) + {\cal O}(\eps), \quad \Label{BasisExp-II}
\eea
where  coefficients of master integrals no longer depend on $\eps$,
but extra contribution -- rational part -- appears. The part
expanded by master integrals is usually called
\emph{cut-constructible} part, because  coefficients
$c_{n,i}(\eps=0)$ can be calculated by the
pure four-dimensional unitarity cut method%
\footnote{If using four-dimensional unitarity method, we would not
obtain pentagon terms. However, the pentagon and box master
integrals are not independent: the scalar pentagon can be expressed
as a sum of five scalar boxes up to ${\cal O}(\eps)$ corrections
\cite{IntegralRecursion}. Therefore, the box terms obtained by
four-dimensional unitarity cut include also the pentagon
contributions up to ${\cal O}(\eps)$.} \cite{BDDK, {BDK-ee4p}}. In
last few years, especially motivated by the twistor string theory
\cite{Witten03}, techniques for one-loop amplitude calculations are
much developed and various amplitudes up to the full six-gluon
cut-constructible part have been calculated \cite{{Cachazo:2004by},
{Bena:2004xu}, {Cachazo:2004dr}, {Britto:2004nj}, {Bern:2004ky},
{Quigley:2004pw}, {Bidder:2004tx}, {oneloop-MHVvertices},
{BCF04-N=4}, {BBCF05-SQCD}, {BFM06-QCD}, {MastroliaTriCut},
{BjerrumBohr:2007vu}}. However, the rational part is totally lost in
the pure four-dimensional unitarity method. It has to appeal to some
other methods to be computed, either by using the improved
traditional tensor reduction methods \cite{XYZ,BGH06}, or the
unitarity bootstrap approach \cite{BDK-Bootstrap, BBDFK-Bootstrap,
{Bootstrap-PhiMHV}}, or other methods. Recently, an automated
package {\tt BlackHat} has been developed to computer the full
one-loop amplitude, by combining the four-dimensional unitarity
method and the bootstrap approach \cite{{Berger:2008sj}}. See
\cite{{BDK-OnShell-Review}, {Risager-PhDThesis}} for a full review.

Another very noticeable development is the method of OPP-reduction
\cite{{OPP}} (based on the technique in \cite{delAguila-Pittau}).
Inspired by this work, several very efficient numerical methods have
been developed \cite{{Ellis:2007br}, {Ossola:2007ax},
{Giele:2008ve}, {Ossola:2008xq}, {Mastrolia:2008jb},
{Ellis:2008ir}}. In \cite{{Giele:2008bc}}, a fully automated
one-loop N-gluon generator --- {\tt Rocket} has been developed by
implementing the OPP-reduction and $D$-dimensional unitarity method.
Some analytic techniques were also developed in \cite{{Forde07},
{Kilgore07}}. The rational part can also be determined in the OPP
approach, by keeping the full $D$-dimensional dependence in all
terms \cite{{Giele:2008ve}, {Ossola:2008xq}, {Giele:2008bc},
{Ellis:2008ir},{Badger:2008cm}}.

The $D$-dimensional unitarity method is an extension of the
four-dimensional unitarity method. This idea origins from the fact
that a null momentum in $(4-2\eps)$-dimension can be equivalent to a
massive momentum in four-dimension \cite{vanNeerven85, Bern-Morgan,
BDDK-N=4}. The power of $D$-dimensional unitarity method is that it
can calculate the full amplitude. In another word, it makes the
rational parts (also ${\cal O}(\eps)$-terms)
\emph{cut-constructible} too. This can be understood by the
following expansion of the amplitude
\bea {\cal A} &=& c_{5,i}(\mu^2)\otimes I_5^{D,(i)}[1] +
c_{4,i}(\mu^2)\otimes I_4^{D,(i)}[1] + c_{3,i}(\mu^2)\otimes
I_3^{D,(i)}[1] \nonumber\\ && + ~c_{2,i}(\mu^2)\otimes
I_2^{D,(i)}[1] + c_{1,i}(\mu^2)\otimes I_1^{D,(i)}[1], \quad
\Label{BasisExp-III} \eea
where the coefficients $c_{n,i}(\mu^2)$ can be obtained by using
$D$-dimensional unitarity method \cite{Brandhuber05, {ABFKM06},
{BF06}, {BF07}, {BFM08}, {BFG08}} (see also \cite{Giele:2008ve,
{Ellis:2008ir}}). The operation ``$\otimes$'' is defined as
\bea f(\mu^2)\otimes I_n^D[1] \equiv I_n^D[f(\mu^2)]. \eea
The coefficients $c_{n,i}(\mu^2)$ are polynomials of $\mu^2$
\cite{{delAguila-Pittau}, BFG08,BFM08}\footnote{A complete proof is
given in \cite{BFG08} for massless case, while using this proof, in
Appendix of \cite{BFM08} has briefly discussed the generalization to
massive case. Although \cite{BFM08} appeared in arXiv earlier than
\cite{BFG08}, as mentioned in citation of \cite{BFM08}, \cite{BFG08}
should be taken first logically. We emphasize that it's necessary to
extract the pentagon terms so that the box coefficients can be
polynomial of $\mu^2$.}, where $\vec\mu$ is the $-2\eps$ component
of the $(4-2\eps)$-dimensional internal loop momentum $p$. By using
following relation \cite{Bern-Chalmers,Bern-Morgan}
\bea I_n^D[(\mu^2)^k] = \frac{\Gamma(k-\eps)}{\Gamma(-\eps)}
I_n^{D+2k}[1] = -\eps~\Gamma(k)~I_n^{D+2k}[1]+ {\cal O}(\eps), \quad
\Label{dim-shift-basis} \eea
all  $\mu^2$-dependent terms can be converted to  terms of
\emph{dimensionally shifted master integrals}, i.e. $I_n^{D+2k}[1]$.
So the expansion can be reexpressed as
\bea {\cal A} &=& c_{5,i}(\mu^2=0) I_5^{D,(i)}[1] + c_{4,i}(\mu^2=0)
I_4^{D,(i)}[1] + c_{3,i}(\mu^2=0) I_3^{D,(i)}[1] \nonumber\\ &&  +~
c_{2,i}(\mu^2=0) I_2^{D,(i)}[1] + c_{1,i}(\mu^2=0) I_1^{D,(i)}[1] +
(\textrm{dimensionally shifted integrals}). \quad
\Label{BasisExp-IV} \eea
The coefficients $c_{n,i}(\mu^2=0)$ are the same as the coefficients
$c_{n,i}(\eps=0)$ which can be obtained by pure four-dimensional
unitarity method. The terms of dimensionally shifted integrals will
produce the rational part and ${\cal O}(\eps)$ terms. The reason is
that the coefficients of the dimensionally shifted master integrals
are always of ${\cal O}(\eps)$-order (which is easy to see from
(\ref{dim-shift-basis})), so only the divergent parts of the
integrals could give finite contributions, which are the rational
parts.

In the series of work \cite{{BF06}, {BF07}, {BFM08}, {BFG08}},
formulas have been given to compute the coefficients
$c_{n,i}(\mu^2)$ (except the tadpoles and massless bubbles)%
\footnote{The coefficients of cut-free functions like tadpoles and
massless bubbles can't be obtained via double-cut unitarity. In the
massless case, these coefficients identically vanish. But in massive
case, other methods are needed to evaluate them, for example, by
considering the known divergent behavior of the amplitude
\cite{Bern-Morgan}, or with alternative techniques \cite{{OPP},
{Ellis:2007br}, {Ossola:2007ax}, {Kilgore07}, {Giele:2008ve},
{Mastrolia:2008jb}, {Ellis:2008ir}}.}. By using these formulas, the
pentagon, box, triangle and bubble coefficients can be read directly
from the tree level input without evaluating any integrals. However,
a potential weakness for these formulas is that {\bf the tree level
inputs are required to have no spurious pole}. This is in contrary
with a general observation: while the tree level inputs obtained by
BCFW on-shell recursion relation \cite{BCFW} are in a very compact
form, they usually contain spurious poles. To take full advantage of
the power of on-shell recursion relation, it is necessary to
generalize the formulas so that they are directly applicable for the
tree level input in the presence of spurious poles. This is the main
goal of the present paper. While solving this problem, we are able
to give very general algebraic expressions for most general tree
level inputs.

The paper is organized as follows. In the next subsection, we list
our main results. In Section 2, we give a brief review of the
$D$-dimensional unitarity method and define various forms of tree
level input. In Section 3, we study some properties of the formulas
from which we can gain some perspective on how to make
generalization. Then in Section 4, we present the generalized
compact formulas, with a rigorous proof. Two further problems are
solved in Section 5. First we give a general pentagon formula by
using the quintuple-cut method. Then we simplify the $u$-dependence
of the box formula based on the previous result. In Section 6, we
implement the new formulas to compute the full five gluon amplitude
$A(1^-,2^+,3^+,4^+,5^+)$. A summary is given in Section 7. In
Appendix A, we summarize the formulas in previous works. In Appendix
B, we show the new pentagon formula is consistent with the previous
one. Then in Appendix C, an equivalent expression is given for
$\W\ell_{ij}$ which is used in box formula. Finally, we give
explicitly the $\mu^2$ terms of the box coefficients in Section 6.

%%%%%%%%%%%%%%%%%%%%%%%%%%%%%%%%%%%%%%%%%%%%%%%%%%%%%%%%%%%%%%%%%
\subsection{\label{final-formulas} Summary of our results}
%%%%%%%%%%%%%%%%%%%%%%%%%%%%%%%%%%%%%%%%%%%%%%%%%%%%%%%%%%%%%%%%%

In this part, we list our main results in this paper. The
definitions of various variables and functions can be found in later
sections.

The starting point is following double cut integral
\bea - i (4\pi)^{D/2} \int {d^{D} p \over (2\pi)^{D}}
~\delta^{(+)}(p^2-M_1^2)~\delta^{(+)}((p-K)^2-M_2^2)~{\cal
T}^{(N)}(\W\ell) , \eea
where
\bea {\cal T}^{(N)}(\W\ell) = A_L^{\textrm{tree}}(\W\ell)\times
A_R^{\textrm{tree}}(\W\ell). \eea
The ${\cal T}^{(N)}(\W\ell)$ can be calculated by any method, for
example Feynman diagram, off-shell recursion relation \cite{BG} or
BCFW method \cite{BCFW}, and allow the presence of spurious poles.
$N$ is the degree for $p$ and will be defined in Section 2. In our
notation, $p=\W \ell+ \vec\mu$ and
\bea  \W\ell &=& {K^2 \over \gb{\ell|K|\ell}}\left[ -\b\sqrt{1-u}
\left( P_{\la \W\la} -{ K\cdot P_{\la \W\la}\over K^2} K\right)- \a
{ K\cdot P_{\la \W\la}\over K^2} K \right]
,~~~~\Label{w-ell-form}\eea
where $P_{\lambda\W\lambda}= \ket{\ell}|\ell]$ and $u,\a,\b$ are
given by (\ref{u-def}) and (\ref{ab-def}). We can also rewrite
$\gb{\ell|K|\ell}=-2K\cdot P_{\la \W\la}$. We should treat
$\ket{\ell}$ and $|\ell]$ as independent variables when we make
replacements given below.

Now we list our results (Figure 1 shows the graphs that correspond
to various formulas):

\begin{itemize}

\item (a) {\bf Pentagon:} It is given by
\bea \textrm{Pen}[K_i,K_j,K_r,K] &=& {\cal
T}^{(N)}(\W\ell_{(i,j,r)}) \cdot D_i(\W\ell_{(i,j,r)})
D_j(\W\ell_{(i,j,r)}) D_r(\W\ell_{(i,j,r)}). \quad \Label{Pen-final}
\eea
where $\W\ell_{(i,j,r)}$ is given by (\ref{pen-ell-solution}) and
$D_i(\W\ell)$ is given by (\ref{Di-def}).

\item (b) {\bf Box:} It is given by
\bea \mathrm{Box}[K_{i},K_{j},K] & = & \frac{1}{2}\left({\cal
T}^{(N)}(\W\ell_{ij})\cdot D_{i}(\W\ell_{ij})D_{j}(\W\ell_{ij}) -
\sum_{r\neq i,j}{\textrm{Pen}[K_i,K_j,K_r,K]\over D_r(\W\ell_{ij})}
\right)
\Bigg|{\begin{matrix} \\
\left\{\scriptsize \begin{matrix} \bket{\ell} & \rightarrow &
\bket{P_{ji,2}}
\\ \ket{\ell} & \rightarrow  & \ket{P_{ji,1}}
\end{matrix}\right. \end{matrix}}    \nonumber\\ && + \{P_{ji,1} \leftrightarrow P_{ji,2}\}
\quad \Label{Box-final} \eea
where $\W\ell_{ij}$ are given by (\ref{box-ellij-I}) or
(\ref{box-ellij-II}), while $\ket{P_{ji,a}}$ and $\bket{P_{ji,a}}$
are given by (\ref{u-free-Pij}).

\item (c) {\bf Triangle:} The triangle coefficient is
\bea \textrm{Tri}[K_s,K] & = &
\frac{1}{2}\frac{(K^{2})^{N+1}}{(-\b\sqrt{1-u})^{N+1}(\sqrt{-4q_s^2
K^2})^{N+1}} \frac{1}{(N+1)! \vev{P_{s,1}~P_{s,2}}^{N+1}} \nonumber
\\ & & \frac{d^{N+1}}{d\tau^{N+1}}\left({\left\langle
\ell|K|\ell\right]^{N+1} \over (K^{2})^{N+1}} {\cal
T}^{(N)}(\W\ell)\cdot D_{s}(\W\ell)
\Bigg|{\scriptsize \begin{matrix} \\ \left\{\begin{matrix}
\bket{\ell} & \rightarrow & |Q_s(u)\ket{\ell} \quad\quad
\\ \ket{\ell} & \rightarrow  & \ket{P_{s,1}-\tau P_{s,2}}
\end{matrix}\right. \end{matrix}} +
\{P_{s,1}\leftrightarrow P_{s,2}\} \right)\Big|_{\tau\to 0} \qquad
\Label{Tri-final} \eea
where $\W\ell$ is given by (\ref{w-ell-form}), $Q_s(u)$ is given by
(\ref{R-Q-massive}), and $P_{s,a}$ is given by (\ref{ufree-spinor}).

It is worth to emphasize the ordering of replacement. We need to
replace $\bket{\ell} \rightarrow  |Q_s(u)\ket{\ell}$ first where
$\ket{\ell}$ shows up. Then we replace $\ket{\ell}  \rightarrow
\ket{P_{s,1}-\tau P_{s,2}}$. To denote this ordering, we have put
the first replacement above the second one. Similar understanding
should apply for later formulas.

\begin{figure}[t]
\centerline{\includegraphics[height=4cm]{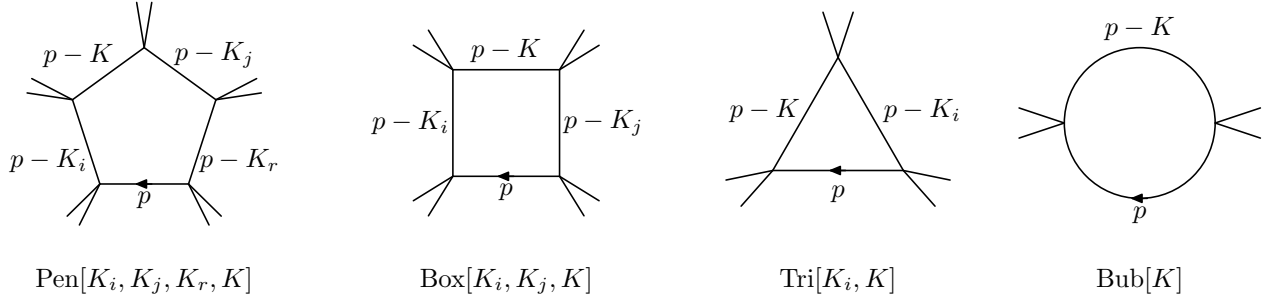}} \caption{The
figures corresponding to various formulas are shown. Notice the
order of the propagators can be changed. Our convention is that all
external momenta are taken outgoing.}
\end{figure}

\item (d) {\bf Bubble:}The bubble coefficient is\footnote{Here we
give the formula with general choice of auxiliary null momentum
$\eta$. For some special choice, the formula may be modified, which
can be found at the end of Section 4.}
\bea \textrm{Bub}[K] &=& (K^{2})^{N+1}\sum_{q=0}^N {(-1)^q \over q!}
{d^q \over d s^q} \left( {\cal B}_{N,N-q}^{(0)}(s) + \sum_{r=1}^k
\sum_{a=q}^N \left[ {\cal B}_{N,N-a}^{(r,a-q,1)}(s) - {\cal
B}_{N,N-a}^{(r,a-q,2)}(s)\right] \right) \Bigg|_{s\rightarrow0}
\qquad \Label{Bub-final} \eea
\bea {\cal B}_{N,t}^{(0)}(s) &=& {d^N \over d\tau^N}
\left[{(2\eta\cdot K)^{t+1}\over (t+1)(K^2)^{t+1}} {1\over N!
[\eta|\W\eta K|\eta]^N \vev{\ell~\eta}^{N+1}}
\left(\frac{\left\langle \ell|K|\ell\right]^{N}}{(K^{2})^{N}} {\cal
T}^{(N)}(\W\ell) \right)
\Bigg|{\scriptsize
\begin{matrix} \\ \left\{\begin{matrix} \bket{\ell} & \rightarrow &
|K+s\eta\ket{\ell}
\\ \ket{\ell} & \rightarrow  & |K-\tau\W\eta|\eta]
\end{matrix}\right. \end{matrix}} \right]
\Bigg|_{\tau\rightarrow0} \Label{Bnt0-final}  \eea
\bea {\cal B}_{N,t}^{(r,b,1)}(s) &=& {1 \over b!~
(\b\sqrt{1-u})^{b+1}~ (\sqrt{-4q_r^2
K^2})^{b+1}\vev{P_{r,1}~P_{r,2}}^b} {d^b \over d\tau^b}\left[
{1\over t+1} {\gb{\ell |\eta| P_{r,1}}^{t+1}\over\gb{\ell |K|
P_{r,1}}^{t+1}}{\vev{\ell |Q_r(u)\eta| \ell}^b \over \vev{\ell |\eta
K| \ell}^{N+1}} \right. \qquad \nonumber \\
&& \left. \times  \left(\frac{\left\langle
\ell|K|\ell\right]^{N+1}}{(K^{2})^{N+1}} {\cal T}^{(N)}(\W\ell)
\cdot D_{r}(\W\ell) \right)
\Bigg|{\scriptsize \begin{matrix} \\ \left\{\begin{matrix}
\bket{\ell} & \rightarrow & |K+s\eta\ket{\ell}~~~
\\ \ket{\ell} & \rightarrow  & \ket{P_{r,1}-\tau P_{r,2}}
\end{matrix}\right. \end{matrix}}
\right]\Bigg|_{\tau\rightarrow0}  \Label{Bnt1-final} \qquad \eea
\bea {\cal B}_{N,t}^{(r,b,2)}(s) &=& {1 \over b!~
(\b\sqrt{1-u})^{b+1}~ (\sqrt{-4q_r^2
K^2})^{b+1}\vev{P_{r,1}~P_{r,2}}^b} {d^b \over d\tau^b} \left[
{1\over t+1} {\gb{\ell |\eta| P_{r,2}}^{t+1}\over\gb{\ell |K|
P_{r,2}}^{t+1}} {\vev{\ell |Q_r(u)\eta| \ell}^b \over \vev{\ell
|\eta K| \ell}^{N+1}}  \right. \qquad \nonumber \\
&& \left. \times \left(\frac{\left\langle
\ell|K|\ell\right]^{N+1}}{(K^{2})^{N+1}} {\cal T}^{(N)}(\W\ell)
\cdot D_{r}(\W\ell) \right)
\Bigg|{\scriptsize \begin{matrix} \\ \left\{\begin{matrix}
\bket{\ell} & \rightarrow & |K+s\eta\ket{\ell}~~~
\\ \ket{\ell} & \rightarrow  & \ket{P_{r,2}-\tau P_{r,1}}
\end{matrix}\right. \end{matrix}}
\right]\Bigg|_{\tau\rightarrow0}  \Label{Bnt2-final} \qquad \eea
where $\W\ell$ is given by (\ref{w-ell-form}), $Q_r(u)$ is given by
(\ref{R-Q-massive}), $P_{r,a}$ is given by (\ref{ufree-null}), and
$\eta, \W\eta$ are arbitrary, generically chosen null vectors. The
summation over $r$ includes all the $D_r(\W\ell)$ that appear in the
denominator of ${\cal T}(\W\ell)$.
\end{itemize}

If $N\leq -2$, there are  contributions only from boxes and
pentagons. If $N\geq -1$, contributions from triangles will kick in,
and finally if $N\geq 0$, bubble contributions show up as well.
Similar observation has been made in \cite{BDDK}.

Our formula should be suitable for both analytic and numerical
calculations. For numerical calculation, since our result should be
polynomial of $u$, so we should have a way to get coefficients of
$u^a$ numerically. This can be done by using discrete Fourier
transformation method\cite{Mastrolia:2008jb}.

%%%%%%%%%%%%%%%%%%%%%%%%%%%%%%%%%%%%%%%%%%%%%%%%%%%%%%%%%%%%%%%%%
\section{D-dimensional unitarity method}
%%%%%%%%%%%%%%%%%%%%%%%%%%%%%%%%%%%%%%%%%%%%%%%%%%%%%%%%%%%%%%%%%

In this section, we briefly review the $D$-dimensional unitarity
method \cite{ABFKM06} and define various forms of tree level input.
In the process, we also establish our conventions and give
definitions of some variables and functions. Some preliminary
knowledge about spinor formalism \cite{ChineseMagic} and color
decompositions \cite{ColorDecomposition} can be found in two classic
reviews \cite{Mangano-Parke,DixonTASI}. Here we only emphasize that,
we use the ``twistor'' convention for the square bracket $[i~j]$, so
that $2 k_i\cdot k_j = \vev{i~j}[i~j]$.

%To switch to the QCD-literature convention for our discussion, we
%only need to use that
%
%\bea (K_i\cdot K_j)_{\textrm{QCD}} = -(K_i\cdot
%K_j)_{\textrm{twistor}}, \quad (s_{ij...k})_{\textrm{QCD}} =
%-(s_{ij...k})_{\textrm{twistor}}. \eea
%

\bigskip

Let's start with the following expression for a double-cut
($K^2$-channel) integral:
\bea  C[{\cal A}] \big|_{K^2-\textrm{channel}} & = &  - i
(4\pi)^{D/2} \int {d^{D} p \over (2\pi)^{D}}~ \delta(p^2-M_1^2)~
\delta((p-K)^2-M_2^2) ~ {\cal T}(p) . \quad \Label{cut-int-I} \eea
%
%In practical use, ${\cal T}(p)$ is the tree level input as
%
%\bea {\cal T}(p) = A_L^{\textrm{tree}}\times A_R^{\textrm{tree}}.
%\quad \eea
%
%To compare with the full amplitude, we can replace the
%$\delta$-function, which impose on-shellness, with propagators
%
%\bea  A |_{K-\textrm{cut}} & = &  \int {d^{4-2\eps}p}~ {1\over
%p^2-M_1^2}~{1\over (p-K)^2-M_2^2} ~ {\cal T}(p)  \qquad \eea
%
%Notice that this equation is valid only on the $K$-cut.
Compared with the full amplitude expressed in (\ref{BasisExp-III}),
i.e.
\bea {\cal A} = \sum_{n,i}~ c_{n,i}(\mu^2)\otimes I_n^{D,(i)}[1],
\quad \eea
the cut integral (\ref{cut-int-I}) can be understood as
\bea C[{\cal A}] \big|_{K^2-\textrm{channel}} = \sum_{n,i}~
c_{n,i}(\mu^2)\otimes C[I_n^{D,(i)}[1]] \big|_{K^2-\textrm{channel}}
, \quad \Label{cut-integral} \eea
where the cut master integral is
\bea C\left[I_n^D[1]\right] \big|_{K^2-\textrm{channel}} = - i
(4\pi)^{D/2} \int {d^{D} p \over (2\pi)^{D}}~{\delta(p^2-M_1^2)~
\delta((p-K)^2-M_2^2)  \over
 \prod_{i=1}^{n-2} ((p-K_i)^2-m_{i}^2)}.
\quad \Label{cut-n-scalar} \eea
Those master integral, which don't contain  two propagators that
correspond to the two $\delta$-functions, will not appear in the cut
integral (\ref{cut-integral}). By evaluating  different cut
integrals, we can get all coefficients $c_{n,i}(\mu^2)$.

%The advantage of the unitarity method is that we only need to
%evaluate the cut integral, in which the integral freedom is reduced
%by two $\delta$-functions. Then it is possible to develop some
%method to calculate the coefficient efficiently.

In  next subsection, we will briefly review how to simplify double
cut phase-space integration. This technique is the foundation of our
method. After these transformations, we can describe the structure
of  cut integrands, from which it is possible to read off
coefficients directly.

%%%%%%%%%%%%%%%%%%%%%%%%%%%%%%%%%%%%%%%%%%%%%%%%%%%%%%%%%%%%%
\subsection{Simplify  phase-space integral}
%%%%%%%%%%%%%%%%%%%%%%%%%%%%%%%%%%%%%%%%%%%%%%%%%%%%%%%%%%%%%

The internal loop momentum $p$ is $(D=4-2\eps)$-dimensional and can
be decomposed as
\bea p = \W\ell + \vec\mu ; \qquad \int d^{4-2\eps} p = \int
d^{-2\eps}\mu  \int d^4 \W \ell \ , \quad \eea
where $\W\ell$ is 4-dimensional and $\vec\mu$ is
$(-2\eps)$-dimensional. We can further decompose $\W\ell$ as
\bea \W\ell = \ell + z K, \qquad \ell^2 = 0 ; \qquad \int d^4\W\ell
= \int dz~d^4\ell~ \delta^{(+)}(\ell^2)~(2\ell\cdot K) , \qquad \eea
where $K$ is the momentum across the cut. $\ell$ is a massless
4-momentum, and can be expressed with spinor variables as
{\cite{CSW04-MHV}}
\bea \ell=t P_{\lambda\W\lambda}, \qquad P_{\la\W\la} = \la \W\la =
\ket{\ell}\bket{\ell} ; \qquad  \int d^4\ell \ \delta^{(+)}(\ell^2)
&=& \int \vev{\ell~d\ell}[\ell~d\ell] \int t ~dt . \eea
The corresponding measure has also been given.

The Lorentz-invariant phase-space (LIPS) of a double cut is defined
by inserting two $\delta$-functions representing the cut conditions:
\bea \int d^{4-2\eps} \Phi & \equiv & \int {d^{4-2\eps}p}~
\delta(p^2-M_1^2) \delta((p-K)^2-M_2^2) \\ &=&
\frac{(4\pi)^\eps}{\Gamma(-\eps)}~\int d\mu^2 ~(\mu^2)^{-1-\eps}
\int {d^4\W\ell}~ \delta(\W\ell^2-\mu^2-M_1^2) ~\delta((\W\ell-K)^2-\mu^2-M_2^2) \nonumber\\
&=& \frac{(4\pi)^\eps}{\Gamma(-\eps)}~\int d\mu^2 ~(\mu^2)^{-1-\eps}
\int dz~d^4\ell~ \delta^{(+)}(\ell^2)~(2\ell\cdot K) \nonumber\\
&& \qquad \times \delta(z(1-z)K^2+z(M_1^2-M_2^2)-M_1^2-\mu^2)
~\delta(-2\ell\cdot K + (1-2z)K^2+M_1^2-M_2^2). \nonumber \eea
We can firstly perform the integral over $z$ with $\delta$-function
$\delta(z(1-z)K^2+z(M_1^2-M_2^2)-M_1^2-\mu^2)$ to reach
\bea \int d^{4-2\eps} \Phi &=&
\frac{(4\pi)^\eps}{\Gamma(-\eps)}~\int d\mu^2 ~(\mu^2)^{-1-\eps}
\int d^4\ell~ \delta^{(+)}(\ell^2) ~\delta(-2\ell\cdot K +
(1-2z)K^2+M_1^2-M_2^2) \\
&=& \frac{(4\pi)^\eps}{\Gamma(-\eps)}~\int d\mu^2 ~(\mu^2)^{-1-\eps}
\int \vev{\ell~d\ell}[\ell~d\ell] \int t ~dt ~\delta(-2t K\cdot
P_{\lambda\W\lambda} + (1-2z)K^2+M_1^2-M_2^2) , \nonumber \eea
where $z$ is fixed by solving the $\delta$-function as
\bea z = { (K^2+M_1^2-M_2^2)- \sqrt{\Delta[K, M_1, M_2]- 4 K^2
\mu^2}\over 2 K^2}~, \qquad \Label{solve-z} \eea
with
\bea \Delta[K,M_1,M_2]\equiv (K^2)^2+(M_1^2)^2+(M_2^2)^2-2 K^2 M_1^2
-2 K^2 M_2^2- 2M_1^2 M_2^2 \ . \quad \Label{Delta-KMM} \eea
Then we can perform the integral over $t$ with the remaining
$\delta$-function to reach
\bea \int d^{4-2\eps} \Phi &=&
\frac{(4\pi)^\eps}{\Gamma(-\eps)}~\int d\mu^2 ~(\mu^2)^{-1-\eps}
\int \vev{\ell~d\ell}[\ell~d\ell] {(1-2z)K^2+M_1^2-M_2^2\over (2
K\cdot P_{\lambda\W\lambda})^2} ~,\quad \eea
and $t$ is
\bea t={(1-2z)K^2+M_1^2-M_2^2\over 2 K\cdot P_{\lambda\W\lambda}} .
\quad \Label{solve-t} \eea
For convenience, one can redefine the $\mu^2$-integral measure as
\bean \int d\mu^2 (\mu^2)^{-1-\eps} = \left( {
\Delta[K,M_1,M_2]\over 4 K^2}\right)^{-\eps} \int_0^1 du \
u^{-1-\eps}, \eean
where the relation between $u$ and $\mu^2$ is
given by
\bea u\equiv {4 K^2\mu^2 \over \Delta[K,M_1,M_2]}. \quad
\Label{u-def}\eea
Then we can rewrite $z,t$ as
\bea z ={ \a -\b \sqrt{1-u}\over 2}, \qquad t = {\b\sqrt{1-u} ~ {K^2
\over 2 K\cdot P_{\lambda\W\lambda}}} , \quad \Label{zt-sol-u} \eea
where
\bea \a = { K^2+M_1^2-M_2^2\over K^2}, \qquad \b={\sqrt{\Delta[K,
M_1, M_2]}\over K^2}. \quad \Label{ab-def}\eea
Notice that when $M_1=M_2=0$ we have $\a=\b=1$, thus reproducing the
massless case. A useful relation between $z$ and $u$ is the
following:
\bea (1-2z) +{ M_1^2-M_2^2\over K^2} = \b \sqrt{1-u}. \quad
\Label{z-rel-1}\eea

After these evaluations, the cut integral (\ref{cut-int-I}) is
transformed to
\bea  C[{\cal A}] &=& \frac{(4\pi)^\eps}{i \pi^{D/2}
\Gamma(-\eps)}~\int d\mu^2 ~(\mu^2)^{-1-\eps} \int
\vev{\ell~d\ell}[\ell~d\ell] {(1-2z)K^2+M_1^2-M_2^2\over (2 K\cdot
P_{\lambda\W\lambda})^2} {\cal
T}(p) \quad \Label{cut-int-II} \nonumber\\
&=& \frac{(4\pi)^\eps}{i \pi^{D/2} \Gamma(-\eps)}~\left( {
\Delta[K,M_1,M_2]\over 4 K^2}\right)^{-\eps} \int_0^1 du \
u^{-1-\eps} \int \vev{\ell~d\ell}[\ell~d\ell] ~\b\sqrt{1-u} ~ {K^2
\over (2 K\cdot P_{\lambda\W\lambda})^2} {\cal T}(p) . \qquad \eea
Here ${\cal T}(p)$ should be interpreted as
\bea {\cal T}(p) = {\cal T}(\W\ell~,\mu^2) = {\cal
T}(tP_{\la\W\la}+zK ~,\mu^2) = {\cal
T}(\ket{\ell},\bket{\ell},\mu^2) , \quad \eea
with $z$ and $t$ defined as in (\ref{zt-sol-u}). In the following,
we may write ${\cal T}(p)$ as ${\cal T}(\W\ell)$ where the
dependence on $\mu^2$ is implicitly included, with $\W\ell$ being
interpreted as
\bea \W\ell = tP_{\la\W\la}+zK = {K^2 \over \gb{\ell|K|\ell}}\left[
-\b\sqrt{1-u} \left( P_{\la \W\la} -{ K\cdot P_{\la \W\la}\over K^2}
K\right)- \a { K\cdot P_{\la \W\la}\over K^2} K \right] . \quad
\Label{well} \eea
%

%%%%%%%%%%%%%%%%%%%%%%%%%%%%%%%%%%%%%%%%%%%%%%%%%%%%%%%
\subsection{Standard form of the input}
%%%%%%%%%%%%%%%%%%%%%%%%%%%%%%%%%%%%%%%%%%%%%%%%%%%%%%%

For a double cut of  physical amplitudes, it is always possible (for
example from Feynman rule) to write ${\cal T}(p)$ as a sum of terms
of the following form
\bea {\cal T}^{(n)}_{sf}(\W\ell) = \frac{ \prod_{j=1}^{n+k}
(-2\W\ell\cdot P_j)}{\prod_{i=1}^k D_i(\W\ell) } , \quad
\Label{calT-Standard} \eea
where $D_i(\W\ell)$ is the propagator
\bea D_i(\W\ell) \equiv (\W\ell-K_i)^2-\mu^2-m_i^2 = -2 \W\ell\cdot
K_i+K_i^2+M_1^2-m_i^2 . \quad ~~~~~\Label{Di-def}\eea
We emphasize that the relation $\W\ell^2=M_1^2+\mu^2$ from the cut
constraint is always assumed in the input.

We will call the form of ${\cal T}(p)$ like (\ref{calT-Standard}) as
the \emph{standard form}, where there are only $D_i(\W\ell)$ in the
denominator. We will frequently use an important quantity -- the
\emph{degree} for $p$ (or equivalently for $\W\ell$). The degree of
${\cal T}(p)$ is defined as the degree of numerator minus the degree
of denominator. For the standard form (\ref{calT-NSP}), the degree
is $n$, which was shown explicitly in the superscript.

After simplifying the phase-space integration, the cut integral for
the standard form can be written in the form of (\ref{cut-int-II}):
\bea  \frac{(4\pi)^\eps}{i \pi^{D/2}\Gamma(-\eps)}~\left( {
\Delta[K,M_1,M_2]\over 4 K^2}\right)^{-\eps} \int_0^1 du \
u^{-1-\eps} \int \vev{\ell~d\ell}[\ell~d\ell]~\b\sqrt{1-u}~{
(K^2)^{n+1}\over \gb{\ell|K|\ell}^{n+2}} {\prod_{j=1}^{n+k}
\gb{\ell|R_j|\ell}\over \prod_{i=1}^k \gb{\ell|Q_i|\ell}} ~, \qquad
\Label{cut-int-sdform} \eea
where
\bean R_j(u) & \equiv & -\b(\sqrt{1-u}) \left(P_j-{P_j\cdot K\over
K^2}K \right) -\a{(P_j\cdot K)\over K^2} K, \\ Q_i(u) & \equiv & -\b
(\sqrt{1-u}) \left(K_i-{K_i\cdot K\over K^2}K \right) -
\left(\a{(K_i\cdot K)\over K^2}- {K_i^2+M_1^2-m_i^2\over K^2}
\right) K , \quad \eean
and $\a, \b$ are given by (\ref{ab-def}). We should notice the
equivalent relations at the integrand level as
\bea -2\W\ell\cdot P_j &=& {K^2 \over \gb{\ell|K|\ell}}
\gb{\ell|R_j|\ell} , \qquad
 D_i(\W\ell) = {K^2 \over \gb{\ell|K|\ell}}
\gb{\ell|Q_i|\ell}, \quad \Label{RQ-relation} \eea
and therefore
\bea {\cal T}^{(n)}_{sf}(\W\ell) &=& {(K^2)^{n}\over
\gb{\ell|K|\ell}^{n}} {\prod_{j=1}^{n+k} \gb{\ell|R_j|\ell}\over
\prod_{i=1}^k \gb{\ell|Q_i|\ell}}. \quad \eea
%
%With the input as standard form, we have formulas for calculation
%various coefficients, which are listed in Appendix \ref{}

For the cut integral with standard form input
(\ref{cut-int-sdform}), we can read off  coefficients directly by
using several formulas, which are reviewed in the Appendix
\ref{rev-formulas}.

%%%%%%%%%%%%%%%%%%%%%%%%%%%%%%%%%%%%%%%%%%%%%%%%%%%%%%%
\subsection{\label{generalT} General form of tree level input and spurious pole}
%%%%%%%%%%%%%%%%%%%%%%%%%%%%%%%%%%%%%%%%%%%%%%%%%%%%%%%

We now consider the general form of  inputs. In unitarity method,
the input needed is a collection of (on-shell) tree level
amplitudes. By using the on-shell recursion relations, very compact
expressions for the tree level input can be obtained. However, such
compact expressions usually contain some spurious poles.

A general form of ${\cal T}(p)$ that with a spurious pole can be
written as (for simplicity, we only consider the case of single
spurious pole, multi-spurious pole is a trivial generalization)
\bea {\cal T}_{sp}^{(N)}(\W\ell) = \sum_t c_t \frac{
\prod_{j=1}^{n_t+k_t+d} (-2\W\ell\cdot P_j)}{S_d \prod_{i=1}^{k_t}
D_i(\W\ell) } , \quad \Label{calT-SP-I} \eea
where $c_t$ are  coefficients that do not depend on $\W\ell$, but
may depend on $\mu^2$. $S_d$ is the spurious pole with degree $d$
and can be generally expressed as
\bea S_d & = & s_0+ \sum_{i}s_{i} (-2\W \ell\cdot V_{1,i})+
\sum_{i_1, i_2} s_{i_1,i_2}(-2\W \ell\cdot V_{2,i_1})(-2\W \ell\cdot
V_{2,i_2})+... \quad \Label{Sd-def}\\ & & +\sum_{i_1,\dots,i_d}
s_{i_1,...,i_d} (-2\W \ell\cdot V_{d,i_1})(-2\W \ell\cdot
V_{d,i_2})...(-2\W \ell\cdot V_{d,i_d}) . \nonumber \eea
% as
%
%\bea S_d = \sum_{t,\{i_t\}} { s_{i_1,i_2,...,i_t}(-2\W \ell\cdot
%V_{t,i_1})(-2\W \ell\cdot V_{t,i_2})...(-2\W \ell\cdot V_{t,i_t})}
%\eea
%
%and $\{i_t\}$ means the set $\{i_1,...,i_t\}$
The degree of ${\cal T}_{sp}^{(N)}(\W\ell)$ is $N$, which is defined
as the maximum of $n_t$
\bea N=\textrm{Max}\{n_t\}. \eea
After putting into (\ref{cut-int-sdform}), ${\cal
T}_{sp}^{(N)}(\W\ell)$ can be rewritten by using the relation
(\ref{RQ-relation}) as:
\bea {\cal T}_{sp}^{(N)}(\W\ell) = \sum_t c_t {(K^2)^{n_t}\over
\gb{\ell|K|\ell}^{n_t}} {\prod_{j=1}^{n_t+k_t+d}
\gb{\ell|R_j|\ell}\over {\WH S}_d \prod_{i=1}^{k_t}
\gb{\ell|Q_i|\ell}} , \qquad \Label{calT-SP-II} \eea
where
\bea {\WH S}_d & = & s_0 {\gb{\ell|K|\ell}^{d}\over (K^2)^{d}} +
\sum_{i} {\gb{\ell|K|\ell}^{d-1}\over (K^2)^{d-1}} s_{i}\gb{\ell|\W
V_{1,i}|\ell} + \sum_{i_1, i_2} {\gb{\ell|K|\ell}^{d-2}\over
(K^2)^{d-2}} s_{i_1,i_2} \gb{\ell|\W V_{2,i_1}|\ell} \gb{\ell|\W
V_{2,i_2}|\ell} +... \quad \Label{WHSd}\\ & & +\sum_{i_1,...,i_d}
s_{i_1,...,i_d} \gb{\ell|\W V_{d,i_1}|\ell}\gb{\ell|\W
V_{d,i_2}|\ell}...\gb{\ell|\W V_{d,i_d}|\ell} , \nonumber \eea
and $\W V_{t,i_z}$ is obtained from $V_{t,i_z}$ just as that $R_j$
obtained from $P_j$.

For a physical amplitude,  poles  other than $D_i(\W\ell)$ are all
\emph{spurious}. It is always possible to regroup  terms into an
expression free of spurious poles, i.e. a sum of standard forms:
\bea {\cal T}_{nsp}^{(N)}(\W\ell) = \sum_r b_r \frac{
\prod_{j=1}^{n_r+k_r} (-2\W\ell\cdot P_j)}{\prod_{i=1}^{k_r}
D_i(\W\ell) } = \sum_r b_r {(K^2)^{n_r}\over \gb{\ell|K|\ell}^{n_r}}
{\prod_{j=1}^{n_r+k_r} \gb{\ell|R_j|\ell}\over \prod_{i=1}^{k_r}
\gb{\ell|Q_i|\ell}} . \quad \Label{calT-NSP-III} \eea
$b_r$ are coefficients that do not depend on $\W\ell$, but may
depend on $\mu^2$. The degree, now the maximum of $n_r$, should also
be $N$.

An important observation is that: to go from ${\cal
T}_{sp}^{(N)}(\W\ell)$ to ${\cal T}_{nsp}^{(N)}(\W\ell)$, we only
need to apply some spinor identities to remove the spurious pole,
while $\W\ell$ must satisfy the two on-shell conditions:
\bea \W\ell^2-\mu^2-M_1^2 = 0 , \qquad (\W\ell-K)^2-\mu^2-M_2^2 = 0.
\quad \Label{cut-constraints}\eea
This observation will be important in the proof of the
generalization in Section \ref{general-formula}.

%%%%%%%%%%%%%%%%%%%%%%%%%%%%%%%%%%%%%%%%%%%%%%%%%%%%%%%%%%%%%%%%%
\section{Looking at the formulas}
%%%%%%%%%%%%%%%%%%%%%%%%%%%%%%%%%%%%%%%%%%%%%%%%%%%%%%%%%%%%%%%%%

The formulas we have known (collected in Appendix
\ref{rev-formulas}) are applicable for the standard form  ${\cal
T}_{sf}(p)$, in which the key requirement is that there is no
spurious pole, i.e. only propagators $D(\W\ell)$ appear in the
denominator. One may expect to remove the spurious pole in a general
input to reach the standard form, since it's possible in principle.
However, in general practice, this procedure is very complicated and
we do not have well-defined algorithm to do that. Besides, the most
compact expression of the tree level input usually has spurious
poles. After removing the spurious poles, the expression may expand
quite a lot, which leads the computation to be less efficient.
Therefore, it is important to be able to generalize formulas to the
case with spurious pole, while keeping the compactness of the tree
input.

In the following, we will study the previous known formulas for the
standard forms input from a new point of view, by which we can gain
some insights on how to generalize the formulas to the case with
spurious poles.

%%%%%%%%%%%%%%%
\subsection{Splitting of the integrand}
%%%%%%%%%%%%%%%

We start from the integrand in (\ref{cut-int-sdform})
\bea I & = &  {(K^2)^{n+1}\over \gb{\ell|K|\ell}^{n+2}}
{\prod_{j=1}^{k+n} \gb{\ell|R_j |\ell} \over \prod_{i=1}^k
\gb{\ell|Q_i |\ell}}~. \qquad\qquad \Label{Integrand}\eea
Then we do the splitting to reach \cite{{BF06}}
\bea I & = & \sum_{i=1}^k F_i(\la) {1\over
\gb{\ell|K|\ell}\gb{\ell|Q_i|\ell}} +\sum_{j=0}^n G_j(\la,\W\la)
{1\over \gb{\ell|K|\ell}^{2+j}}~, \qquad \Label{I-split}\eea
where
\bea F_i(\la) = {(K^2)^{n+1}\over \vev{\ell|K
Q_i|\ell}^{n+1}}{\prod_{j=1}^{k+n} \vev{\ell|R_j Q_i|\ell}\over
\prod_{t=1,t\neq i}^k \vev{\ell|Q_t Q_i |\ell}}~, \qquad
\Label{Fi}\eea
\bea \sum_{j=0}^n G_j(\la,\W\la) {1\over \gb{\ell|K|\ell}^{2+j}}=
 (K^2)^{n+1}\sum_{q=0}^n {(-)^q\over q!} {d^q
B_{n,n-q}(s)\over ds^q}\Big|_{s=0}~, \qquad
\Label{bubble-ourcase}\eea
with
\bea B_{n,t}(s) \equiv {\gb{\ell|\eta|\ell}^t\over
\gb{\ell|K|\ell}^{2+t}}{ \prod_{j=1}^{n+k}
\vev{\ell|R_j(K+s\eta)|\ell}\over \vev{\ell|\eta K|\ell}^n
\prod_{p=1}^k \vev{\ell| Q_p(K+s\eta)|\ell}}~. \qquad
\Label{Bnt-ourcase}\eea

The box and triangle contributions will come from the $F_i$ term in
(\ref{I-split}), while the $G_j$ term will contribute to bubble.
Taking the integrand into the cut integral, and extracting the
residues of various poles, we can get the final coefficients. With
the above splitting, we will know all coefficients by putting them
into following expressions:
\bea C[Q_i,Q_j,K] & = & {(K^2)^{n+2}\over 2}\left({\prod_{s=1}^{k+n}
\gb{P_{ji,1}|R_s |P_{ji,2}}\over \gb{P_{ji,1}|K
|P_{ji,2}}^{n+2}\prod_{t=1,t\neq i,j}^k \gb{P_{ji,1}|Q_t
|P_{ji,2}}}+ \{P_{ji,1}\leftrightarrow P_{ji,2}\}
\right),~~\Label{beg-box-exp}\eea
for boxes defined by $Q_i, Q_j, K$,
\bea C[Q_s,K] & = & {(K^2)^{n+1}\over
2(\sqrt{\Delta_s})^{n+1}}\frac{1}{(n+1)!
\vev{P_{s,1}~P_{s,2}}^{n+1}} \nonumber
\\ & & \frac{d^{n+1}}{d\tau^{n+1}}\left({\prod_{j=1}^{k+n}
\vev{P_{s,1}-\tau P_{s,2} |R_j Q_s|P_{s,1}-\tau P_{s,2}}\over
\prod_{t=1,t\neq s}^k \vev{P_{s,1}-\tau P_{s,2}|Q_t Q_s
|P_{s,1}-\tau P_{s,2}}} + \{P_{s,1}\leftrightarrow
P_{s,2}\}\right)\Bigg|_{\tau\to 0}, \qquad \Label{beg-tri-exp}\eea
for triangles defined by $Q_s, K$ and finally for bubble
\bea
 C[K] = (K^2)^{n+1} \sum_{q=0}^n {(-1)^q\over q!} {d^q \over
ds^q}\left.\left( {\cal B}_{n,n-q}^{(0)}(s)+\sum_{r=1}^k\sum_{a=q}^n
\left({\cal B}_{n,n-a}^{(r;a-q;1)}(s)-{\cal
B}_{n,n-a}^{(r;a-q;2)}(s)\right)\right)\right|_{s=0},~~~~~\Label{beg-Re-gen-n}
\eea
where definition of various variables and functions can be found in
Appendix \ref{rev-formulas}. The key observation is that {\bf all
coefficients are just the sum or difference of relative residues of
pole.}  In $F_i(\la)$, the poles from $\vev{\ell|Q_t Q_i |\ell}$
contribute to box, and those from $\vev{\ell|K Q_i|\ell}$ contribute
to triangle. Bubbles are from the poles in $G_j$, or more precisely,
${\cal B}_{n,t}^{(0)}(s)$ is from the poles $\vev{\ell|\eta
K|\ell}$, and ${\cal B}_{n,t}^{(r,b,1)}(s), {\cal
B}_{n,t}^{(r,b,2)}(s)$ are from the poles $\vev{\ell | Q_p (K+s\eta)
| \ell}$. Details of the derivations can be found in
\cite{BF06,BF07}

Let us recall how do we get above formulas. Starting from standard
input, we algebraically simplify it by using splitting technique.
Then {\sl we perform phase space integrations carefully and add all
contributions from physical poles together}. If now the input has
spurious poles, the same method tells us that we need to perform
phase space integration with spurious poles. However, we can not do
that in general. The reason is following: (1) The form of spurious
poles can be arbitrary and we do not know how to write them down;
(2) To perform the phase space integration, we need to write it into
total derivative and then take residues. For a given spurious pole,
we do not know how to perform both manipulations.

Because these reasons, it seems impossible to find results by
standard method. However, considering these  poles are all
\emph{spurious}, this means that even we are able to extract the
residues of  spurious poles term by term, their sum should vanish
from physical point of view. In this sense, we can expect that the
formulas can be generalized to the general form without many
modifications. Then the question is how to do it? Which new method
will enable us to do it?

To be able to proceed, we find that we need to take a new
perspective in understanding above results. Remembering that  when
the spurious poles show up, there are at least two terms with
spurious pole, so the added result will cancel out. In another word,
we will have following equations (we assume there is only one term
of standard form for simplicity):
\bea {\cal T}(p) & \equiv &  { \prod_{j=1}^{n+k} (-2\W\ell\cdot P_j)
\over \prod_{i=1}^k D_i(\W\ell) }
 \qquad \Label{I-form-1}\\
& = & \sum_{t,\{i_t\}} { s_{i_1,i_2,...,i_t}(-2\W \ell\cdot
V_{t,i_1})(-2\W \ell\cdot V_{t,i_2})...(-2\W \ell\cdot
V_{t,i_t})\prod_{j=1}^{n+k} (-2\W\ell\cdot P_j) \over S_d
\prod_{i=1}^k
D_i(\W\ell) } \qquad \Label{I-form-2} \\
& = & \sum_{t} c_t { \prod_{j=1}^{n_t+k_t} (-2\W\ell\cdot
P_{j})\over S_d \prod_{i=1}^{k_t} D_{i}(\W\ell) } \qquad
\Label{I-form-3}\eea
where $S_d$ is given in (\ref{Sd-def})and $\{i_t\}$ means the set
$\{i_1,...,i_t\}$. Among these three forms, the first form is the
one without the spurious pole, while the third form is the one
obtained by using on-shell tree level recursion relation. The second
one is the bridge bring the first one to the third one.

After putting into (\ref{cut-int-sdform}),  we have  following three
forms for the integrand
\bea I[\ket{\ell},\bket{\ell}] & \equiv & {(K^2)^{n+1} \over
\gb{\ell|K|\ell}^{n+2}} {\prod_{j=1}^{n+k} \gb{\ell|R_j |\ell} \over
\prod_{i=1}^k \gb{\ell|Q_i |\ell}} \qquad \Label{I-t-form-1}\\
& = & \sum_{t,\{i_t\}} s_{i_1,i_2,...,i_t} {(K^2)^{n+1+t-d} \over
\gb{\ell|K|\ell}^{n+2+t-d}} {\prod_{z=1}^{t} \gb{\ell|\W V_{t,i_z}
|\ell}\prod_{j=1}^{n+k} \gb{\ell|R_j |\ell} \over
\widehat{S}_d[\ket{\ell},|\ell] ]\prod_{i=1}^k \gb{\ell|Q_i |\ell}}
\qquad \Label{I-t-form-2}
\\
& = & \sum_{t} c_t {(K^2)^{n_t+1-d} \over
\gb{\ell|K|\ell}^{n_t+2-d}} {\prod_{j=1}^{n_t+k_t} \gb{\ell|R_j
|\ell} \over \widehat{S}_d[\ket{\ell},|\ell]]\prod_{i=1}^{k_t}
\gb{\ell|Q_i |\ell}} \qquad \Label{I-t-form-3}\eea
which should be considered as function of $\ket{\ell}$ and
$\bket{\ell}$. {\bf It is important that we have treated
$\ket{\ell}$ and $\bket{\ell}$ as independent variables}.

%%%%%%%%%%%%%%%%%
\subsection{The splitting result for box and triangle}
%%%%%%%%%%%%%%%%%

The next step is to find the expression contributing to various
coefficients.

Let us start from the box and triangle. From the form
(\ref{Integrand}) we can see that the relative expression (i.e., for
box and triangle having $Q_i$) is
\bea  F_i^{(I)} & = &  { (K^2)^{n+1} \over \vev{\ell|K
Q_i|\ell}^{n+1}}{\prod_{j=1}^{n+k} \vev{\ell|R_j Q_i|\ell}\over
\prod_{t=1,t\neq i}^k \vev{\ell|Q_t Q_i |\ell}}~. \qquad
\Label{Fi-form-1}\eea
Comparing (\ref{Fi-form-1}) and (\ref{Integrand}) we found that it
is obtained from (\ref{Integrand}) by following manipulations:
\begin{itemize}

\item (a) Multiplying $\gb{\ell|K|\ell}$ and $\gb{\ell|Q_i|\ell}$ at
(\ref{Integrand});

\item (b) Then replacing $\bket{\ell}\to Q_i\ket{\ell}$.

\end{itemize}

Using this observation, we can get two results from
(\ref{I-t-form-2}) and (\ref{I-t-form-3}) as following
\bea F_i^{(II)} & = & \sum_{t,\{i_t\}} s_{i_1,i_2,...,i_t}
{(K^2)^{n+1+t-d} \over \vev{\ell|K Q_i|\ell}^{n+1+t-d}}
{\prod_{z=1}^{t} \vev{\ell|\W V_{t,i_z} Q_i|\ell}\prod_{j=1}^{n+k}
\vev{\ell|R_j Q_i|\ell} \over
\widehat{S}_d[\ket{\ell},Q_i\ket{\ell}]\prod_{\W i=1,\W i\neq i}^k
\vev{\ell|Q_{\W i} Q_i|\ell}}~~~\Label{Fi-form-2}\eea
and
\bea F_i^{(III)} & = & \sum_{t} c_t {(K^2)^{n_t+1-d} \over
\vev{\ell|K Q_i|\ell}^{n_t+1-d}} {\prod_{j=1}^{n_t+k_t}
\vev{\ell|R_j Q_i |\ell}\vev{\ell|Q_i Q_i|\ell} \over
\widehat{S}_d[\ket{\ell},Q_i\ket{\ell}]\prod_{\W i=1}^{k_t}
\vev{\ell|Q_{\W i} Q_i |\ell}}~~~\Label{Fi-form-3}\eea
It is worth to give a remark regarding the expression
(\ref{Fi-form-3}). In this formula we have factor $\vev{\ell|Q_i
Q_i|\ell}=0$ in numerator. If there is same factor $\vev{\ell|Q_i
Q_i|\ell}$ in denominator, then both factors will cancel each other,
but if there is no such factor in denominator, we know immediately
that this term contributes zero. A subtle point for box formula will
be discussed in Section \ref{subtle-box}.

Since three expressions (\ref{I-t-form-1}), (\ref{I-t-form-2}) and
(\ref{I-t-form-3}) are equal to each other at the algebraic level.
After the same algebraic manipulation, we should have that three
expressions (\ref{Fi-form-1}), (\ref{Fi-form-2}) and
(\ref{Fi-form-3}) are equal to each other too.

Now we can read out the expression for the box and triangle
coefficients.

%%%%%%%%%%%%%%%%%%%%
\subsection{Read the coefficients}
%%%%%%%%%%%%%%%%%%%%

The coefficients of box from (\ref{Fi-form-1}) is
\bea C[Q_i, Q_j,K]^{(I)} & = & {1\over 2} \left.{(K^2)^{n+2} \over
\vev{\ell|K Q_i|\ell}^{n+2}}{\prod_{s=1}^{n+k} \vev{\ell|R_s
Q_i|\ell}\over \prod_{t=1,t\neq i,j}^k \vev{\ell|Q_t Q_i
|\ell}}\right|_{\ket{\ell}\to \ket{P_{ji,1}}}\nonumber \\
& & + {1\over 2} \left.{(K^2)^{n+2} \over \vev{\ell|K
Q_i|\ell}^{n+2}}{\prod_{s=1}^{n+k} \vev{\ell|R_s Q_i|\ell}\over
\prod_{t=1,t\neq i,j}^k \vev{\ell|Q_t Q_i
|\ell}}\right|_{\ket{\ell}\to
\ket{P_{ji,2}}}~~~\Label{Box-form-1}\eea
which is equivalent to (\ref{beg-box-exp}) by noticing that
\bea \frac{\vev{P_{ji,1}|X Q_i(u)|P_{ji,1}}}{\vev{P_{ji,1}|X'
Q_i(u)|P_{ji,1}}}
=\frac{\gb{P_{ji,1}|X|P_{ji,2}}}{\gb{P_{ji,1}|X'|P_{ji,2}}}. \eea
Compared to (\ref{Fi-form-1}), the coefficient can be obtained by
following algebraic actions:
\begin{itemize}

\item (a) Multiplying $ {K^2\vev{\ell|Q_j Q_i|\ell}\over 2 \vev{\ell|K
Q_i|\ell}}$;

\item (b) Sum up two terms with $\ket{\ell}\to \ket{P_{ji,1}(u)}$ and $\ket{\ell}\to
\ket{P_{ji,2}(u)}$.

\end{itemize}

Now it is obviously we should apply same actions to
(\ref{Fi-form-2}) and (\ref{Fi-form-3}) to get
\bea & & C[Q_i, Q_j,K]^{(II)}  =  \sum_{t,\{i_t\}}
{s_{i_1,i_2,...,i_t}\over 2} {(K^2)^{n+2+t-d}\over \vev{\ell|K
Q_i|\ell}^{n+2+t-d}} \quad \Label{Box-form-2}\\
& &\left\{ \left.\left({\prod_{z=1}^{t} \vev{\ell|\W V_{t,i_z}
Q_i|\ell}\prod_{s=1}^{n+k} \vev{\ell|R_s Q_i |\ell} \over
\widehat{S}_d[\ket{\ell},Q_i\ket{\ell}]\prod_{\W i=1,\W i\neq i,j}^k
\vev{\ell|Q_{\W i} Q_i|\ell}}\right)\right|_{\ket{\ell}\to
\ket{P_{ji,1}}}+\left.\left({\prod_{z=1}^{t} \vev{\ell|\W V_{t,i_z}
Q_i|\ell}\prod_{s=1}^{n+k} \vev{\ell|R_s Q_i |\ell} \over
\widehat{S}_d[\ket{\ell},Q_i\ket{\ell}]\prod_{\W i=1,\W i\neq i,j}^k
\vev{\ell|Q_{\W i} Q_i|\ell}}\right)\right|_{\ket{\ell}\to
\ket{P_{ji,2}}}\right\} \nonumber \eea
and
\bea C[Q_i, Q_j,K]^{(III)}  &=& \sum_{t} {c_t \over2}
{(K^2)^{n_t+2-d}\over \vev{\ell|K Q_i|\ell}^{n_t+2-d}} \left\{
\left.\left({\prod_{s=1}^{n_t+k_t} \vev{\ell|R_s Q_i
|\ell}\vev{\ell|Q_i Q_i|\ell} \vev{\ell|Q_j Q_i|\ell}\over
\widehat{S}_d[\ket{\ell},Q_i\ket{\ell}]\prod_{\W i=1}^{k_t}
\vev{\ell|Q_{\W i} Q_i |\ell}}\right)\right|_{\ket{\ell}\to
\ket{P_{ji,1}}} \right. \quad \nonumber\\ &&
\left. +\left.\left({\prod_{s=1}^{n_t+k_t} \vev{\ell|R_s Q_i
|\ell}\vev{\ell|Q_i Q_i|\ell} \vev{\ell|Q_j Q_i|\ell}\over
\widehat{S}_d[\ket{\ell},Q_i\ket{\ell}]\prod_{\W i=1}^{k_t}
\vev{\ell|Q_{\W i} Q_i |\ell}}\right)\right|_{\ket{\ell}\to
\ket{P_{ji,2}}}\right\} \quad  \Label{Box-form-3} \eea

Formula (\ref{Box-form-3}) is the algebraic expression for box in
the presence of spurious poles. We got it by various algebraic
replacements starting from input tree-level amplitudes. We can do
similar algebraic replacements to get algebraic expressions for
triangles and bubbles too. For later two cases,
 there are two complicities compared to box.  First  there is operation of
taking derivatives. Second,  formulas depend explicitly on the
degree of the tree level inputs. These two issues can be solved with
some considerations. In the next section, we will give a rigorous
proof for the generalization of these formulas. The final formulas
will be equivalent to those obtained by naive substitutions. Before
going to that, let's have a look at an example demonstrating some
subtle point regarding the box coefficients.

%%%%%%%%%%%%%%%%%%%
\subsection{\label{subtle-box} Subtle point regarding to box coefficients}
%%%%%%%%%%%%%%%%%%%

In the standard form we know that if one term contributes to
triangle coefficient $C[Q_i,K]$, it must have $D_i$ in its
denominator. Similarly if one term contributes to box coefficients
$C[Q_i,Q_j,K]$, it must have $D_i, D_j$ in its denominator.

If the form is not standard, i.e., having spurious pole, above
observation will be different. For triangle, above observation is
not modified, i.e., if one term contributes to triangle coefficient
$C[Q_i,K]$, it must have $D_i(\W\ell)$ in its denominator. However,
for the box, the above observation is not true anymore: one term can
contribute to $C[Q_i,Q_j,K]$ even it has only  $D_i$ or  $D_j$ in
its denominator (but at least one of two).

Let us consider one simple example. It is given by
\bea {\cal T} & = & {1\over (\ell-K_1)^2 (\ell+K_4)^2},~~~~
K_1^2=K_4^2=0 \eea
for cut $K_{12}$. Here we do everything in pure 4D so we do not have
$u$ floating around. It is obviously the only contribution is box
 and no others. However we can rewrite it as
\bea {\cal T} & = & {1\over (2\ell\cdot K_{41})(\ell-K_1)^2}-
{1\over
(2\ell\cdot K_{41})(\ell+K_4)^2} ~, \nonumber \\
{\cal T}_1 & = &-{1\over (-2\ell\cdot K_{41})(\ell-K_1)^2},~~~{\cal
T}_2= {1\over (-2\ell\cdot K_{41})(\ell+K_4)^2}~, \qquad \eea
where $(2\ell\cdot K_{41})$ is the spurious pole. Now let us apply
our general method to find various coefficients. Before doing so,
let us list following quantities (we have set $z=u=0$)
\bean Q_1 & = & - K_1, \qquad Q_2=K_4, \qquad \W V=-K_{41}. \qquad
\eean

First let us start from box. Spurious pole shows up in two terms,
one with propagator $(\ell-K_1)^2$ and another one with propagator
$(\ell+K_4)^2$, thus {\bf we need to consider all possible
combinations of boxes in these two sets.} In our simple case, these
is only one option: box with $(\ell-K_1)^2$ and $(\ell+K_4)^2$.
There are two terms, both with $k_t=1, d=1$, and $n_t=-2$.

Let us apply formula (\ref{Box-form-3}) with the choice $i=1, j=2$.
For the second term ${\cal T}_2$, the contribution is
\bean \textrm{Box}_2 & = & {1\over 2}\left({\vev{\ell|Q_1
Q_1|\ell}\over \vev{\ell|\W V Q_1|\ell}}\right)_{\ket{\ell}\to
\ket{P_1}}+{1\over 2} \left({\vev{\ell|Q_1 Q_1|\ell}\over
\vev{\ell|\W V Q_1|\ell}}\right)_{\ket{\ell}\to \ket{P_2}}=0\eean
This can be seen immediately since the denominator does not have
$D_1$ propagator. For the second term ${\cal T}_1$, the contribution
is
\bean \textrm{Box}_1 & = & -{1\over 2} \left({\vev{\ell|Q_2
Q_1|\ell}\over \vev{\ell|\W V Q_1|\ell}}\right)_{\ket{\ell}\to
\ket{P_1}}- {1\over 2}\left({\vev{\ell|Q_2 Q_1|\ell}\over
\vev{\ell|\W V Q_1|\ell}}\right)_{\ket{\ell}\to \ket{P_2}}=1 \eean
Adding two contributions together we see that indeed we have
reproduced the right box coefficient.

However, there is an subtle point regarding the calculation of
${\cal T}_1$ in this example. In fact, when we put in the solution
$P_1, P_2$ into $\vev{\ell|Q_2 Q_1|\ell}$ and $\vev{\ell|\W V
Q_1|\ell}$, both are zero, so we need to take proper limit. {\sl The
point is that even $\ell$ be arbitrary we have $\vev{\ell|Q_2
Q_1|\ell}=\vev{\ell|\W V Q_1|\ell}$ and $\vev{\ell|Q_1
Q_1|\ell}=0$}.

Now let us move to triangle, since $n=-2$ we see that triangle
coefficients are zero as we familiar with (when the power of
derivative is negative, we should take it as inserting
$\tau^{|a|}\to 0$). For bubble, we have $n=-2$, again the
contribution is zero.

It is worth to emphasize again that above subtle point is related to
box coefficients only.

%%%%%%%%%%%%%%%%%%%%%%%%%%%%%%%%%%%%%%%%%%%%%%%%%%%%%%%%%%%%%%%%%
\section{\label{general-formula} Formulas with spurious pole}
%%%%%%%%%%%%%%%%%%%%%%%%%%%%%%%%%%%%%%%%%%%%%%%%%%%%%%%%%%%%%%%%%

Now we generalize the formulas with a rigorous proof. As a
byproduct, we will obtain more compact expressions for the formulas.
We will take two steps. First we reformulate the formulas for the
standard form input into a more compact form, in which the tree
level input is inserted directly. Then with this new presentation,
it will be easy to generalize the formulas to the tree level input
in the presence of spurious poles.

%%%%%%%%%%%%%%%
\subsection{Reformulate the formulas}
%%%%%%%%%%%%%%%

We first reformulate the formulas for the tree level input being
standard form. Using the relation (\ref{RQ-relation})
\bea -2\W\ell\cdot P_j &=& {K^2 \over \gb{\ell|K|\ell}}
\gb{\ell|R_j|\ell} , \qquad
 D_i(\W\ell) = {K^2 \over \gb{\ell|K|\ell}}
\gb{\ell|Q_i|\ell}, \quad \nonumber \eea
we can reformulate the box formula (\ref{box-exp}) as
\bea C[Q_i,Q_j,K] & = & {1\over 2}\left({(K^2)^{n+2}
\prod_{s=1}^{k+n} \gb{\ell|R_s |\ell}\over \gb{\ell|K
|\ell}^{n+2}\prod_{t=1,t\neq
i,j}^k \gb{\ell|Q_t |\ell}} \Bigg|{\scriptsize \begin{matrix} \\
\left\{\begin{matrix} \bket{\ell} & \rightarrow & \bket{P_{ji,2}}
\\ \ket{\ell} & \rightarrow  & \ket{P_{ji,1}}
\end{matrix}\right. \end{matrix}} + \{P_{ji,1}\leftrightarrow
P_{ji,2}\} \right) \nonumber\\
& = & {1\over 2}\left({\cal
T}^{(n)}_{sf}(\W\ell)\cdot D_i(\W\ell)\cdot D_j(\W\ell) \Bigg|{\begin{matrix} \\
\left\{\scriptsize \begin{matrix} \bket{\ell} & \rightarrow &
\bket{P_{ji,2}}
\\ \ket{\ell} & \rightarrow  & \ket{P_{ji,1}}
\end{matrix}\right. \end{matrix}} +
\{P_{ji,1}\leftrightarrow P_{ji,2}\} \right) \qquad \Label{Box-new}
\eea
Similarly, for the triangle formula (\ref{tri-exp}), we can get
\bea C[Q_s,K] & = & {(K^2)^{n+1}\over
2(\sqrt{\Delta_s})^{n+1}}\frac{1}{(n+1)!
\vev{P_{s,1}~P_{s,2}}^{n+1}} \nonumber
\\ & & \frac{d^{n+1}}{d\tau^{n+1}}\left( \frac{\left\langle
\ell|K|\ell\right]^{n+1}}{(K^{2})^{n+1}} {\cal T}^{(n)}_{sf}(\W\ell)
\cdot D_{s}(\W\ell)
\Bigg|{\scriptsize \begin{matrix} \\ \left\{\begin{matrix}
\bket{\ell} & \rightarrow & |Q_s\ket{\ell} \quad\quad~~~
\\ \ket{\ell} & \rightarrow  & \ket{P_{s,1}-\tau P_{s,2}}
\end{matrix}\right. \end{matrix}} +
\{P_{s,1}\leftrightarrow P_{s,2}\} \right)\Big|_{\tau\to 0}, \qquad
\Label{Tri-new} \eea
and for bubble (\ref{bub-exp})--(\ref{cal-B-r-2}), we have
\bea {\cal B}_{n,t}^{(0)}(s) &=& {d^n \over d\tau^n}
\left[{(2\eta\cdot K)^{t+1}\over (t+1)(K^2)^{t+1}} {1\over n!
[\eta|\W\eta K|\eta]^n \vev{\ell~\eta}^{n+1}}
\left(\frac{\left\langle \ell|K|\ell\right]^{n}}{(K^{2})^{n}} {\cal
T}^{(n)}_{sf}(\W\ell) \right)
\Bigg|{\scriptsize
\begin{matrix} \\ \left\{\begin{matrix} \bket{\ell} & \rightarrow &
|K+s\eta\ket{\ell}
\\ \ket{\ell} & \rightarrow  & |K-\tau\W\eta|\eta]
\end{matrix}\right. \end{matrix}} \right]
\Bigg|_{\tau\rightarrow0} \Label{Bnt0-new}  \eea
\bea {\cal B}_{n,t}^{(r,b,1)}(s) &=& {(-1)^{b+1} \over b!
(\sqrt{\Delta_r})^{b+1}\vev{P_{r,1}~P_{r,2}}^b}
{d^b \over d\tau^b} \nonumber\\
&& \left[ {1\over t+1} {\gb{\ell |\eta| P_{r,1}}^{t+1}\over\gb{\ell
|K| P_{r,1}}^{t+1}} {\vev{\ell |Q_r\eta| \ell}^b \over \vev{\ell
|\eta K| \ell}^{n+1}} \left(\frac{\left\langle
\ell|K|\ell\right]^{n+1}}{(K^{2})^{n+1}} {\cal T}^{(n)}_{sf}(\W\ell)
\cdot D_{r}(\W\ell) \right)
\Bigg|{\scriptsize \begin{matrix} \\ \left\{\begin{matrix}
\bket{\ell} & \rightarrow & |K+s\eta\ket{\ell}~~~
\\ \ket{\ell} & \rightarrow  & \ket{P_{r,1}-\tau P_{r,2}}
\end{matrix}\right. \end{matrix}}
\right]\Bigg|_{\tau\rightarrow0}  \Label{Bnt1-new} \eea
\bea {\cal B}_{n,t}^{(r,b,2)}(s) &=& {(-1)^{b+1} \over b!
(\sqrt{\Delta_r})^{b+1}\vev{P_{r,1}~P_{r,2}}^b}
{d^b \over d\tau^b} \nonumber\\
&& \left[ {1\over t+1} {\gb{\ell |\eta| P_{r,2}}^{t+1}\over\gb{\ell
|K| P_{r,2}}^{t+1}} {\vev{\ell |Q_r\eta| \ell}^b \over \vev{\ell
|\eta K| \ell}^{n+1}} \left(\frac{\left\langle
\ell|K|\ell\right]^{n+1}}{(K^{2})^{n+1}} {\cal T}^{(n)}_{sf}(\W\ell)
\cdot D_{r}(\W\ell) \right)
\Bigg|{\scriptsize \begin{matrix} \\ \left\{\begin{matrix}
\bket{\ell} & \rightarrow & |K+s\eta\ket{\ell}~~~
\\ \ket{\ell} & \rightarrow  & \ket{P_{r,2}-\tau P_{r,1}}
\end{matrix}\right. \end{matrix}}
\right]\Bigg|_{\tau\rightarrow0}  \Label{Bnt2-new} \eea

We can find that the formulas actually preserve the structure of
tree level input ${\cal T}^{(n)}_{sf}(\W\ell)$ quite well. The main
structure of the formulas is that: first make a substitution for
$\bket{\ell}$, then for $\ket{\ell}$ , and finally do some
operations at algebraic level.

The substitution for $\ket{\ell}$ and $\bket{\ell}$ is equivalent to
a substitution for $\W\ell$, by using the relation (\ref{well})
\bea \W\ell = {K^2 \over \gb{\ell|K|\ell}}\left[ -\b\sqrt{1-u}
\left( P_{\la \W\la} -{ K\cdot P_{\la \W\la}\over K^2} K\right)- \a
{ K\cdot P_{\la \W\la}\over K^2} K \right] . \quad \eea
It is not difficult to find that, with these substitutions, $\W\ell$
 satisfies the two on-shell conditions (\ref{cut-constraints}),
because $P_{\la\W\la}$ under the substitution satisfies the massless
condition. For box this is obvious. For triangle and bubble, we have
\bea P_{\la\W\la}^2 \propto \vev{\ell|\ell} &=& \vev{P_{s,1(2)}-\tau
P_{s,2(1)}|P_{s,1(2)}-\tau P_{s,2(1)}} = -\tau
(\vev{P_{s,1}~P_{s,2}}+\vev{P_{s,2}~P_{s,1}}) = 0 . \eea
%

%%%%%%%%%%%%%%%%%%%%%%%%%%%%%%%%%%%%%%%%%%%%%%%%%%%%%%%%%%%%%%%%%
\subsection{\label{idea-proof} Generalize to general tree level input}
%%%%%%%%%%%%%%%%%%%%%%%%%%%%%%%%%%%%%%%%%%%%%%%%%%%%%%%%%%%%%%%%%

To generalize the formulas, a naive conjecture is that: for a
general form of ${\cal T}(p)$ with degree $N$, the formulas
(\ref{Box-new})--(\ref{Bnt2-new}) are unchanged, but only with the
following substitution
\bean {\cal T}^{(n)}_{sf}(\W\ell) \rightarrow {\cal T}(p), \qquad n
\rightarrow N , \quad \Label{substitute-calT} \eean
where
\bea {\cal T}(p) = {\cal T}_{sp}^{(N)}(\W\ell) = \sum_t c_t \frac{
\prod_{j=1}^{n_t+k_t+d} (-2\W\ell\cdot P_j)}{S_d \prod_{i=1}^{k_t}
D_i(\W\ell) } = \sum_t c_t {(K^2)^{n_t}\over \gb{\ell|K|\ell}^{n_t}}
{\prod_{j=1}^{n_t+k_t+d} \gb{\ell|R_j|\ell}\over {\WH S}_d
\prod_{i=1}^{k_t} \gb{\ell|Q_i|\ell}}, \quad \Label{calT-SP} \eea
or equivalently in a form without spurious pole,
\bea {\cal T}(p) = {\cal T}_{nsp}^{(N)}(\W\ell) = \sum_r b_r {\cal
T}^{(n_r)}_{sf}(\W\ell) = \sum_r b_r \frac{ \prod_{j=1}^{n_r+k_r}
(-2\W\ell\cdot P_j)}{\prod_{i=1}^{k_r} D_i(\W\ell) } = \sum_r b_r
{(K^2)^{n_r}\over \gb{\ell|K|\ell}^{n_r}} {\prod_{j=1}^{n_r+k_r}
\gb{\ell|R_j|\ell}\over \prod_{i=1}^{k_r} \gb{\ell|Q_i|\ell}} .
\quad \Label{calT-NSP} \eea
The degree $N$ of ${\cal T}(p)$ is defined as the maximum of $n_t$
or $n_r$:
\bea N=\textrm{Max}\{n_t\}=\textrm{Max}\{n_r\}. \eea
As mentioned at the end of Section \ref{generalT}, the important
observation is that: to go from ${\cal T}_{sp}^{(N)}(\W\ell)$ to
${\cal T}_{nsp}^{(N)}(\W\ell)$, we only need to apply some spinor
identities to remove the spurious pole, while $\W\ell$ must satisfy
the two on-shell conditions (\ref{cut-constraints}).

We now want to prove the conjecture. First we can see that if the
formulas is true for the substitution
\bea {\cal T}^{(n)}_{sf}(\W\ell) \rightarrow {\cal
T}_{nsp}^{(N)}(\W\ell), \qquad n \rightarrow N , \quad
\Label{sub-nsp} \eea
then it will also be true for the substitution
\bea {\cal T}^{(n)}_{sf}(\W\ell) \rightarrow {\cal
T}_{sp}^{(N)}(\W\ell), \qquad n \rightarrow N . \quad \eea
This is because in the formulas the substitutions for $\W\ell$ do
not break the on-shell conditions (\ref{cut-constraints}) as shown
in the end of last subsection, and also by the following simple
observation: if two functions $f_1(p)$ and $f_2(p)$ are equivalent
by some algebraic operations, then it would also be true that
\bea {d^n \over ds^n}[f_1(p)|_{p\rightarrow
q(s)}]\Big|_{s\rightarrow0} = {d^n \over ds^n}[f_2(p)|_{p\rightarrow
q(s)}]\Big|_{s\rightarrow0}. \eea

Thus to prove the conjecture, we only need to prove for the case
(\ref{sub-nsp}). Since ${\cal T}_{nsp}^{(N)}(\W\ell)$ is a sum of
the standard forms with different degrees, and $N$ is the highest
degree, what we need to prove is  actually that the formulas with
lower degree can be reexpressed with a higher degree. We will prove
this case by case.

%%%%%%%%%%%%%%%
\subsubsection{Formula for box}
%%%%%%%%%%%%%%%

It is trivially true for box formula, because the structure of the
formula doesn't depend on the degree of input. So we get the formula
directly by using (\ref{Box-new}) as
\bea C[Q_i,Q_j,K] & = & {1\over 2}\left({\cal
T}(\W\ell) \cdot D_i(\W\ell) \cdot D_j(\W\ell) \Bigg|{\begin{matrix} \\
\left\{\scriptsize \begin{matrix} \bket{\ell} & \rightarrow &
\bket{P_{ji,2}}
\\ \ket{\ell} & \rightarrow  & \ket{P_{ji,1}}
\end{matrix}\right. \end{matrix}} +
\{P_{ji,1}\leftrightarrow P_{ji,2}\} \right)  \qquad
\Label{Box-proof} \eea
for a general form of ${\cal T}(\W\ell)$.

The formula (\ref{Box-proof}) can be understood from another point
of view, as the box coefficient obtained by using the generalized
unitarity method of quadruple cut, where the substitutions for
$\W\ell$ just correspond to the two solutions that are solved from
the constraints of quadruple cut.

%%%%%%%%%%%%%%%
\subsubsection{Formula for triangle}
%%%%%%%%%%%%%%%

For triangle, it is not so trivial as box, because there are
operations of taking derivatives that depend on the degree of input.
From (\ref{sub-nsp}), we need to prove that
\bea C[Q_s,K] & = & {(K^2)^{N+1}\over
2(\sqrt{\Delta_s})^{N+1}}\frac{1}{(N+1)!
\vev{P_{s,1}~P_{s,2}}^{N+1}} \nonumber
\\ & & \frac{d^{N+1}}{d\tau^{N+1}}\left( \frac{\left\langle
\ell|K|\ell\right]^{N+1}}{(K^{2})^{N+1}} {\cal
T}_{nsp}^{(N)}(\W\ell) \cdot D_{s}(\W\ell)
\Bigg|{\scriptsize \begin{matrix} \\ \left\{\begin{matrix}
\bket{\ell} & \rightarrow & |Q_s\ket{\ell} \quad\quad~~~
\\ \ket{\ell} & \rightarrow  & \ket{P_{s,1}-\tau P_{s,2}}
\end{matrix}\right. \end{matrix}} +
\{P_{s,1}\leftrightarrow P_{s,2}\} \right)\Big|_{\tau\to 0}
\nonumber\\
& = & \sum_r b_r {(K^2)^{{n_r}+1}\over
2(\sqrt{\Delta_s})^{{n_r}+1}}\frac{1}{({n_r}+1)!
\vev{P_{s,1}~P_{s,2}}^{{n_r}+1}} \nonumber
\\ & & \frac{d^{{n_r}+1}}{d\tau^{{n_r}+1}}\left( \frac{\left\langle
\ell|K|\ell\right]^{{n_r}+1}}{(K^{2})^{{n_r}+1}} {\cal
T}^{({n_r})}_{sf}(\W\ell) \cdot D_{s}(\W\ell)
\Bigg|{\scriptsize \begin{matrix} \\ \left\{\begin{matrix}
\bket{\ell} & \rightarrow & |Q_s\ket{\ell} \quad\quad~~~
\\ \ket{\ell} & \rightarrow  & \ket{P_{s,1}-\tau P_{s,2}}
\end{matrix}\right. \end{matrix}} +
\{P_{s,1}\leftrightarrow P_{s,2}\} \right)\Big|_{\tau\to 0}. \qquad
\eea

We first do the following calculation:
\bea \gb{\ell|K|\ell}\Bigg|{\scriptsize \begin{matrix} \\
\left\{\begin{matrix} \bket{\ell} & \rightarrow & |Q_s\ket{\ell}
\quad\quad~~~
\\ \ket{\ell} & \rightarrow  & \ket{P_{s,1}-\tau P_{s,2}}
\end{matrix}\right. \end{matrix}} & = & \tau  \vev{P_{s,1}~P_{s,2}} \sqrt{\Delta_s} ~. \qquad \Label{ellKell-tri} \eea
Then by using the relation that
\bea {d ( \tau^m f(\tau))\over d\tau^n} \Big|_{\tau\to 0} = {
n!\over (n-m)!}{d ( f(\tau))\over d\tau^{n-m}} \Big|_{\tau\to 0}~,
\quad  \Label{derivative-eq} \eea
we can find that
\bea && {(K^2)^{n+1}\over 2(\sqrt{\Delta_s})^{n+1}}\frac{1}{(n+1)!
\vev{P_{s,1}~P_{s,2}}^{n+1}}
\frac{d^{n+1}}{d\tau^{n+1}}\left({\gb{\ell|K|\ell}^m \over (K^2)^m}
f(P_{\la\W\la})
\Bigg|{\scriptsize \begin{matrix} \\ \left\{\begin{matrix}
\bket{\ell} & \rightarrow & |Q_s\ket{\ell} \quad\quad~~~
\\ \ket{\ell} & \rightarrow  & \ket{P_{s,1}-\tau P_{s,2}}
\end{matrix}\right. \end{matrix}} +
\{P_{s,1}\leftrightarrow P_{s,2}\} \right)\Big|_{\tau\to 0} \quad
\nonumber\\
&=& {(K^2)^{n-m+1}\over 2(\sqrt{\Delta_s})^{n-m+1}}\frac{1}{(n-m+1)!
\vev{P_{s,1}~P_{s,2}}^{n-m+1}} \nonumber\\
&& \times~ \frac{d^{n-m+1}}{d\tau^{n-m+1}}\left( f(P_{\la\W\la})
\Bigg|{\scriptsize \begin{matrix} \\ \left\{\begin{matrix}
\bket{\ell} & \rightarrow & |Q_s\ket{\ell} \quad\quad~~~
\\ \ket{\ell} & \rightarrow  & \ket{P_{s,1}-\tau P_{s,2}}
\end{matrix}\right. \end{matrix}} +
\{P_{s,1}\leftrightarrow P_{s,2}\} \right)\Big|_{\tau\to 0}
\nonumber \eea
where the function $f(P_{\la\W\la})$ should be general but without
factor $\gb{\ell|K|\ell}$ in its denominator. So by substituting
${\cal T}_{nsp}^{(N)}(\W\ell)$ in the formula and changing $n$ to
$N$, we have

\bea C[Q_s,K] & = & {(K^2)^{N+1}\over
2(\sqrt{\Delta_s})^{N+1}}\frac{1}{(N+1)!
\vev{P_{s,1}~P_{s,2}}^{N+1}}\frac{d^{N+1}}{d\tau^{N+1}} \nonumber
\\ & & \left({\left\langle
\ell|K|\ell\right]^{N+1} \over (K^{2})^{N+1}} \sum_r b_r
{(K^2)^{n_r+1}\over \gb{\ell|K|\ell}^{n_r+1}} {\prod_{j=1}^{n_r+k_r}
\gb{\ell|R_j|\ell}\gb{\ell|Q_s|\ell}\over \prod_{i=1}^{k_r}
\gb{\ell|Q_i|\ell}}
\Bigg|{\scriptsize \begin{matrix} \\ \left\{\begin{matrix}
\bket{\ell} & \rightarrow & |Q_s\ket{\ell} \quad\quad~~~
\\ \ket{\ell} & \rightarrow  & \ket{P_{s,1}-\tau P_{s,2}}
\end{matrix}\right. \end{matrix}} +
\{P_{s,1}\leftrightarrow P_{s,2}\} \right)\Big|_{\tau\to 0}
\nonumber \\ & = & {(K^2)^{N+1}\over
2(\sqrt{\Delta_s})^{N+1}}\frac{1}{(N+1)!
\vev{P_{s,1}~P_{s,2}}^{N+1}} \nonumber
\\ & & \frac{d^{N+1}}{d\tau^{N+1}}\left( \sum_r b_r
{\left\langle \ell|K|\ell\right]^{N-n_r} \over (K^{2})^{N-n_r}}
{\prod_{j=1}^{n_r+k_r} \gb{\ell|R_j|\ell}\gb{\ell|Q_s|\ell}\over
\prod_{i=1}^{k_r} \gb{\ell|Q_i|\ell}}
\Bigg|{\scriptsize \begin{matrix} \\ \left\{\begin{matrix}
\bket{\ell} & \rightarrow & |Q_s\ket{\ell} \quad\quad~~~
\\ \ket{\ell} & \rightarrow  & \ket{P_{s,1}-\tau P_{s,2}}
\end{matrix}\right. \end{matrix}} +
\{P_{s,1}\leftrightarrow P_{s,2}\} \right)\Big|_{\tau\to 0}
\nonumber\\ & = & \sum_r  b_r ~ {(K^2)^{n_r+1}\over
2(\sqrt{\Delta_s})^{n_r+1}}\frac{1}{(n_r+1)!
\vev{P_{s,1}~P_{s,2}}^{n_r+1}} \nonumber
\\ & & \frac{d^{n_r+1}}{d\tau^{n_r+1}}\left(
{\prod_{j=1}^{n_r+k_r} \gb{\ell|R_j|\ell}\gb{\ell|Q_s|\ell}\over
\prod_{i=1}^{k_r} \gb{\ell|Q_i|\ell}}
\Bigg|{\scriptsize \begin{matrix} \\ \left\{\begin{matrix}
\bket{\ell} & \rightarrow & |Q_s\ket{\ell} \quad\quad~~~
\\ \ket{\ell} & \rightarrow  & \ket{P_{s,1}-\tau P_{s,2}}
\end{matrix}\right. \end{matrix}} +
\{P_{s,1}\leftrightarrow P_{s,2}\} \right)\Big|_{\tau\to 0}, \qquad
\eea
which is just what we want to prove.  Notice that if one term in
${\cal T}(\W\ell)$ doesn't has $D_s(\W\ell)$ in the denominator, its
contribution would be zero since
\bea \gb{\ell|Q_s|\ell} \Big|{\scriptsize \begin{matrix} \\
\left\{\begin{matrix} \bket{\ell} & \rightarrow & |Q_s\ket{\ell}
\quad\quad~~~
\\ \ket{\ell} & \rightarrow  & \ket{P_{s,1}-\tau P_{s,2}}
\end{matrix}\right. \end{matrix}} & = & 0 ~. \qquad \eea
Therefore we have
\bea C[Q_s,K] & = & {(K^2)^{N+1}\over
2(\sqrt{\Delta_s})^{N+1}}\frac{1}{(N+1)!
\vev{P_{s,1}~P_{s,2}}^{N+1}} \nonumber
\\ & & \frac{d^{N+1}}{d\tau^{N+1}}\left({\left\langle
\ell|K|\ell\right]^{N+1} \over (K^{2})^{N+1}} {\cal
T}^{(N)}(\W\ell)\cdot D_{s}(\W\ell)
\Bigg|{\scriptsize \begin{matrix} \\ \left\{\begin{matrix}
\bket{\ell} & \rightarrow & |Q_s\ket{\ell} \quad\quad~~~
\\ \ket{\ell} & \rightarrow  & \ket{P_{s,1}-\tau P_{s,2}}
\end{matrix}\right. \end{matrix}} +
\{P_{s,1}\leftrightarrow P_{s,2}\} \right)\Big|_{\tau\to 0} \qquad
\Label{Tri-proof} \eea
for general ${\cal T}^{(N)}(p)$.
%Notice that it's necessary to keep
%the factor $\gb{\ell|K|\ell}^{N+1}$ in the parentheses, before using
%the relation (\ref{ellKell-tri}) for $\gb{\ell|K|\ell}$, in order to
%cancel those that appear in the denominator of the tree level input.

%%%%%%%%%%%%%%%
\subsubsection{Formulas for bubble}
%%%%%%%%%%%%%%%

For bubble, the proof is similar to the case of triangle. Consider
the bubble formulas
\bea C[K] &=& (K^{2})^{n+1}\sum_{q=0}^n {(-1)^q \over q!} {d^q \over
d s^q} \left( {\cal B}_{n,n-q}^{(0)}(s) + \sum_{r=1}^k \sum_{a=q}^n
\left[ {\cal B}_{n,n-a}^{(r,a-q,1)}(s) - {\cal
B}_{n,n-a}^{(r,a-q,2)}(s)\right] \right) \Bigg|_{s\rightarrow0}
\quad \Label{Bub-forproof} \eea
where ${\cal B}_{n,n-q}^{(0)}(s)$, ${\cal
B}_{n,n-a}^{(r,a-q,1)}(s)$, and ${\cal B}_{n,n-a}^{(r,a-q,2)}(s)$
are given by (\ref{Bnt0-new}), (\ref{Bnt1-new}) and
(\ref{Bnt2-new}). We want to generalize the formulas to the input
${\cal T}_{nsp}^{(N)}(\W\ell)$.

We first consider ${\cal B}_{n,t}^{(0)}(s)$:
\bea {\cal B}_{n,t}^{(0)}(s) &=& {d^n \over d\tau^n}
\left[{(2\eta\cdot K)^{t+1}\over (t+1)(K^2)^{t+1}} {1\over n!
[\eta|\W\eta K|\eta]^n \vev{\ell~\eta}^{n+1}}
\left(\frac{\left\langle \ell|K|\ell\right]^{n}}{(K^{2})^{n}} {\cal
T}^{(n)}_{sf}(\W\ell) \right)
\Bigg|{\scriptsize
\begin{matrix} \\ \left\{\begin{matrix} \bket{\ell} & \rightarrow &
|K+s\eta\ket{\ell}
\\ \ket{\ell} & \rightarrow  & |K-\tau\W\eta|\eta]
\end{matrix}\right. \end{matrix}} \right]
\Bigg|_{\tau\rightarrow0} ~. \qquad \eea
As in the case of
triangle, we need the calculation
\bean \gb{\ell|K|\ell}\Big|{\scriptsize
\begin{matrix} \\ \left\{\begin{matrix} \bket{\ell} & \rightarrow &
|K+s\eta\ket{\ell}
\\ \ket{\ell} & \rightarrow  & |K-\tau\W\eta|\eta]
\end{matrix}\right. \end{matrix}}
& = & -\tau s ~[\eta| {\W\eta} K \eta ] \vev{\ell~\eta} \big|_{
\ket{\ell} \rightarrow |K-\tau\W\eta|\eta] }~. \qquad \eean
Then by using twice the relation (\ref{derivative-eq}), we have
\bea && (K^{2})^{n+1}\sum_{q=0}^n {(-1)^q \over q!} {d^q \over d
s^q} \left\{ {d^n \over d\tau^n} \left[{(2\eta\cdot K)^{n-q+1}\over
(n-q+1)(K^2)^{n-q+1}} \right. \right.
\nonumber\\ && \qquad \qquad \qquad \times \left. \left. {1\over n!
[\eta|\W\eta K|\eta]^n \vev{\ell~\eta}^{n+1}} \left(
{\gb{\ell|K|\ell}^m \over (K^2)^m} f(P_{\la\W\la}) \right)
\Bigg|{\scriptsize
\begin{matrix} \\ \left\{\begin{matrix} \bket{\ell} & \rightarrow &
|K+s\eta\ket{\ell}
\\ \ket{\ell} & \rightarrow  & |K-\tau\W\eta|\eta]
\end{matrix}\right. \end{matrix}} \right]
\Bigg|_{\tau\rightarrow0} \right\} \Bigg|_{s\rightarrow0}
\nonumber\\ &=&  (K^{2})^{n-m+1}\sum_{q=0}^n {(-1)^q \over q!} {d^q
\over d s^q} \left\{ (-s)^m {d^{n-m} \over d\tau^{n-m}}
\left[{(2\eta\cdot K)^{n-q+1}\over (n-q+1)(K^2)^{n-q+1}}  \right.
\right.
\nonumber\\ && \qquad \qquad \qquad \times \left. \left. {1\over
(n-m)! [\eta|\W\eta K|\eta]^{n-m} \vev{\ell~\eta}^{n-m+1}} \left(
f(P_{\la\W\la}) \right)
\Bigg|{\scriptsize
\begin{matrix} \\ \left\{\begin{matrix} \bket{\ell} & \rightarrow &
|K+s\eta\ket{\ell}
\\ \ket{\ell} & \rightarrow  & |K-\tau\W\eta|\eta]
\end{matrix}\right. \end{matrix}} \right]
\Bigg|_{\tau\rightarrow0} \right\} \Bigg|_{s\rightarrow0}
\nonumber\\ &=&  (K^{2})^{n-m+1}\sum_{q=0}^{n-m} {(-1)^q \over q!}
{d^q \over d s^q} \left\{ {d^{n-m} \over d\tau^{n-m}}
\left[{(2\eta\cdot K)^{n-m-q+1}\over (n-m-q+1)(K^2)^{n-m-q+1}}
\right. \right.
\nonumber\\ && \qquad \qquad \qquad \times \left. \left. {1\over
(n-m)! [\eta|\W\eta K|\eta]^{n-m} \vev{\ell~\eta}^{n-m+1}} \left(
f(P_{\la\W\la}) \right)
\Bigg|{\scriptsize
\begin{matrix} \\ \left\{\begin{matrix} \bket{\ell} & \rightarrow &
|K+s\eta\ket{\ell}
\\ \ket{\ell} & \rightarrow  & |K-\tau\W\eta|\eta]
\end{matrix}\right. \end{matrix}} \right]
\Bigg|_{\tau\rightarrow0} \right\} \Bigg|_{s\rightarrow0} \qquad
\nonumber \eea
So if we make the substitute (\ref{sub-nsp}) for the formula of
${\cal B}_{n,t}^{(0)}(s)$ and put into (\ref{Bub-forproof}), we have
\bea && (K^{2})^{N+1}\sum_{q=0}^N {(-1)^q \over q!} {d^q \over d
s^q}
\left\{ {d^N \over d\tau^N} \left[{(2\eta\cdot K)^{N-q+1}\over
(N-q+1)(K^2)^{N-q+1}} \right.\right.
\nonumber\\ && \left.\left. \times {1\over N! [\eta|\W\eta K|\eta]^N
\vev{\ell~\eta}^{N+1}} \left( {\gb{\ell|K|\ell}^N \over (K^2)^N}
\sum_r {(K^2)^{n_r}\over \gb{\ell|K|\ell}^{n_r}}
{\prod_{j=1}^{n_r+k_r} \gb{\ell|R_j|\ell}\over \prod_{i=1}^{k_r}
\gb{\ell|Q_i|\ell}} \right)
\Bigg|{\scriptsize
\begin{matrix} \\ \left\{\begin{matrix} \bket{\ell} & \rightarrow &
|K+s\eta\ket{\ell}
\\ \ket{\ell} & \rightarrow  & |K-\tau\W\eta|\eta]
\end{matrix}\right. \end{matrix}} \right]
\Bigg|_{\tau\rightarrow0} \right\} \Bigg|_{s\rightarrow0}
\nonumber\\ &=& (K^{2})^{N+1}\sum_{q=0}^N {(-1)^q \over q!} {d^q
\over d s^q}
\left\{ {d^N \over d\tau^N} \left[{(2\eta\cdot K)^{N-q+1}\over
(N-q+1)(K^2)^{N-q+1}} \right.\right.
\nonumber\\ && \left.\left. \times {1\over N! [\eta|\W\eta K|\eta]^N
\vev{\ell~\eta}^{N+1}} \left( \sum_r {\gb{\ell|K|\ell}^{N-n_r} \over
(K^2)^{N-n_r}} {\prod_{j=1}^{n_r+k_r} \gb{\ell|R_j|\ell}\over
\prod_{i=1}^{k_r} \gb{\ell|Q_i|\ell}} \right)
\Bigg|{\scriptsize
\begin{matrix} \\ \left\{\begin{matrix} \bket{\ell} & \rightarrow &
|K+s\eta\ket{\ell}
\\ \ket{\ell} & \rightarrow  & |K-\tau\W\eta|\eta]
\end{matrix}\right. \end{matrix}} \right]
\Bigg|_{\tau\rightarrow0} \right\} \Bigg|_{s\rightarrow0}
\nonumber\\
&=& \sum_r ~(K^{2})^{{n_r}+1}\sum_{q=0}^{n_r} {(-1)^q \over q!} {d^q
\over d s^q}
\left\{ {d^{n_r} \over d\tau^{n_r}} \left[{(2\eta\cdot
K)^{n_r-q+1}\over (n_r-q+1)(K^2)^{n_r-q+1}} \right.\right.
\nonumber\\ && \left.\left. \times  {1\over {n_r}! [\eta|\W\eta
K|\eta]^{n_r} \vev{\ell~\eta}^{{n_r}+1}} \left(
{\prod_{j=1}^{n_r+k_r} \gb{\ell|R_j|\ell}\over \prod_{i=1}^{k_r}
\gb{\ell|Q_i|\ell}} \right)
\Bigg|{\scriptsize
\begin{matrix} \\ \left\{\begin{matrix} \bket{\ell} & \rightarrow &
|K+s\eta\ket{\ell}
\\ \ket{\ell} & \rightarrow  & |K-\tau\W\eta|\eta]
\end{matrix}\right. \end{matrix}} \right]
\Bigg|_{\tau\rightarrow0} \right\} \Bigg|_{s\rightarrow0} \nonumber
\eea
which is just what we need to prove for the generalization.

With the same procedure, we can also prove it for ${\cal
B}_{n,t}^{(r,b,1)}(s)$ and ${\cal B}_{n,t}^{(r,b,2)}(s)$, by using
the calculation
\bean \gb{\ell|K|\ell} \Big|{\scriptsize \begin{matrix} \\
\left\{\begin{matrix} \bket{\ell} & \rightarrow &
|K+s\eta\ket{\ell}~~~
\\ \ket{\ell} & \rightarrow  & \ket{P_{r,2}-\tau P_{r,1}}
\end{matrix}\right. \end{matrix}}
&=& - s \vev{\ell| \eta K |\ell} \Big|_{ \ket{\ell}  \rightarrow
\ket{P_{r,2}-\tau P_{r,1}} } . \eean
Notice that there is a summation over $r$ in (\ref{Bub-forproof}),
i.e. sum over all the $D_r$ that appear in the denominator of tree
level input. We also need to show that if the input has no $D_r$ in
denominator, it vanish, i.e.
\bea \sum_{q=0}^n {(-1)^q \over q!} {d^q \over d s^q}  \sum_{a=q}^n
{\cal B}_{n,n-a}^{(r,a-q,1(2))}(s) \Bigg|_{s\rightarrow0} = 0 ~,
\quad \eea
if ${\cal T}^{(n)}_{sf}(\W\ell)$ has no $D_r$ in denominator. By
substituting (\ref{Bnt1-new}) or (\ref{Bnt2-new}) in it,  and using
the relation that
\bea \vev{\ell|Q_r(K+s\eta)|\ell} \Big|{\scriptsize \begin{matrix}
\\ \left\{\begin{matrix} \bket{\ell} & \rightarrow &
|K+s\eta\ket{\ell}~~~
\\ \ket{\ell} & \rightarrow  & \ket{P_{r,2}-\tau P_{r,1}}
\end{matrix}\right. \end{matrix}}
= -\tau \vev{P_{r,1}~P_{r,2}}\sqrt{\Delta_r} +
s\vev{\ell|Q_r\eta|\ell}\Big|{\scriptsize \begin{matrix}
\\ \left\{\begin{matrix} \bket{\ell} & \rightarrow &
|K+s\eta\ket{\ell}~~~
\\ \ket{\ell} & \rightarrow  & \ket{P_{r,2}-\tau P_{r,1}}
\end{matrix}\right. \end{matrix}}, \qquad \Label{box-cancel} \eea
we can find this is indeed true, due to the cancellation between the
two terms in the right-hand side of above relation.

Therefore, we finally have the bubble formulas
\bea C[K] &=& (K^{2})^{N+1}\sum_{q=0}^N {(-1)^q \over q!} {d^q \over
d s^q} \left( {\cal B}_{N,N-q}^{(0)}(s) + \sum_{r=1}^k \sum_{a=q}^N
\left[ {\cal B}_{N,N-a}^{(r,a-q,1)}(s) - {\cal
B}_{N,N-a}^{(r,a-q,2)}(s)\right] \right) \Bigg|_{s\rightarrow0}
\Label{Bub-proof} \eea
\bea {\cal B}_{N,t}^{(0)}(s) &=& {d^N \over d\tau^N}
\left[{(2\eta\cdot K)^{t+1}\over (t+1)(K^2)^{t+1}} {1\over N!
[\eta|\W\eta K|\eta]^N \vev{\ell~\eta}^{N+1}}
\left(\frac{\left\langle \ell|K|\ell\right]^{N}}{(K^{2})^{N}} {\cal
T}^{(N)}(\W\ell) \right)
\Bigg|{\scriptsize
\begin{matrix} \\ \left\{\begin{matrix} \bket{\ell} & \rightarrow &
|K+s\eta\ket{\ell}
\\ \ket{\ell} & \rightarrow  & |K-\tau\W\eta|\eta]
\end{matrix}\right. \end{matrix}} \right]
\Bigg|_{\tau\rightarrow0} \Label{Bnt0-proof}  \eea
\bea {\cal B}_{N,t}^{(r,b,1)}(s) &=& {(-1)^{b+1} \over b!
(\sqrt{\Delta_r})^{b+1}\vev{P_{r,1}~P_{r,2}}^b}
{d^b \over d\tau^b} \\
&& \left[ {1\over t+1} {\gb{\ell |\eta| P_{r,1}}^{t+1}\over\gb{\ell
|K| P_{r,1}}^{t+1}} {\vev{\ell |Q_r\eta| \ell}^b \over \vev{\ell
|\eta K| \ell}^{N+1}} \left(\frac{\left\langle
\ell|K|\ell\right]^{N+1}}{(K^{2})^{N+1}} {\cal T}^{(N)}(\W\ell)
\cdot D_{r}(\W\ell) \right)
\Bigg|{\scriptsize \begin{matrix} \\ \left\{\begin{matrix}
\bket{\ell} & \rightarrow & |K+s\eta\ket{\ell}~~~
\\ \ket{\ell} & \rightarrow  & \ket{P_{r,1}-\tau P_{r,2}}
\end{matrix}\right. \end{matrix}}
\right]\Bigg|_{\tau\rightarrow0}  \Label{Bnt1-proof} \nonumber \eea
\bea {\cal B}_{N,t}^{(r,b,2)}(s) &=& {(-1)^{b+1} \over b!
(\sqrt{\Delta_r})^{b+1}\vev{P_{r,1}~P_{r,2}}^b}
{d^b \over d\tau^b} \\
&& \left[ {1\over t+1} {\gb{\ell |\eta| P_{r,2}}^{t+1}\over\gb{\ell
|K| P_{r,2}}^{t+1}} {\vev{\ell |Q_r\eta| \ell}^b \over \vev{\ell
|\eta K| \ell}^{N+1}} \left(\frac{\left\langle
\ell|K|\ell\right]^{N+1}}{(K^{2})^{N+1}} {\cal T}^{(N)}(\W\ell)
\cdot D_{r}(\W\ell) \right)
\Bigg|{\scriptsize \begin{matrix} \\ \left\{\begin{matrix}
\bket{\ell} & \rightarrow & |K+s\eta\ket{\ell}~~~
\\ \ket{\ell} & \rightarrow  & \ket{P_{r,2}-\tau P_{r,1}}
\end{matrix}\right. \end{matrix}}
\right]\Bigg|_{\tau\rightarrow0}  \Label{Bnt2-proof} \nonumber \eea
for general ${\cal T}^{(N)}(p)$. The summation of $r$ is for all the
$D_r(\W\ell)$ that appear in the denominator of ${\cal T}(\W\ell)$.

%%%%%%%%%%%%%%%
\paragraph{A special choice of $\eta$.}
%%%%%%%%%%%%%%%

As discussed in the Appendix B.3.1 of \cite{BF07}, we can use a
special choice of $\eta$: choosing $\eta=K_1$ in the case where
$K_1^2=0$. By this choice, we have
\bea {1\over \vev{\ell|Q_1 (K+s\eta)|\ell}} & = &  - {1\over
\vev{\ell|\eta K|\ell}} { 1\over \b\sqrt{1-u} +s \left( \b\sqrt{1-u}
{\eta\cdot K\over K^2}+\a_1 \right)}~. \quad \eea
where $\a_1$ is given by (\ref{alpha-mass}).

The general bubble formula for this special choice, i.e. $\eta=K_1$,
becomes%
\footnote{Notice the summation over $r$ no longer includes $r=1$.}
\bea C[K] &=& (K^{2})^{N+1}\sum_{q=0}^N {(-1)^q \over q!} {d^q \over
d s^q} \left( {\cal \hat B}_{N,N-q}^{(0)}(s) + \sum_{r=2}^k
\sum_{a=q}^N \left[ {\cal \hat B}_{N,N-a}^{(r,a-q,1)}(s) - {\cal
\hat B}_{N,N-a}^{(r,a-q,2)}(s)\right] \right) \Bigg|_{s\rightarrow0}
\Label{spe-Bub} \eea
where
\bea {\cal \hat B}_{N,t}^{(0)}(s) &=& -{ 1\over \b\sqrt{1-u} +s
\left( \b\sqrt{1-u} {\eta\cdot K\over K^2}+\a_1 \right)} {d^{N+1}
\over d\tau^{N+1}} \left[{(2\eta\cdot K)^{t+1}\over
(t+1)(K^2)^{t+1}} \right. \nonumber \\ && \left. {1\over (N+1)!
[\eta|\W\eta K|\eta]^{N+1} \vev{\ell~\eta}^{N+2}}
\left(\frac{\left\langle \ell|K|\ell\right]^{N+1}}{(K^{2})^{N+1}}
{\cal T}^{(N)}(\W\ell) \cdot D_1(\W\ell) \right)
\Bigg|{\scriptsize
\begin{matrix} \\ \left\{\begin{matrix} \bket{\ell} & \rightarrow &
|K+s\eta\ket{\ell}
\\ \ket{\ell} & \rightarrow  & |K-\tau\W\eta|\eta]
\end{matrix}\right. \end{matrix}} \right]
\Bigg|_{\tau\rightarrow0} \quad \Label{spe-Bnt0} \eea
\bea {\cal \hat B}_{N,t}^{(r,b,1)}(s) &=& -{ 1\over \b\sqrt{1-u} +s
\left( \b\sqrt{1-u} {\eta\cdot K\over K^2}+\a_1 \right)} {(-1)^{b+1}
\over b! (\sqrt{\Delta_r})^{b+1}\vev{P_{r,1}~P_{r,2}}^b} {d^b \over
d\tau^b}  \left[ {1\over t+1} {\gb{\ell |\eta|
P_{r,1}}^{t+1}\over\gb{\ell |K| P_{r,1}}^{t+1}} \right. \qquad
\nonumber \\ && \left. {\vev{\ell |Q_r\eta| \ell}^b \over \vev{\ell
|\eta K| \ell}^{N+2}} \left(\frac{\left\langle
\ell|K|\ell\right]^{N+2}}{(K^{2})^{N+2}} {\cal T}^{(N)}(\W\ell)
\cdot D_1(\W\ell)\cdot D_{r}(\W\ell) \right)
\Bigg|{\scriptsize \begin{matrix} \\ \left\{\begin{matrix}
\bket{\ell} & \rightarrow & |K+s\eta\ket{\ell}~~~
\\ \ket{\ell} & \rightarrow  & \ket{P_{r,1}-\tau P_{r,2}}
\end{matrix}\right. \end{matrix}}
\right]\Bigg|_{\tau\rightarrow0}  \Label{spe-Bnt1} \eea
\bea {\cal \hat B}_{N,t}^{(r,b,2)}(s) &=& -{ 1\over \b\sqrt{1-u} +s
\left( \b\sqrt{1-u} {\eta\cdot K\over K^2}+\a_1 \right)} {(-1)^{b+1}
\over b! (\sqrt{\Delta_r})^{b+1}\vev{P_{r,1}~P_{r,2}}^b} {d^b \over
d\tau^b}  \left[ {1\over t+1} {\gb{\ell |\eta|
P_{r,2}}^{t+1}\over\gb{\ell |K| P_{r,2}}^{t+1}} \right. \qquad
\nonumber \\ && \left. {\vev{\ell |Q_r\eta| \ell}^b \over \vev{\ell
|\eta K| \ell}^{N+2}} \left(\frac{\left\langle
\ell|K|\ell\right]^{N+2}}{(K^{2})^{N+2}} {\cal T}^{(N)}(\W\ell)
\cdot D_1(\W\ell)\cdot D_{r}(\W\ell) \right)
\Bigg|{\scriptsize \begin{matrix} \\ \left\{\begin{matrix}
\bket{\ell} & \rightarrow & |K+s\eta\ket{\ell}~~~
\\ \ket{\ell} & \rightarrow  & \ket{P_{r,2}-\tau P_{r,1}}
\end{matrix}\right. \end{matrix}}
\right]\Bigg|_{\tau\rightarrow0} . \Label{spe-Bnt2} \eea
%

%%%%%%%%%%%%%%%%%%%%%%%%%%%%%%%%%%%%%%%%%%%%%%%%%%%%%%%%%%%%%%%%%
\section{On pentagon and box formulas}
%%%%%%%%%%%%%%%%%%%%%%%%%%%%%%%%%%%%%%%%%%%%%%%%%%%%%%%%%%%%%%%%%

There are two complexities for  box formulas that deserve more
study. First, the box formula that we discussed before is not a
polynomial of $u$, because it contains  also pentagon contributions,
indicated by a linear factor $(au+b)$ in the denominator. We need to
separate the pentagon part from the box, so that the true box
coefficient is a polynomial of $u$. The second complexity is that
the null momenta $P_{ji;1(2)}(u)$ depend on $u$ in a very nontrivial
way (as $Q_j(u)+ x_a Q_i(u)$), unlike the cases of triangles and
bubbles.

For the formulas with a standard input form, these two problems have
been solved in \cite{BFG08, BFM08}. We review these simplified
formulas in Appendix \ref{u-simp}. Now we want to deal with the
general form of tree level input.

In the following two subsections, we first use a ``quintuble-cut''
method to calculate the pentagon coefficients%
\footnote{The idea of using ``quintuble-cut'' to determined pentagon
coefficients has appeared in \cite{Giele:2008ve}.}. The true box
coefficients can be obtained by subtracting the pentagon
contributions. Then in the second subsection, we will give a way to
simplify the $u$ dependence for the box formula, by generalizing the
result in the case of standard form.

%%%%%%%%%%%%%%%%%%%%%%%%%%%%%%%%%%%%%%%%%%%%%%%%%%%
\subsection{Pentagon coefficient}
%%%%%%%%%%%%%%%%%%%%%%%%%%%%%%%%%%%%%%%%%%%%%%%%%%%

The pentagon master integral is
\bea I_5^D[1] =  \int d^D p ~{1\over (p^2-M_1^2) ((p-K)^2-M_2^2)
((p-K_i)^2-m_i^2) ((p-K_j)^2-m_j^2) ((p-K_r)^2-m_r^2) }~.  \quad
\Label{pen-scalar} \eea
The quintuple-cut for the master integral is given by replacing the
five propagators with five $\delta$-functions:
\bea  && \textrm{Cut}\left[ I_5^D [1] \right]\Big|_{\textrm{quintuble-cut}} \nonumber\\
& \equiv & \int d^D p ~{\delta(p^2-M_1^2) \delta((p-K)^2-M_2^2)
\delta((p-K_i)^2-m_i^2) \delta((p-K_j)^2-m_j^2)
\delta((p-K_r)^2-m_r^2)} \qquad \nonumber
\\
&=& \int d^4 \W\ell ~d^{-2\eps} \mu ~\delta(\W\ell^2-\mu^2-M_1^2)~
\delta(-2\W\ell\cdot K + K^2 +M_1^2-M_2^2) ~\delta(-2\W\ell\cdot K_i
+ K_i^2 +M_1^2-m_i^2)~ \nonumber\\ &&  \qquad \qquad \quad \times
\delta(-2\W\ell\cdot K_j + K_j^2 +M_1^2-m_j^2)~\delta(-2\W\ell\cdot
K_r + K_r^2 +M_1^2-m_r^2) ~. \eea
The integral is totally fixed by five $\delta$-functions. So to get
the pentagon coefficient, it is possible to use the quintuple-cut
method to read the coefficient directly.

As we have discussed before, the coefficient can be accepted as a
function of $\mu^2$. In another word, we only need to fix the
four-dimensional component of $p$, i.e. $\W\ell$, and leave the
integral for $\mu$. This can be done by using the latter four
$\delta$-functions.

The latter four $\delta$-functions give four equations for $\W\ell$
(we have also used the first $\delta$-function that $\W\ell^2 =
M_1^2+ \mu^2$)
\bea \left\{ ~ \begin{matrix}
-2\W\ell\cdot K + K^2 +M_1^2-M_2^2 = 0
\\ -2\W\ell\cdot K_i + K_i^2 +M_1^2-m_i^2 = 0
\\ -2\W\ell\cdot K_j + K_j^2 +M_1^2-m_j^2 = 0
\\ -2\W\ell\cdot K_r + K_r^2 +M_1^2-m_r^2 = 0
\end{matrix} \right. \quad
\eea
by which $\W\ell$ can be solved as
\bea \W\ell_{(i,j,r)} = l_0 K + l_i K_i + l_j K_j + l_r K_r , \quad
\Label{pen-ell-solution} \eea
where
\bea \begin{pmatrix} l_0 \cr l_i \cr l_j \cr l_r
\end{pmatrix} = {1\over 2} \,\begin{pmatrix}
K^2 & K_i\cdot K & K_j\cdot K & K_r\cdot K \cr
K\cdot K_i & K_i^2 & K_j\cdot K_i & K_r\cdot K_i \cr
K\cdot K_j & K_i\cdot K_j & K_j^2 & K_r\cdot K_j \cr
K\cdot K_r & K_i\cdot K_r & K_j\cdot K_r & K_r^2
\end{pmatrix}^{-1} \cdot
\begin{pmatrix}
K^2 +M_1^2-M_2^2 \cr K_i^2 +M_1^2-m_i^2 \cr K_j^2 +M_1^2-m_j^2 \cr
K_r^2 +M_1^2-m_r^2
\end{pmatrix}. \quad \Label{l-exp0ijr} \eea

For a general input ${\cal T}(\W\ell)$ of double-cut integral, the
pentagon coefficient can be written as
\bea \textrm{Pen}[K_i,K_j,K_r,K] &=& \left[{\cal T}(\W\ell) \cdot
D_i(\W\ell) D_j(\W\ell) D_r(\W\ell) \right]
\big|_{\W\ell\rightarrow\W\ell_{(i,j,r)}} \nonumber\\ &=& {\cal
T}(\W\ell_{(i,j,r)}) \cdot D_i(\W\ell_{(i,j,r)})
D_j(\W\ell_{(i,j,r)}) D_r(\W\ell_{(i,j,r)}) . \quad \Label{Pen} \eea
In the appendix \ref{compare-pen}, we show that for the standard
tree level input, this formula is equivalent with the previous-known
formula. We emphasize that the solution $\W\ell_{(i,j,r)}$ can also
be used in generalized unitarity with multi-cut. For example, for
five-cut (with another three cuts across $D_i,D_j,D_r$), we have the
corresponding pentagon coefficient simply as
\bea \textrm{Pen}[K_i,K_j,K_r,K] = A_1^{\textrm{tree}}(\W\ell)\times
A_2^{\textrm{tree}}(\W\ell)\times A_3^{\textrm{tree}}(\W\ell)\times
A_4^{\textrm{tree}}(\W\ell)\times A_5^{\textrm{tree}}(\W\ell)
\big|_{\W\ell\rightarrow\W\ell_{(i,j,r)}}. \eea

There is one point about the solution (\ref{pen-ell-solution}):
there is matrix inverse we need to do. If $\W\ell_{(i,j,r)}$ is
contracted with any momentum, we may use (\ref{beta-proof}) to
calculate, which is simpler.

%%%%%%%%%%%%%%%%%%%%%%%%%%%%%%%%%%%%%%%%%%%%%%%%%%%
\subsection{\label{box-u-sim-I} Simplify the $u$ dependence of the box formula}
%%%%%%%%%%%%%%%%%%%%%%%%%%%%%%%%%%%%%%%%%%%%%%%%%%%

Now we want to simplify the $u$-dependence of the general box
formula. As reviewed in Appendix \ref{u-simp}, for the standard
input form, we have known how to simplify the $u$-dependence of the
box formula as
\bea C[Q_i,Q_j,K] & = & {1\over 2}\left({(K^2)^{n+2}\over \gb{\ell|K
|\ell}^{n+2}} { \prod_{s=1}^{k+n} \gb{\ell|R_s(u) |\ell}
\over\prod_{t=1,t\neq
i,j}^k \gb{\ell|Q_t(u) |\ell}} \Bigg|{\scriptsize \begin{matrix} \\
\left\{\begin{matrix} \bket{\ell} & \rightarrow & \bket{P_{ji,2}(u)}
\\ \ket{\ell} & \rightarrow  & \ket{P_{ji,1}(u)}
\end{matrix}\right. \end{matrix}} + \{P_{ji,1}(u)\leftrightarrow
P_{ji,2}(u)\} \right)
\nonumber\\ & = & {1\over 2}\left({(K^2)^{n+2}\over \gb{\ell|K
|\ell}^{n+2}} { \prod_{s=1}^{k+n} \gb{\ell|\W R_s(u) |\ell}
\over\prod_{t=1,t\neq
i,j}^k \gb{\ell|\W Q_t(u) |\ell}} \Bigg|{\scriptsize \begin{matrix} \\
\left\{\begin{matrix} \bket{\ell} & \rightarrow &
\bket{P_{ji,2}(u=0)}
\\ \ket{\ell} & \rightarrow  & \ket{P_{ji,1}(u=0)}
\end{matrix}\right. \end{matrix}} + \{P_{ji,1}(u=0)\leftrightarrow
P_{ji,2}(u=0)\} \right).  \nonumber \eea
It is worth to emphasize that {\sl in the first line, $u$-dependence
is everywhere}: in the spinor component of $P_{ji}$ as well as
inside the square roots $\sqrt{ (2Q_i(u)\times Q_j(u))^2-4 Q_i(u)^2
q_j(u)^2}$. Because of this, no matter analytically or numerically,
it will become very tedious and complicated. To avoid this problem,
as analyzed in \cite{BFG08}, another equivalent expression is given
as in the second line of above formula, where null vector
$P_{ji,1(2)}(u=0)$  as well as square root do not depend on $u$
anymore. The only $u$-dependence is following replacement:
\bea R_s(u) \rightarrow \W R_s(u), \qquad Q_t(u) \rightarrow \W
Q_t(u), \qquad \Label{box-rule-I} \eea
where $\W R_s(u)$ and $\W Q_t(u)$ are given by (\ref{W-R-S-1}) and
(\ref{W-Q-t-1}).
%
%\bea  {\W R}_s(u) &=& {p_s\cdot q_0^{(q_i,q_j,K)}\over
%(q_0^{(q_i,q_j,K)})^2}
%(\a^{(q_i,q_j)}(u)-1)(-\b q_0^{(q_i,q_j,K)}) + R_s(u=0), \quad \\
%\W Q_t(u) &= &{q_t\cdot q_0^{(q_i,q_j,K)}\over
%(q_0^{(q_i,q_j,K)})^2} (\a^{(q_i,q_j)}(u)-1)(-\b
%q_0^{(q_i,q_j,K)})+Q_t(u=0), \qquad \eea
%
%The $\a^{(q_i,q_j)}$ and $q_0^{(q_i,q_j,K)}$ are given by
%(\ref{def-alpha-box}) and (\ref{q-0})\footnote{We have emphasized
%this simplification since it is easy to be overlooked.}.

We want to generalize the formula, so that it can be used for the
general tree level input. To realize this, we need to generalize the
above rule, which should not be confined to the special form with
$R(u)$ and $Q(u)$. This can be achieved if we find a substitution
for $\W\ell$. Using (\ref{W-R-S-1}) and noticing that
(\ref{R-Q-massive})
\bea   R_s(u=0) = -\b P_s + (\b-\a){P_s \cdot K\over K^2}K,  \quad
\Label{Ru=0} \eea
where $\a, \b$ are given by (\ref{ab-def}), we can find a relation
for $\W R(u)$ that
\bea \frac{K^{2}}{\left\langle \ell|K|\ell\right]}\gb{\ell|\W
R_s|\ell} & = & \frac{K^{2}}{\left\langle
\ell|K|\ell\right]}\gb{\ell \left|{P_s\cdot q_0^{(q_i,q_j,K)}\over
(q_0^{(q_i,q_j,K)})^2} (\a^{(q_i,q_j)}(u)-1)(-\b q_0^{(q_i,q_j,K)})
-\b P_s + (\b-\a){P_s \cdot K\over K^2}K
\right|\ell} \nonumber\\
& = & -2 \left[-\b \frac{K^{2}}{\left\langle
\ell|K|\ell\right]}\left([\a^{(q_i,q_j)}(u)-1]
{q_0^{(q_i,q_j,K)}\cdot P_{\la \W\la}\over (q_0^{(q_i,q_j,K)})^2}
q_0^{(q_i,q_j,K)} + P_{\la \W\la} \right)-{1\over2}(\b-\a)K \right]\cdot P_s \nonumber\\
& \equiv & -2 \W\ell_{ij} \cdot P_s  \quad \Label{reform-wR} \eea
where we have defined
\bea \W\ell_{ij} \equiv - \b \frac{K^{2}}{\left\langle
\ell|K|\ell\right]}\left([\a^{(q_i,q_j)}(u)-1]
{q_0^{(q_i,q_j,K)}\cdot P_{\la \W\la}\over (q_0^{(q_i,q_j,K)})^2}
q_0^{(q_i,q_j,K)} +P_{\la \W\la} \right)-{1\over2}(\b-\a)K. \quad
\Label{box-ellij-I} \eea
With the same $\W\ell_{ij}$, it is easy to find that
\bea {K^2 \over \gb{\ell|K|\ell}} \gb{\ell|\W Q_t|\ell} =
-2\W\ell_{ij}\cdot K_t+ K_t^2+M_1^2-m_t^2 = D_t(\W\ell_{ij}) . \quad
\Label{reform-wQ} \eea
Comparing (\ref{reform-wR}) and (\ref{reform-wQ}) with the relation
\bea -2 \W\ell\cdot P_s = \frac{K^{2}}{\left\langle
\ell|K|\ell\right]}\gb{\ell|R_s|\ell} , \qquad
 D_t(\W\ell) = {K^2 \over \gb{\ell|K|\ell}}
\gb{\ell|Q_t|\ell},  \eea
we can find the rule (\ref{box-rule-I}) is equivalent to
\bea \W\ell \rightarrow \W\ell_{ij} . \qquad  \eea
In Appendix \ref{box-u-sim-II}, we give another equivalent
expression for $\W\ell_{ij}$, which can avoid the appearance of
$q_0^{(q_i,q_j,K)}$.

Just following the argument of Section \ref{idea-proof}, this new
rule can be generalized directly to the formula (\ref{Box-proof})
with general tree level input, with only one condition that:
$\W\ell_{ij}$, with the substitution for $P_{\la \W\la}$ in the
formulas, should satisfy the two on-shell conditions
\bea \W\ell_{ij}^2-\mu^2-M_1^2 = 0 , \qquad
(\W\ell_{ij}-K)^2-\mu^2-M_2^2 = 0. \quad \Label{ellij-onshell} \eea
%.
We will proof this is indeed true.

We first proof the first condition. Using (\ref{box-ellij-I})
directly, we have that (for simplicity we omit the superscript of
$q_0^{(q_i,q_j,K)}$)
\bea \W\ell_{ij}^2 & = &
%\b^2 \frac{(K^{2})^2}{\left\langle
%\ell|K|\ell\right]^2}\left(\left(\a^{(q_i,q_j)}(u)^2-1\right)
%{\left(q_0\cdot P_{\la \W\la}\right)^2 \over q_0^2}
%\right)+\b(\b-\a)\frac{K^{2}}{\left\langle \ell|K|\ell\right]}K\cdot
%P_{\la \W\la} + {1\over4}(\b-\a)^2 K^2 \qquad
%
%\nonumber\\ & = &  \b^2 \frac{(K^{2})^2}{4 q_0^2}
%{\gb{\ell|q_0|\ell}^2 \over \left\langle \ell|K|\ell\right]^2}
%\left(\a^{(q_i,q_j)}(u)^2-1\right) +{1\over4}(\a^2-\b^2) K^2
%
%
%\nonumber\\ & = &
-\mu^2 \b^4 \frac{K^{2}}{q_0^2}
{\gb{\ell|q_0|\ell}^2 \over \left\langle \ell|K|\ell\right]^2}
{(2q_i\cdot q_j)^2-4 q_i^2 q_j^2 \over \Delta(Q_i,Q_j)} +M_1^2
\qquad \Label{ell-square} \eea
where we have used (\ref{u-def}) that $u=4\mu^2/(\b^2K^2)$, and
define
\bea \Delta(Q_i,Q_j) & = & \b^2 \left\{ \b^2[ (2q_i\cdot q_j)^2-4
q_i^2 q_j^2]+4K^2[ \a_i\a_j(2q_i\cdot q_j)-\a_i^2 q_j^2-\a_j^2
q_i^2] \right\}~. \quad \eea
By using the relations%
\footnote{These relations can be found in \cite{BFG08}, below the
equation (5.16) in \cite{BFG08}. Here we have made a direct
generalization to the massive case, and also let $u=0$.}:
\bean \gb{P_{ji,1}|K|P_{ji,2}}\gb{P_{ji,2}|K|P_{ji,1}} & = & \b^4
{K^2\over Q_i^2} [ (2q_i\cdot q_j)^2-4 q_i^2 q_j^2]~, \\
\gb{P_{ji,1}|q_0|P_{ji,2}}\gb{P_{ji,2}|K|P_{ji,1}} &=& 2i \b^2
{ K^2\over Q_i^2} {q_0^2 \sqrt{\Delta(Q_i,Q_j)}}~, \\
\gb{P_{ji,1}|K|P_{ji,2}}\gb{P_{ji,2}|q_0|P_{ji,1}} &=& -2i \b^2 {
K^2\over Q_i^2} {q_0^2 \sqrt{\Delta(Q_i,Q_j)}}~, \qquad \eean
we have
\bea {\gb{\ell|q_0|\ell}^2 \over \left\langle \ell|K|\ell\right]^2}
\Bigg|{\scriptsize \begin{matrix} \\
\left\{\begin{matrix} \bket{\ell} & \rightarrow & \bket{P_{ji,2}}
\\ \ket{\ell} & \rightarrow  & \ket{P_{ji,1}}
\end{matrix}\right. \end{matrix}} = {\gb{\ell|q_0|\ell}^2 \over
\left\langle \ell|K|\ell\right]^2}\Bigg|{\scriptsize \begin{matrix}
\\ \left\{\begin{matrix} \bket{\ell} & \rightarrow & \bket{P_{ji,1}}
\\ \ket{\ell} & \rightarrow  & \ket{P_{ji,2}}
\end{matrix}\right. \end{matrix}} = {-4 (q_0^2)^2 \Delta(Q_i,Q_j)
\over \b^4 [ (2q_i\cdot q_j)^2-4 q_i^2 q_j^2]^2} ~. \qquad \eea
Substituting this back to (\ref{ell-square}), we can find that to
have $\W\ell_{ij}^2-\mu^2-M_1^2 = 0$, we only need the relation
\bea 4 q_0^2 K^2 = (2q_i\cdot q_j)^2-4 q_i^2 q_j^2~. \qquad
\Label{q0-square} \eea
To proof this relation, we notice
\bea \eps_{\mu_1 \mu_2 \mu_3 \mu_4} \eps^{\nu_1 \nu_2 \nu_3 \nu_4} =
- \sum_{p\in S_4} sign(p)
\delta_{\mu_1}^{\nu_{p(1)}}\delta_{\mu_2}^{\nu_{p(2)}}\delta_{\mu_3}^{\nu_{p(3)}}\delta_{\mu_4}^{\nu_{p(4)}}
~~~~\Label{eps-identity}\eea
where $p$ is permutation of $\nu_1,\nu_2, \nu_3,\nu_4$.
%The extra sign in
%the right hand side is because when we write $\eps_{0123}=1$ and
%move index up to $\eps^{0123}$ there is extra sign, since we are in
%metric $(+---)$ or $(-+++)$. To show it is right we check several
%situations. First assume $i_1=i_2$, then there is permutation pair
%with different sign to cancel each other. Similarly for the case of
%$j_1=j_2$. Now the only nonzero case is all $i_k$ different as well
%as all $j_k$ different. Then the right hand side has only one term
%is zero which is same as the left hand side.
Using this we can calculate that for  $q_0={ \eps_{\mu \nu \rho\xi}
q_i^\nu q_j^\rho K^\xi\over K^2}$ as
\bean q_0^2 & = & {1\over (K^2)^2} \sum_\mu \eps_{\mu \nu \rho\xi}
q_i^\nu q_j^\rho K^\xi \eps^{\mu \W\nu \W \rho \W \xi} (q_i)_{\W\nu}
(q_j)_{\W\rho} K_{\W \xi} \\ %~. \qquad \eean
%
%Since the sum of $\mu$ and totally anti-symmetric, we see that the
%nonzero sum of (\ref{eps-identity})
%
%\bean  (K^2)^2 q_0^2
& = &  {-1\over (K^2)^2 }[ K^2( q_i^2 q_j^2- (q_i\cdot q_j)^2)-
q_i^2 (K\cdot q_j)^2- q_j^2(q_i\cdot K)^2+2 (q_i\cdot q_j) (q_i\cdot
K)(q_j\cdot K)] ~. \qquad \eean
Noticing that $q_i\cdot k=q_j\cdot K=0$, we get
\bea  q_0^2 K^2=  (q_i\cdot q_j)^2 - q_i^2 q_j^2 ~, \qquad \eea
which is just the relation (\ref{q0-square}). Thus we prove that
it's true for the first on-shell condition in (\ref{ellij-onshell}).
%
%This is equivalent to the relation
%
%\bea  && (\eps_{\mu\nu\rho\sigma}K_i^\mu K_j^\nu K^\rho)\cdot
%(\eps^{\mu'\nu'\rho'\sigma}{K_i}_{\mu'} {K_j}_{\nu'} K_{\rho'})
%\nonumber\\
%&=& K_i^2(K_j\cdot K)^2 + K_j^2(K_i\cdot K)^2 + K^2(K_i\cdot K_j)^2
%- 2(K_i\cdot K_j)(K_i\cdot K)(K_j\cdot K)-K_i^2 K_j^2 K^2. \quad
%\eea
%

Then for the second condition, the proof is trivial since we have
\bea 2\W\ell_{ij}\cdot K =
\b{K^2\over\gb{\ell|K|\ell}}(-2P_{\la\W\la}\cdot K) -(\b-\a)K^2 = \a
K^2 =  K^2+M_1^2-M_2^2 . \qquad \eea

Therefore, the general box coefficient (\ref{Box-proof}) is
\bea C[Q_i,Q_j,K] & = & {1\over 2}\left({\cal
T}(\W\ell_{ij}) \cdot D_i(\W\ell_{ij}) \cdot D_j(\W\ell_{ij}) \Bigg|{\begin{matrix} \\
\left\{\scriptsize \begin{matrix} \bket{\ell} & \rightarrow &
\bket{P_{ji,2}}
\\ \ket{\ell} & \rightarrow  & \ket{P_{ji,1}}
\end{matrix}\right. \end{matrix}} +
\{P_{ji,1}\leftrightarrow P_{ji,2}\} \right) ,  \qquad
\Label{Box-improved} \eea
and the true box coefficient is obtained by subtracting the pentagon
contributions:
\bea \mathrm{Box}[K_{i},K_{j},K] & = & \frac{1}{2}\left({\cal
T}^{(N)}(\W\ell_{ij})\cdot D_{i}(\W\ell_{ij})D_{j}(\W\ell_{ij})
-\sum_r {\textrm{Pen}[K_i,K_j,K_r,K]\over D_r(\W\ell_{ij})} \right)
\Bigg|{\begin{matrix} \\
\left\{\scriptsize \begin{matrix} \bket{\ell} & \rightarrow &
\bket{P_{ji,2}}
\\ \ket{\ell} & \rightarrow  & \ket{P_{ji,1}}
\end{matrix}\right. \end{matrix}}    \nonumber\\ && + \{P_{ji,1} \leftrightarrow
P_{ji,2}\} ~. \quad  \eea
Notice that the null vectors $P_{ji,1(2)}$ are independent of $u$
now.

%%%%%%%%%%%%%%%%%%%%%%%%%%%%%%%%%%%%%%%%%%%%%%%%%%%
\subsubsection{Other methods to deal with $u$-dependence}
%%%%%%%%%%%%%%%%%%%%%%%%%%%%%%%%%%%%%%%%%%%%%%%%%%%

In above we have given one way to make $u$-dependence simpler. In
this part we will give other ways to deal with $u$-dependence.

We can expand any spinor into two independent ones. Sometimes we
need to use the spinor or anti-spinor components of $P_1, P_2$. For
this we can expand into arbitrary null momenta $a,b$, i.e. $\la_{P}=
{\ket{a}+ y \ket{b}\over \sqrt{t}}$ and $\W \la_{P}={|a]+ \O y
|b]\over \sqrt{t}}$ where $t$ is necessary normalization factor. We
can solve that
\bean t={\gb{b|a|b}\over \gb{b|P|b}}, \qquad y= {- \gb{a|P|b}\over
\gb{b|P|b}} ~~ \Longrightarrow ~~ \ket{P}= {-|P|b] \vev{a~b}\over
\sqrt{\gb{b|P|b} \gb{b|a|b}}}, \qquad |P]={ -|P|b\rangle [a~b]\over
\sqrt{\gb{b|P|b} \gb{b|a|b}}}~.~~~~\Label{P-spinor} \eean

When applying this idea to box coefficients,  we  use $a,b$ as the
$P_{ji,1}(u=0)$ and $P_{ji,2}(u=0)$. For the box, factor
$\sqrt{\gb{b|P|b} \gb{b|a|b}}$ will cancel out eventually as well as
factor $\vev{a~b},~[a~b]$ since numerator and denominator are same
degree polynomial. Thus we can get final replacement rule as
\bea
\ket{P_{ji,1}(u)}=|P_{ji,1}(u)|P_{ji,1}(u=0)],~~~~|P_{ji,1}(u)]=|
P_{ji,1}(u)\ket{P_{ji,1}(u=0)}, \qquad \eea
and
\bea
\ket{P_{ji,2}(u)}=|P_{ji,2}(u)|P_{ji,2}(u=0)],~~~~|P_{ji,2}(u)]=|
P_{ji,2}(u)\ket{P_{ji,2}(u=0)} . \qquad \eea
%

%%%%%%%%%%%%%%%%%%%%%%%%%%%%%%%%%%%%%%%%%%%%%%%%%%%%%%%%%%%%%%%%%
\section{Gluon example: $A(1^-,2^+,3^+,4^+,5^+)$}
%%%%%%%%%%%%%%%%%%%%%%%%%%%%%%%%%%%%%%%%%%%%%%%%%%%%%%%%%%%%%%%%%

In this part we use this simple five-gluon example to demonstrate our method%
\footnote{We emphasize that the main purpose of choosing this simple
example is to illustrate the using of our formulas. Since the tree
level inputs will have spurious poles, it can also serve as a good
example to check our formulas. The same procedure should be
implemented directly to compute more complicated cases.}. The
implementation of the formulas into automatic tools is
straightforward. In the following example, we do  all calculations
analytically with the Mathematica package {\tt S@M} \cite{S@M}.

By supersymmetric identities\cite{Grisaru1977}, the computation is
equivalent to one with a scalar field circulating in the loop
\cite{BDK-FiveGluon}. In the sense of four-dimensional unitarity,
this amplitude is cut free, i.e. it has only rational part and
${\cal O}(\eps)$ contribution. We will use our formulas to calculate
whole results including the ${\cal O}(\eps)$ contribution, which
according to our knowledge, should be the first time. The ${\cal
O}(\eps)$ contribution would be important in the higher loop
calculations, such as discussed in \cite{Bern:1998sc}, in order to
calculate $n$-parton two-loop amplitude to ${\cal O}(\eps^0)$, the
$(n+1)$-parton one-loop amplitude needs to be evaluated to ${\cal
O}(\eps^2)$.

This five-gluon amplitude can be expanded as a linear combination of
one pentagon, five boxes, five one-mass triangles, five two-mass
triangles and five bubbles. To obtain various coefficients, we need
all the five kinds of double cuts. We will illustrate the use of the
formulas by giving a detail discussion on the calculation of the
$K_{23}$-cut. For the other cuts, we list the final results
directly.

%%%%%%%%%%%%%%%%%%%%%%%%%%%%%%%%%%%%%%%%%%%%%%%%%%%
\subsection{$K_{23}$-cut}
%%%%%%%%%%%%%%%%%%%%%%%%%%%%%%%%%%%%%%%%%%%%%%%%%%%

The needed tree level input  can be obtained from the on-shell
recursion relation as \cite{BGKS-MassiveOnShell}
\bea {\cal T}_{K_{23}}(\W\ell) &=& A(\ell_1,4^+,5^+,1^-,\ell_2)
A(-\ell_2,2^+,3^+,-\ell_1)
\nonumber\\ &=& \left({\gb{1|\ell_2 K_{45}\ell_1|4}^2 \over
\vev{4~5}\vev{5~1}((\ell_1+k_4)^2-\mu^2)((\ell_1+K_{45})^2-\mu^2)[1|K_{45}\ell_1|4]}
-~{\mu^2 [4~5]^3 \over K_{451}^2 [5~1][1|K_{45}\ell_1|4]}\right)
\qquad
 \nonumber\\ & & \times \left({\mu^2 [2~3] \over \vev{2~3}
((\ell_1-k_3)^2-\mu^2)}\right)
\nonumber\\ &=& {\mu^2 [2~3] \over \vev{4~5}\vev{5~1}\vev{2~3}}
\cdot {\langle 1|(K_{23}-\W\ell) K_{45}\W\ell|4]^2 \over
D_{1}(\W\ell)D_{2}(\W\ell)D_{3}(\W\ell)~[1|K_{45}\W\ell|4]} - {\mu^4
[2~3] [4~5]^3 \over K_{451}^2 \vev{2~3}[5~1]} \cdot {1 \over
D_{1}(\W\ell)[1|K_{45}\W\ell|4]} \qquad \Label{Int-K23} \eea
where we have defined
\bea \W\ell \equiv \ell_1 = K_{23}-\ell_2, \eea
and %The input quantities for the calculation are:
\bea K=K_{23}, \qquad K_1=k_3, \qquad K_2=-k_4, \qquad K_3=-K_{45},
\qquad \eea
\bea D_{1}(\W\ell) = \langle 3|\W\ell|3], \qquad D_{2}(\W\ell) =
-\langle 4|\W\ell|4], \qquad D_{3}(\W\ell) = K_{45}^2 +2\W\ell\cdot
K_{45}. \qquad \Label{D123} \eea
Notice that there is a spurious pole in the tree level input:
\bea [1|K_{45}\W\ell|4]=[1~4]\langle 4|\W\ell|4]+[1~5]\langle
5|\W\ell|4], \qquad \Label{SP-K23} \eea
so it can serve as good example for checking our generalized
formulas. The first term of the input (\ref{Int-K23}) has degree
$0$, while the second term has degree $-2$, which does not
contribute to triangle and bubble. Furthermore, by comparing
(\ref{D123}) and (\ref{SP-K23}), we can find that there's no subtle
relation between the spurious pole and $D_{2}(\W\ell)$ or
$D_{3}(\W\ell)$, as that mentioned in Section \ref{subtle-box},  so
this second term has no box (and no pentagon) contributions either.
Therefore, we only need to consider the first term in
(\ref{Int-K23}) in the calculation, i.e. we can let
\bea {\cal T}_{K_{23}}(\W\ell) &=& {\mu^2 [2~3] \over
\vev{4~5}\vev{5~1}\vev{2~3}} \cdot {\langle 1|(K_{23}-\W\ell)
K_{45}\W\ell|4]^2 \over
D_{1}(\W\ell)D_{2}(\W\ell)D_{3}(\W\ell)~[1|K_{45}\W\ell|4]} .\qquad
\eea

From this cut we can calculate the coefficients of the pentagon,
three boxes, one one-mass triangle, two two-mass triangles, and one
bubble, as shown in Figure 2. We calculate them case by case. All
the needed formulas are collected in Section \ref{final-formulas}.

\begin{figure}[t]
\centerline{\includegraphics[height=9cm]{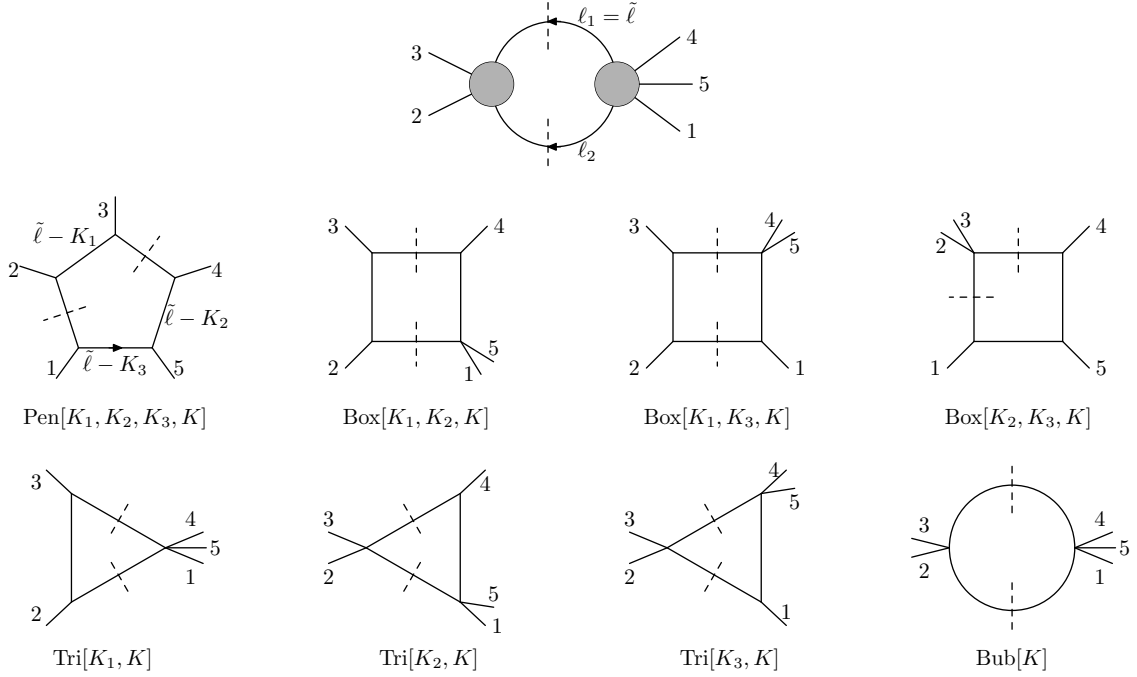}} \caption{The
first figure illustrates the $K_{23}$-cut. And the other figures
show all the scalar bases of which the coefficients can be obtained
from this cut.}
\end{figure}

%%%%%%%%%%%%%%%%%%%%%%%%%%%%%%%%%%%%%%%%%%%%%%%%%%%
\subsubsection{Coefficients of pentagon}
%%%%%%%%%%%%%%%%%%%%%%%%%%%%%%%%%%%%%%%%%%%%%%%%%%%

There is only one pentagon. Using (\ref{Pen-final}), we have
\bea \textrm{Pen}[K_1,K_2,K_3,K] &=& \left[{\cal T}_{K_{23}}(\W\ell)
\cdot D_1(\W\ell) D_2(\W\ell) D_3(\W\ell) \right]
\Big|_{\W\ell\rightarrow \W\ell_{(1,2,3)}}
\nonumber\\ &=& {\mu^2 [2~3] \over \vev{4~5}\vev{5~1}\vev{2~3}}
\cdot {\langle 1|(K_{23}-\W\ell_{(1,2,3)})
K_{45}\W\ell_{(1,2,3)}|4]^2 \over [1|K_{45}\W\ell_{(1,2,3)}|4]}
%
%\nonumber\\
%&=& -\frac{K_{23}{}^3 K_{45}{}^3 \langle 1|2\rangle ^2 \langle
%1|5\rangle ^2 \langle 3|4\rangle ^2 [2|1]  [4|3]
%[5|1]}{\text{$\Delta $}^3}\mu^2
%
\nonumber \\&=& \frac{s_{23}^3 s_{45}^3 s_{12} s_{15} s_{34}
\vev{1~2}\vev{3~4}\vev{1~5}}{\text{$\Delta $}^3}\mu^2\eea
where $\Delta$ is defined as
\bea \Delta =\vev{2~3} (\left\langle
4|k_{1}k_{2}k_{3}k_{1}|5\right\rangle +\left\langle
4|k_{1}k_{5}k_{2}k_{3}|5\right\rangle +\left\langle
4|k_{2}k_{3}k_{4}k_{1}|5\right\rangle) ~. \qquad \eea
Obviously, the pentagon contribution is ${\cal O}(\eps)$.

%%%%%%%%%%%%%%%%%%%%%%%%%%%%%%%%%%%%%%%%%%%%%%%%%%%
\subsubsection{Coefficients of box}
%%%%%%%%%%%%%%%%%%%%%%%%%%%%%%%%%%%%%%%%%%%%%%%%%%%

There are three boxes. Using (\ref{Box-final}), the box from
$K_1,K_2$ is
\bea c_{[51|2|3|4]} & = & \mathrm{Box}[K_1,K_2,K] \nonumber\\
& = & \frac{1}{2}\left[{\cal T}_{K_{23}}(\W\ell_{12})\cdot
D_{1}(\W\ell_{12})D_{2}(\W\ell_{12}) - {\textrm{Pentagon}\over
D_3(\W\ell_{12})} \right]
\Bigg|{\begin{matrix} \nonumber\\
\left\{\scriptsize \begin{matrix} \bket{\ell} & \rightarrow &
\bket{P_{21,2}}
\\ \ket{\ell} & \rightarrow  & \ket{P_{21,1}}
\end{matrix}\right. \end{matrix}}+ \{P_{21,1} \leftrightarrow P_{21,2}\} \quad\qquad
\nonumber\\ &=& -\frac{\langle 4~5\rangle  (\langle
1|2|4|3|2|1|5]+\langle 1|3|2|4|3|1|5]+\langle
1|5|4|2|3|4|5])^2}{s_{15} s_{23}{}^2  \langle 2~4\rangle  \langle
3~4\rangle \Delta } (\mu^2)^2 + (\textrm{$\mu^2$-term}) \eea
where we have changed $u$ to $\mu^2$, by using the relation
\bea u = {4 \mu^2 \over K_{23}^2} . \quad \eea
Similarly, we can get another two box coefficients
\bea c_{[23|4|5|1]} & = & \mathrm{Box}[K_2,K_3,K] = \frac{\langle
1~4\rangle \langle 1~5\rangle  [3~2] [5~4]^2}{
 \Delta } (\mu^2)^2 + (\textrm{$\mu^2$-term})
\\
c_{[45|1|2|3]} & = & \mathrm{Box}[K_1,K_3,K] = \frac{\langle
1~2\rangle \langle 1~3\rangle  [3~2]^2 [5~4]}{ \Delta } (\mu^2)^2 +
(\textrm{$\mu^2$-term}) \qquad \eea
As expected, after subtracting the pentagon contribution, the box
coefficient is a polynomial of $\mu^2$. The complete box
coefficients with $\mu^2$-term will be given in Appendix
\ref{mu2-box}.

%%%%%%%%%%%%%%%%%%%%%%%%%%%%%%%%%%%%%%%%%%%%%%%%%%%
\subsubsection{Coefficients of triangle}
%%%%%%%%%%%%%%%%%%%%%%%%%%%%%%%%%%%%%%%%%%%%%%%%%%%

There are three triangles. Using (\ref{Tri-final}) with $N=0$, the
triangle of $K_1$ is
\bea c_{[451|2|3]} = \mathrm{Tri}[K_1,K] &=&
\frac{1}{2}\frac{(K^{2})}{(-\sqrt{1-u})~\sqrt{-4q_1^2 K^2}}
\frac{1}{ \vev{P_{1,1}~P_{1,2}}} \nonumber
\\ & & \frac{d}{d\tau}\left({\left\langle
\ell|K|\ell\right] \over (K^{2})} {\cal T}_{K_{23}}(\W\ell)\cdot
D_{1}(\W\ell)
\Bigg|{\scriptsize \begin{matrix} \\ \left\{\begin{matrix}
\bket{\ell} & \rightarrow & |Q_1\ket{\ell} \quad\quad~~~
\\ \ket{\ell} & \rightarrow  & \ket{P_{1,1}-\tau P_{1,2}}
\end{matrix}\right. \end{matrix}} +
\{P_{1,1}\leftrightarrow P_{1,2}\} \right)\Big|_{\tau\to 0} \quad \nonumber\\
& = & -\frac{\left(\langle 1~4\rangle ^2 \langle 2~3\rangle ^2-2
\langle 1~3\rangle ^2 \langle 2~4\rangle ^2\right) \langle 1|2|3]}{2
\langle 1~5\rangle  \langle 2~3\rangle ^2 \langle 2~4\rangle ^2
\langle 3~4\rangle  \langle 4~5\rangle } \mu^2 \eea
Similarly, we get another two triangles
\bea c_{[4|51|23]} &=&  \mathrm{Tri}[K_{2},K] = -\frac{\langle
1~2\rangle \langle 1~4\rangle ^2 \langle 4|K_{23}|4]}{2 \langle
1~5\rangle \langle 2~3\rangle  \langle 2~4\rangle ^2 \langle
3~4\rangle \langle 4~5\rangle } \mu^2 \qquad
\\
c_{[1|23|45]} &=& \mathrm{Tri}[K_{3},K] = 0 \eea
%
%Since $n=0$, the coefficients have no $u$-dependence.

%%%%%%%%%%%%%%%%%%%%%%%%%%%%%%%%%%%%%%%%%%%%%%%%%%%
\subsubsection{Coefficient of bubble}
%%%%%%%%%%%%%%%%%%%%%%%%%%%%%%%%%%%%%%%%%%%%%%%%%%%

Using (\ref{Bub-final}) for $n=0$ and $k=3$, we have
\bea \mathrm{Bub}\left[K\right] = K^2 \left[
\mathcal{B}_{0,0}^{(0)}(0) + \sum_{r=1}^{3}
\left(\mathcal{B}_{0,0}^{(r,0,1)}(0) -
\mathcal{B}_{0,0}^{(r,0,2)}(0) \right)\right] \Label{ex-bub} \eea
where
\bea \mathcal{B}_{0,0}^{(0)}(0) & = & \frac{2\eta\cdot
K}{K^2}\frac{1}{\left\langle \ell\ \eta\right\rangle } {\cal
T}(\W\ell) \Bigg|{\scriptsize
\begin{matrix} \\ \left\{\begin{matrix} \bket{\ell} & \rightarrow &
|K\ket{\ell}
\\ \ket{\ell} & \rightarrow  & |K|\eta]
\end{matrix}\right. \end{matrix}} \\
\mathcal{B}_{0,0}^{(r,0,1)}(0) & = & \frac{1}{\sqrt{1-u}~
\sqrt{-4q_r^2 K^2}} \frac{\left\langle
\ell|\eta|P_{r,1}\right]}{\left\langle
\ell|K|P_{r,1}\right]}\frac{1}{\left\langle \ell|\eta
K|\ell\right\rangle } \left(\frac{\left\langle
\ell|K|\ell\right]}{(K^{2})} {\cal T}(\W\ell) \cdot
D_r(\W\ell)\right)
\Bigg|{\scriptsize \begin{matrix} \\
\left\{\begin{matrix} \bket{\ell} & \rightarrow & |K\ket{\ell}
\\ \ket{\ell} & \rightarrow  & \ket{P_{r,1}}
\end{matrix}\right. \end{matrix}} \\
\mathcal{B}_{0,0}^{(r,0,2)}(0) & = & \frac{1}{\sqrt{1-u}
~\sqrt{-4q_r^2 K^2}}\frac{\left\langle
\ell|\eta|P_{r,2}\right]}{\left\langle
\ell|K|P_{r,2}\right]}\frac{1}{\left\langle \ell|\eta
K|\ell\right\rangle } \left(\frac{\left\langle
\ell|K|\ell\right]}{(K^{2})} {\cal T}(\W\ell) \cdot D_r(\W\ell)
\right)
\Bigg|{\scriptsize \begin{matrix} \\
\left\{\begin{matrix} \bket{\ell} & \rightarrow & |K\ket{\ell}
\\ \ket{\ell} & \rightarrow  & \ket{P_{r,2}}
\end{matrix}\right. \end{matrix}}
\eea
The explicit results for each terms are
\bea \mathcal{B}_{0,0}^{(0)}(0) & = & -\frac{\langle 1|K_{23}|\eta
]^3}{s_{23} \langle 1~5\rangle \langle 2~3\rangle ^3 \langle
4~5\rangle  [2~\eta ] [3~\eta ] \langle 4|K_{23}|\eta ]} \mu^2
\nonumber\\
\mathcal{B}_{0,0}^{(1,0,1)}(0) & = & \frac{\langle 1~3\rangle ^3
[3~\eta ]}{s_{23} \langle 1~5\rangle  \langle 2~3\rangle ^3 \langle
3~4\rangle  \langle 4~5\rangle [2~\eta ]} \mu^2
\nonumber\\
\mathcal{B}_{0,0}^{(1,0,2)}(0) & = & -\frac{\langle 1~2\rangle ^3
[2~\eta ]}{s_{23} \langle 1~5\rangle  \langle 2~3\rangle ^3 \langle
2~4\rangle  \langle 4~5\rangle  [3~\eta ]} \mu^2  \nonumber\\
\mathcal{B}_{0,0}^{(2,0,1)}(0) & = & \frac{\langle 1~4\rangle ^3
\langle 5|4|\eta ]}{\langle 1~5\rangle  \langle 2~3\rangle \langle
2~4\rangle  \langle 3~4\rangle  \langle 4~5\rangle ^2 \langle
4|K_{23}|4] \langle 4|K_{23}|\eta ]} \mu^2 \nonumber \eea
and
\bea \mathcal{B}_{0,0}^{(2,0,2)}(0) & = &
\mathcal{B}_{0,0}^{(3,0,1)}(0) = \mathcal{B}_{0,0}^{(3,0,2)}(0) = 0
. \nonumber \qquad \eea
After taking them back in (\ref{ex-bub}) and doing the summation, we
can find
\bea c_{[23|451]} = \mathrm{Bub}\left[K_{23}\right] = -\frac{\langle
1~2\rangle ^3 \langle 3~4\rangle ^2 [4~2]+\langle 1~3\rangle ^3
\langle 2~4\rangle ^2 [4~3]-3 \langle 1~2\rangle \langle 1~3\rangle
\langle 1~5\rangle  \langle 2~4\rangle  \langle 3~4\rangle
[5~4]}{\langle 1~5\rangle  \langle 2~3\rangle ^3 \langle 2~4\rangle
\langle 3~4\rangle  \langle 4~5\rangle  \langle 4|K_{23}|4]} \mu^2
\quad \eea
As expected, $\eta$ doesn't appear in the final result.
% and the result also has no $u$-dependence since $n=0$.

%%%%%%%%%%%%%%%%%%%%%%%%%%%%%%%%%%%%%%%%%%%%%%%%%%%
\subsection{Other cuts}
%%%%%%%%%%%%%%%%%%%%%%%%%%%%%%%%%%%%%%%%%%%%%%%%%%%

For other cuts, we first give the necessary tree level input for the
calculation \cite{BGKS-MassiveOnShell}, then list the final results
directly.

%%%%%%%%%%%%%%%%%%%%%%%%%%%%%%%%%%%%%%%%%%%%%%%%%%%
\subsubsection{Tree level input}
%%%%%%%%%%%%%%%%%%%%%%%%%%%%%%%%%%%%%%%%%%%%%%%%%%%

The needed tree level input obtained from the on-shell recursion
relation for $K_{34}$-cut is
\bea {\cal T}_{K_{34}}(\W\ell) &=& A(\ell_1,5^+,1^-,2^+,\ell_2)
A(-\ell_2,3^+,4^+,-\ell_1)
\nonumber\\ &=& \left({\gb{1|\ell_1|5}^2 \gb{1|\ell_2|2}^2 \over
\vev{5~1}\vev{1~2}((\ell_1+k_5)^2-\mu^2)((\ell_1+K_{51})^2-\mu^2)
[2|K_{51}\ell_1|5]} -{\mu^2 [5~2]^4 \over K_{512}^2
[5~1][1~2][2|K_{51}\ell_1|5]}\right)
\nonumber\\ && \times \left({\mu^2 [3~4] \over \vev{3~4}
((\ell_1-k_4)^2-\mu^2)}\right)
\nonumber\\ &=& {\mu^2 [3~4] \over\vev{3~4} \vev{5~1}\vev{1~2} }
\cdot {\langle 1|\W\ell|5]^2 \langle 1|(K_{34}-\W\ell)|2]^2 \over
((\W\ell-k_4)^2-\mu^2)((\W\ell+k_5)^2-\mu^2)((\W\ell+K_{51})^2-\mu^2)
~[2|K_{51}\W\ell|5]}
\nonumber\\ & &- ~{\mu^4 [3~4] [5~2]^4 \over K_{512}^2
\vev{3~4}[5~1][1~2] }\cdot {1\over ((\W\ell-k_4)^2-\mu^2)
[2|K_{51}\W\ell|5]} \qquad \Label{Int-K34} \eea
and $K_{23}$-cut it is
\bea {\cal T}_{K_{12}}(\W\ell) &=& A(\ell_1,3^+,4^+,5^+,\ell_2)
A(-\ell_2,1^-,2^+,-\ell_1) \nonumber\\ &=& \left({-\mu^2
[5|K_{34}\ell_1|3] \over
\vev{3~4}\vev{4~5}((\ell_1+k_3)^2-\mu^2)((\ell_1+K_{34})^2-\mu^2)}\right)
\left(-{\gb{1|\ell_1|2}^2 \over K_{12}^2
((\ell_1-k_2)^2-\mu^2)}\right)
\nonumber\\
&=& {\mu^2 \over \vev{3~4}\vev{4~5}K_{12}^2 }\cdot
{[5|K_{34}\W\ell|3] ~\langle 1|\W\ell|2 ]^2 \over
((\W\ell-k_2)^2-\mu^2)((\W\ell+k_3)^2-\mu^2)((\W\ell+K_{34})^2-\mu^2)}~.
\quad \Label{Int-K12} \eea
Similar to the case of $K_{23}$-cut, the second term in ${\cal
T}_{K_{34}}(\W\ell)$ can also be neglected. $K_{45}$-cut and
$K_{51}$-cut are not necessary, since they are related to
$K_{23}$-cut and $K_{12}$-cut by symmetry, although we have also
done the calculation independently in order to show the results are
consistent. The coefficients are related by a symmetry operation,
for example
\bea c_{[51|2|3|4]} =
-c_{[12|3|4|5]}\Big|_{2\leftrightarrow5,3\leftrightarrow4}, \qquad
\Label{coef-sym} \eea
which is similar for other coefficients.

%%%%%%%%%%%%%%%%%%%%%%%%%%%%%%%%%%%%%%%%%%%%%%%%%%%
\subsubsection{Other coefficients}
%%%%%%%%%%%%%%%%%%%%%%%%%%%%%%%%%%%%%%%%%%%%%%%%%%%

The other two boxes are
\bea c_{[12|3|4|5]} &=& -\frac{\langle 2~3\rangle (\langle
1|2|3|5|4|3|2]+\langle 1|4|5|3|4|1|2]+\langle
1|5|3|4|5|1|2])^2}{s_{12} s_{45}{}^2 \langle 3~4\rangle  \langle
3~5\rangle \Delta } (\mu^2)^2 +
(\textrm{$\mu^2$-term})  \qquad \qquad   \\
c_{[34|5|1|2]} &=&   \frac{\langle 1~2\rangle  \langle 1~5\rangle
\langle 2~3\rangle \langle 4~5\rangle [4~3] [5~2]^2}{\langle
2~5\rangle \langle 3~4\rangle \Delta } (\mu^2)^2 +
(\textrm{$\mu^2$-term})\eea
The other four one-mass triangles are
\bea c_{[345|1|2]} &=& \frac{\langle 1~2\rangle ^2 \langle
1~5\rangle [2|1]}{2 \langle 2~3\rangle \langle 2~5\rangle ^2 \langle
3~4\rangle \langle 4~5\rangle } \mu^2
\\
c_{[123|4|5]} &=& \frac{\left(2 \langle 1~4\rangle ^2 \langle
3~5\rangle ^2-\langle 1~3\rangle ^2 \langle 4~5\rangle ^2\right)
\langle 1|5|4]}{2 \langle 1~2\rangle  \langle 2~3\rangle  \langle
3~4\rangle \langle 3~5\rangle ^2 \langle 4~5\rangle ^2}
\mu^2  \\
c_{[234|5|1]} &=& \frac{\langle 1~2\rangle  \langle 1~5\rangle ^2
[5~1]}{2 \langle 2~3\rangle  \langle 2~5\rangle ^2 \langle
3~4\rangle \langle 4~5\rangle } \mu^2
\\
c_{[512|3|4]} &=&  { [4~3] \over 2 \langle 1~2\rangle \langle
1~5\rangle  \langle 2~3\rangle \langle 2~4\rangle ^2 \langle
3~4\rangle ^2 \langle 3~5\rangle ^2 \langle 4~5\rangle ^3}
\\ &&  \Big(\langle 1~4\rangle ^4 \langle 3~5\rangle ^2
\left(\langle 2~5\rangle ^2 \langle 3~4\rangle ^2-2 \langle
2~4\rangle ^2 \langle 3~5\rangle ^2\right)-2 \langle 1~4\rangle
\langle 1~5\rangle ^3 \langle 2~4\rangle ^2 \langle 3~4\rangle ^3
\langle 3~5\rangle \nonumber\\ && \left.\left. +2 \langle 1~4\rangle
^3 \langle 1~5\rangle  \langle 2~4\rangle \langle 3~4\rangle \langle
3~5\rangle ^2 (\langle 2~3\rangle \langle 4~5\rangle + \langle
2~4\rangle  \langle 3~5\rangle ) \right.\right.
 +\langle 1~5\rangle ^4 \langle 2~4\rangle ^2 \langle
3~4\rangle ^4 \Big) \mu^2 \qquad \nonumber\eea
The other three two-mass triangles are
\bea c_{[5|12|34]} &=&  \frac{\langle 1~5\rangle ^2 (\langle
1~3\rangle  \langle 2~5\rangle +\langle 1~2\rangle  \langle
3~5\rangle ) \langle 5|K_{12}|5]}{2 \langle 1~2\rangle  \langle
2~5\rangle ^2 \langle 3~4\rangle \langle
3~5\rangle ^2 \langle 4~5\rangle } \mu^2  \qquad \qquad \qquad \qquad   \\
c_{[2|34|51]} &=& \frac{\langle 1~2\rangle ^2 (\langle 1~5\rangle
\langle 2~4\rangle +\langle 1~4\rangle  \langle 2~5\rangle ) \langle
2|K_{51}|2]}{2 \langle 1~5\rangle \langle 2~3\rangle  \langle
2~4\rangle ^2 \langle 2~5\rangle ^2 \langle 3~4\rangle } \mu^2  \\
c_{[3|45|12]} &=& -\frac{\langle 1~3\rangle ^2 \langle 1~5\rangle
(\langle 3|4|3]+\langle 3|5|3])}{2 \langle 1~2\rangle  \langle
2~3\rangle  \langle 3~4\rangle  \langle 3~5\rangle ^2 \langle
4~5\rangle } \mu^2 \qquad \eea
The other four bubbles are
\bea c_{[45|123]} &=&  -\frac{-3 \langle 1~2\rangle  \langle
1~4\rangle \langle 1~5\rangle  \langle 3~4\rangle  \langle
3~5\rangle [3~2]+\langle 1~4\rangle ^3 \langle 3~5\rangle ^2
[4~3]+\langle 1~5\rangle ^3 \langle 3~4\rangle ^2 [5~3]}{\langle
1~2\rangle \langle 2~3\rangle  \langle 3~4\rangle  \langle
3~5\rangle \langle 4~5\rangle ^3 \langle 3|K_{45}|3]} \mu^2  \qquad
\qquad \qquad \qquad
\\ c_{[12|345]} &=& -\left (\frac{\langle 1~3\rangle ^3 (\langle 2|1|5]
(\langle 3~5\rangle (\langle 2|4|1]+2 \langle 2|5|1])+\langle
2~5\rangle  \langle 3|4|1])-\langle 2~3\rangle \langle
2|4|5|2|1])}{\langle 2~3\rangle ^3 \langle 3~4\rangle \langle
3~5\rangle  \langle 4~5\rangle (\langle 2|1|5|3\rangle
-\langle 2|5|4|3\rangle ) \langle 3|K_{45}|3]} \right. \\
&&  \left. +\frac{\langle 1~2\rangle \left(s_{15} \langle 1~3\rangle
\langle 2~3\rangle +2 \langle 1~3\rangle  \langle 2|5|2|3\rangle
-\langle 1~2\rangle \langle 3|4|5|3\rangle \right)}{\langle
2~3\rangle ^2 \langle 2~5\rangle \langle 3~4\rangle \langle
4~5\rangle  (\langle
2|1|5|3\rangle -\langle 2|5|4|3\rangle )}\right) \mu^2  \nonumber \\
c_{[51|234]} &=& -\left(\frac{\langle 1~4\rangle ^3 (\langle 5|1|2]
(\langle 2~5\rangle  \langle 4|3|1]+\langle 2~4\rangle  (2 \langle
5|2|1]+\langle 5|3|1]))-\langle 4~5\rangle \langle
5|3|2|5|1])}{\langle 2~3\rangle  \langle 2~4\rangle  \langle
3|4\rangle  \langle 4~5\rangle ^3 (-\langle 4|2|1|5\rangle +\langle
4|3|2|5\rangle ) \left\langle 4\left|K_{23}\right|4\right]} \right.
\\ && \left.+\frac{\langle 1~5\rangle \left(-s_{12} \langle
1~4\rangle  \langle 4~5\rangle +\langle 1~5\rangle  \langle
4|2|3|4\rangle -2 \langle 1~4\rangle  \langle 4|5|2|5\rangle
\right)}{\langle 2~3\rangle  \langle 2~5\rangle \langle 3~4\rangle
\langle 4~5\rangle ^2 (-\langle 4|2|1|5\rangle +\langle
4|3|2|5\rangle )} \right) \mu^2  \nonumber %\nonumber\\
\eea
\bea
c_{[34|512]} &=&  \frac{-1}{\langle 1~2\rangle \langle 1~5\rangle
\langle 3~4\rangle ^3 }\left(\frac{\langle 1~5\rangle ^2 \langle
3~4\rangle  \langle 1|5|1|5\rangle  \langle 3|4|3|4\rangle }{\langle
2~5\rangle  \langle 3~5\rangle  \langle 4~5\rangle \langle 5|2|1]
\langle 5|K_{12}|5]}+\frac{\langle 1~3\rangle ^2 \langle
1|3|1|3\rangle }{\langle 2~3\rangle  \langle 3~5\rangle [4~1]}
+\frac{\langle 1~4\rangle ^2 \langle 1|4|1|4\rangle }{\langle
2~4\rangle  \langle 4~5\rangle [3~1]}\right. \\ && \left.
-\frac{(\langle 4|1|2|5\rangle -\langle 4|3|1|5\rangle )^4}{\langle
4~5\rangle ^3 \langle 2|5|1] \langle 5|2|1] \langle 5|4|1] [3~1]}
+\frac{\langle 1~2\rangle ^4 \langle 3~4\rangle ^3 (\langle 5|1|2]
[3~1]-\langle 5|2|1] [3~2])}{\langle 2~3\rangle \langle 2~4\rangle
\langle 2~5\rangle (-\langle 4|2|1|5\rangle +\langle 4|3|2|5\rangle
) \langle 2|5|1]}\right)\mu^2
 \qquad\nonumber \eea

As it is obvious, one given coefficient can be calculated from
various cuts. Using Mathematica we calculate them and do find same
results. Also, we check that the coefficients satisfy the
symmetrical relation (\ref{coef-sym}). These are strong tests for
our method.

%%%%%%%%%%%%%%%%%%%%%%%%%%%%%%%%%%%%%%%%%%%%%%%%%%%%%%%%%%%%%%%%%
\subsection{Rational parts}
%%%%%%%%%%%%%%%%%%%%%%%%%%%%%%%%%%%%%%%%%%%%%%%%%%%%%%%%%%%%%%%%%

To extract the rational parts, we need the following relation
\cite{Bern-Chalmers,Bern-Morgan}
\bea I_4^D[(\mu^2)^2] = -{1\over 6}+{\cal O}(\eps) , \quad
I_3^D[\mu^2] ={1\over 2}+{\cal O}(\eps) , \quad  I_2^D[\mu^2] =
-{K^2\over 6}+{\cal O}(\eps) . \eea
The rational part of the amplitude is
\bea R &=& {i\over (4\pi)^2} \Bigg[ -{1\over
6(\mu^2)^2}\left(c_{[12|3|4|5]}+c_{[23|4|5|1]}+c_{[34|5|1|2]}+c_{[45|1|2|3]}+c_{[51|2|3|4]}\right)
\Big|_{\textrm{$(\mu^2)^2$-term}} \nonumber\\
&& + ~~{1\over
2\mu^2}~\left(c_{[123|4|5]}+c_{[234|5|1]}+c_{[345|1|2]}+c_{[451|2|3]}+c_{[512|3|4]}\right.
\nonumber\\
&& \quad\qquad~ \left.+
c_{[12|34|5]}+c_{[23|45|1]}+c_{[34|51|2]}+c_{[45|12|3]}+c_{[51|23|4]}\right)
\nonumber\\
&& - ~~{1\over 6\mu^2}~\left(c_{[12|345]}K_{12}^2
+c_{[23|451]}K_{23}^2 +c_{[34|512]}K_{34}^2 +c_{[45|123]}K_{45}^2
+c_{[51|234]}K_{51}^2 \right) \Bigg]. \qquad \eea
The rational part of the same amplitude has been given in
\cite{BDK-FiveGluon} as
\bea \W R &=& {i\over 48\pi^2} {1\over [1~2]
\vev{2~3}\vev{3~4}\vev{4~5}[5~1]}
\Bigg[-(s_{23}+s_{34}+s_{45})[2~5]^2 - [2~4]\vev{4~3}[3~5][2~5]
\nonumber\\ && \left. - {[1~2][1~5]\over\vev{1~2}\vev{1~5}}
\left(\vev{1~2}^2 \vev{1~3}^2 {[2~3]\over\vev{2~3}}+ \vev{1~3}^2
\vev{1~4}^2 {[3~4]\over\vev{3~4}}+ \vev{1~4}^2 \vev{1~5}^2
{[4~5]\over\vev{4~5}}\right)\right]. \qquad \eea
With the help of Mathematica one can easily check the following
result
\bea R = {1\over2} \W R \eea
The factor 1/2 is because $R$ is for the amplitude with a real
scalar circulating in the loop, while that for $\W R$ it is a
complex scalar. As we can see, the spurious poles, such as $\Delta$
in the box coefficient, are all cancelled.

Our result $R$ looks more complicated than the $\W R$. The reason is
because when we use the Mathematica, we have chosen a basis for
spinors and expressed other quantities by this basis. The choice
will sacrifice some simplicity, but make the result unique for the
chosen basis and easy to compare. Furthermore, it makes the
evaluation straightforward either analytically or numerically.

%%%%%%%%%%%%%%%%%%%%%%%%%%%%%%%%%%%%%%%%%%%%%%%%%%%%%%%%%%%%%%%%%
\section{Summary}
%%%%%%%%%%%%%%%%%%%%%%%%%%%%%%%%%%%%%%%%%%%%%%%%%%%%%%%%%%%%%%%%%

In this paper, we generalize  formulas for  coefficients obtained by
$D$-dimensional unitarity method \cite{BF06, BF07, BFM08, BFG08} to
the case where spurious poles are allowed in tree-level input. While
keeping algebraic expressions explicitly, this generalization makes
sure that we can use the most compact tree-level input to reduce the
complexity of computations. We summarize some main points of our
final formulas as following:

\begin{itemize}

\item
The tree-level input (obtained by any method) can be inserted into
the formulas directly without any modification.

\item
The formulas are for the full coefficients, i.e. including all the
contribution of cut part, rational part and ${\cal O}(\eps)$ part.
If we want only the cut part, we can take the four-dimensional tree
level input, and let $u=0$ in the formulas. This will simplify the
formulas dramatically.

\item
The formulas are for the general case, massive or massless.

%\item
%By using the formulas, we don't need to do any integration or
%solving any equations. The final coefficients can be read directly
%from the tree level input by doing some derivative operations.

\item
The formulas are at the analytical level. It can give compact
analytical expression for some results, such as that shown in the
example of Section 6.

\item
The formulas are suitable for the non-renormalizable case, such as
the one loop calculation of gravity theory.

\end{itemize}
%

%%%%%%%%%%%%%%%%%%%%%%%%%%%%%%%
\acknowledgments
%%%%%%%%%%%%%%%%%%%%%%%%%%%%%%%

We would like to thank Ruth Britto for participation at the early
stage of this project and C.-J. Zhu for helpful comments on the
draft. BF is supported by Qiu-Shi Professor Fellowship from Zhejiang
University, China. GY is supported by funds from the National
Natural Science Foundation of China with grant Nos. 10475104 and
10525522.

%%%%%%%%%%%%%%%%%%%%%%%%%
\appendix
%%%%%%%%%%%%%%%%%%%%%%

%%%%%%%%%%%%%%%%%%%%%%%%%%%%%%%%%%%%%%%%%%%%%%%%%%%%%%%%%%%%%%%%%%%%%
\section{\label{rev-formulas} Review of formulas}
%%%%%%%%%%%%%%%%%%%%%%%%%%%%%%%%%%%%%%%%%%%%%%%%%%%%%%%%%%%%%%%%%%%%%

We review  formulas of various coefficients for the standard input
form \cite{{BF06}, {BF07}, {BFM08}, {BFG08}}
\bea {\cal T}^{(n)}_{sf}(\W\ell) = \frac{ \prod_{j=1}^{n+k}
(-2\W\ell\cdot P_j)}{\prod_{i=1}^k D_i(\W\ell) } = {(K^2)^{n}\over
\gb{\ell|K|\ell}^{n}} {\prod_{j=1}^{n+k} \gb{\ell|R_j|\ell}\over
\prod_{i=1}^k \gb{\ell|Q_i|\ell}}. \quad  \eea
We have defined the quantities:
\bea R_j(u) \equiv -\b(\sqrt{1-u}) p_j +\b_j K, \qquad Q_i(u) \equiv
-\b (\sqrt{1-u}) q_i+\a_i K,~~~\Label{R-Q-massive}\eea
where
\bea p_j & \equiv & \left(P_j-{P_j\cdot K\over K^2}K \right), \qquad
\b_j \equiv -{(P_j\cdot K)\over K^2}\left(1+{M_1^2-M_2^2\over K^2}
\right), ~~~\Label{beta-mass}
\\ q_i & \equiv & \left(K_i-{K_i\cdot
K\over K^2}K \right),   \qquad  \a_i  \equiv  -{(K_i\cdot K)\over
K^2}\left(1+{M_1^2-M_2^2\over K^2} \right)+ {K_i^2+M_1^2-m_i^2\over
K^2}.~~~\Label{alpha-mass}\eea
and $\b$ are given by (\ref{ab-def}).

%%%%%%%%%%%%%%%%%%%%%%%%%%%%%%%%%%%%%%
\subsection{\label{u-unsimp} Formulas}
%%%%%%%%%%%%%%%%%%%%%%%%%%%%%%%%%%%%%%

The box coefficients (pentagons not separated) are given as
\bea C[Q_i,Q_j,K] & = & {(K^2)^{n+2}\over 2}\left({\prod_{s=1}^{k+n}
\gb{P_{ji,1}|R_s |P_{ji,2}}\over \gb{P_{ji,1}|K
|P_{ji,2}}^{n+2}\prod_{t=1,t\neq i,j}^k \gb{P_{ji,1}|Q_t
|P_{ji,2}}}+ \{P_{ji,1}\leftrightarrow P_{ji,2}\}
\right),~~\Label{box-exp}\eea
where
\bea
\Delta_{ji} &=& (2Q_j \cdot Q_i)^2-4 Q_j^2 Q_i^2  \nonumber \\
P_{ji,1} &=& Q_j + \left( {-2Q_j \cdot Q_i + \sqrt{\Delta_{ji}}\over 2Q_i^2} \right) Q_i \nonumber \\
P_{ji,2} &=& Q_j + \left( {-2Q_j \cdot Q_i - \sqrt{\Delta_{ji}}\over
2Q_i^2} \right) Q_i~~~~\Label{box-null} \eea

The triangle coefficients are given as
\bea C[Q_s,K] & = & { (K^2)^{n+1}\over
2}\frac{1}{(\sqrt{\Delta_s})^{n+1}}\frac{1}{(n+1)!
\vev{P_{s,1}~P_{s,2}}^{n+1}} \nonumber
\\ & & \times \frac{d^{n+1}}{d\tau^{n+1}}\left.\left({\prod_{j=1}^{k+n}
\vev{P_{s,1}-\tau P_{s,2} |R_j Q_s|P_{s,1}-\tau P_{s,2}}\over
\prod_{t=1,t\neq s}^k \vev{P_{s,1}-\tau P_{s,2}|Q_t Q_s
|P_{s,1}-\tau P_{s,2}}} + \{P_{s,1}\leftrightarrow
P_{s,2}\}\right)\right|_{\tau=0},~~~~~\Label{tri-exp}\eea
where
\bea
\Delta_{s} &=& (2Q_s \cdot K)^2-4 Q_s^2 K^2 \nonumber \\
P_{s,1} &=& Q_s + \left({-2Q_s \cdot K + \sqrt{\Delta_{s}}\over
2K^2} \right) K
\nonumber \\
P_{s,2} &=& Q_s + \left({-2Q_s \cdot K - \sqrt{\Delta_{s}}\over
2K^2} \right) K ~~~~\Label{tri-null} \eea
Note that the triangle coefficient is present only when $n\geq -1$.

The bubble coefficients are given as
\bea
 C[K] = (K^2)^{n+1} \sum_{q=0}^n {(-1)^q\over q!} {d^q \over
ds^q}\left.\left( {\cal B}_{n,n-q}^{(0)}(s)+\sum_{r=1}^k\sum_{a=q}^n
\left({\cal B}_{n,n-a}^{(r;a-q;1)}(s)-{\cal
B}_{n,n-a}^{(r;a-q;2)}(s)\right)\right)\right|_{s=0},~~~~~\Label{bub-exp}
\eea
where
\bea {\cal B}_{n,t}^{(0)}(s)\equiv {d^n\over d\tau^n}\left.\left( {1
\over n! [\eta|\W \eta K|\eta]^{n}}  {(2\eta\cdot K)^{t+1} \over
(t+1) (K^2)^{t+1}}{\prod_{j=1}^{n+k} \vev{\ell|R_j
(K+s\eta)|\ell}\over \vev{\ell~\eta}^{n+1} \prod_{p=1}^k \vev{\ell|
Q_p(K+s\eta)|\ell}}|_{\ket{\ell}\to |K-\tau \W \eta|\eta]
}\right)\right|_{\tau= 0},~~~\Label{cal-B-0}\eea
\bea & & {\cal B}_{n,t}^{(r;b;1)}(s)  \equiv  {(-1)^{b+1}\over
 b! \sqrt{\Delta_r}^{b+1} \vev{P_{r,1}~P_{r,2}}^b}{d^b \over d\tau^{b}}
\left({1\over (t+1)} {\gb{P_{r,1}-\tau
P_{r,2}|\eta|P_{r,1}}^{t+1}\over \gb{P_{r,1}-\tau
P_{r,2}|K|P_{r,1}}^{t+1}}\right. \nonumber \\ & & \times
\left.\left. {\vev{P_{r,1}-\tau P_{r,2}|Q_r \eta|P_{r,1}-\tau
P_{r,2}}^{b} \prod_{j=1}^{n+k} \vev{P_{r,1}-\tau P_{r,2}|R_j
(K+s\eta)|P_{r,1}-\tau P_{r,2}}\over \vev{P_{r,1}-\tau P_{r,2}|\eta
K|P_{r,1}-\tau P_{r,2}}^{n+1} \prod_{p=1,p\neq r}^k
\vev{P_{r,1}-\tau P_{r,2}| Q_p(K+s\eta)|P_{r,1}-\tau
P_{r,2}}}\right)\right|_{\tau=0},~~~\Label{cal-B-r-1}\eea
\bea & & {\cal B}_{n,t}^{(r;b;2)}(s)  \equiv  {(-1)^{b+1}\over
 b! \sqrt{\Delta_r}^{b+1} \vev{P_{r,1}~P_{r,2}}^{b}}{d^{b} \over d\tau^{b}}
\left({1\over (t+1)} {\gb{P_{r,2}-\tau
P_{r,1}|\eta|P_{r,2}}^{t+1}\over \gb{P_{r,2}-\tau
P_{r,1}|K|P_{r,2}}^{t+1}}\right. \nonumber \\ & & \times
\left.\left. {\vev{P_{r,2}-\tau P_{r,1}|Q_r \eta|P_{r,2}-\tau
P_{r,1}}^{b} \prod_{j=1}^{n+k} \vev{P_{r,2}-\tau P_{r,1}|R_j
(K+s\eta)|P_{r,2}-\tau P_{r,1}}\over \vev{P_{r,2}-\tau P_{r,1}|\eta
K|P_{r,2}-\tau P_{r,1}}^{n+1} \prod_{p=1,p\neq r}^k
\vev{P_{r,2}-\tau P_{r,1}| Q_p(K+s\eta)|P_{r,2}-\tau
P_{r,1}}}\right)\right|_{\tau=0}.~~~\Label{cal-B-r-2}\eea
where $\Delta_r, P_{r,1}, P_{r,2}$ are given by (\ref{tri-null}),
and $\eta, \W\eta$ are arbitrary, generically chosen null vectors.
Note that the bubble coefficient exists only when $n\geq 0$.

%%%%%%%%%%%%%%%%%%%%%%%%%%%%%%%%%%%%%%
\subsection{\label{u-simp} Formulas with $u$ simplified}
%%%%%%%%%%%%%%%%%%%%%%%%%%%%%%%%%%%%%%

The $u$-dependence for  above formulas can be simplified further
\cite{BFG08,BFM08}. Here we collect these simplified formulas. The
pentagon formula is also given. The most important simplification is
for the null spinors. As we have emphasized in main text, in the
simplified formulas,  $u$-dependence becomes minimum and much
simpler, especially the null momenta do not depend on $u$ anymore..
The null spinors $P_{ji,a}$ ($a=1,2$) in box formula are taken as
\bea \ket{P_{ji,a}} \equiv \ket{P_{ji,a}(u=0)}, \qquad |P_{ji,a}]
\equiv |P_{ji,a}(u=0)] .  \qquad \Label{u-free-Pij} \eea
while the null spinors $P_{s,a}$ in triangle and bubble formulas are
taken as,
\bea \ket{P_{s,a}} \equiv \ket{P_{q_s,a}},~~~~~~|P_{s,a}] \equiv
|P_{q_s,a}] , \qquad \Label{ufree-spinor} \eea
where
\bea P_{q_s,a} \equiv q_s \pm \left({\sqrt{-q_s^2 \over K^2}}
\right)K. \qquad \Label{ufree-null} \eea
Notice that for simplicity we use the same symbols for the null
spinors as those in Appendix \ref{u-unsimp}. Whether they depend on
$u$ or not should be clear according to the context. We also use
different notations for the coefficients, such as
$\textrm{Box}[Q_i,Q_j,K], \textrm{Tri}[Q_s,K]$ other than
$C[Q_i,Q_j,K], C[Q_s,K]$.

The pentagon coefficients are given by
\bea \textrm{Pen}[Q_i, Q_j, Q_t,K] = (K^2)^{n+3} {\prod_{s=1}^{n+k}
\b_s^{(K_i,K_j,K_t;P_s)} \over \prod_{w=1,w\neq i,j,t}^k
\gamma_w^{(K_i,K_j,K_w,K_t)}}. \quad \Label{pen-standform} \eea
where
\bea \b_s^{(K_i,K_j,K_t;P_s)} & = & -{(K_i^2+M_1^2-m_i^2)
\eps(P_s,K_j,K, K_t)+(K_j^2+M_1^2-m_j^2) \eps(K_i,P_s,K, K_t)\over
K^2 \eps(K_i,K_j,K, K_t)}\nonumber \\ & & - {(K^2+M_1^2-M_2^2)
\eps(K_i,K_j,P_s, K_t)+(K_t^2+M_1^2-m_t^2) \eps(K_i,K_j,K, P_s)\over
K^2 \eps(K_i,K_j,K, K_t)} \qquad \Label{beta-pen} \eea
\bea \gamma_s^{(K_i,K_j,K_s,K_t)} & = & {(K_i^2+M_1^2-m_i^2)
\eps(K,K_j,K_s, K_t)+(K_j^2+M_1^2-m_j^2) \eps(K_i,K,K_s, K_t)\over
K^2 \eps(K_i,K_j,K, K_t)}\nonumber \\ & & + {(K_s^2+M_1^2-m_s^2)
\eps(K_i,K_j,K, K_t)+(K_t^2+M_1^2-m_t^2) \eps(K_i,K_j,K_s,K)\over
K^2 \eps(K_i,K_j,K, K_t)}\nonumber
\\ & & -{(K^2+M_1^2-M_2^2) \eps(K_i,K_j,K_s, K_t)\over K^2 \eps(K_i,K_j,K,
K_t)} \qquad \Label{gamma-pen}\eea

The box coefficients are given by
\bea \textrm{Box}[Q_i,Q_j,K] & = & {(K^2)^{2+n}\over
2}\left\{{\prod_{s=1}^{k+n} \gb{P_{ji;1}| {\W R}_s(u)
|P_{ji;2}}\over \gb{P_{ji;1}|K |P_{ji;2}}^{n+2} \prod_{t=1,t\neq
i,j}^k\gb{P_{ji;1}|\W Q_t(u) |P_{ji;2}}}\right.\nonumber  \qquad \\
& & \left.-\sum_{t=1,t\neq i,j}^k { \prod_{s=1}^{n+k}
\b_s^{(q_i,q_j,q_t;p_s)} \over \prod_{w=1,w\neq i,j,t}^k
\gamma_w^{(K_i,K_j;K_w,K_t)} } { \gb{P_{ji;1}|K|P_{ji;2}}\over
\gb{P_{ji;1}|\W Q_t(u)|P_{ji;2}} }\right\}\nonumber \\ & & +\{
P_{ji;1}\leftrightarrow P_{ji;2}\} \qquad \Label{k=4-box-equiv}\eea
where
\bea  {\W R}_s(u) &=& {p_s\cdot q_0^{(q_i,q_j,K)}\over
(q_0^{(q_i,q_j,K)})^2}
(\a^{(q_i,q_j)}(u)-1)(-\b q_0^{(q_i,q_j,K)}) + R_s(u=0), \qquad \Label{W-R-S-1}\\
\W Q_t(u) &= &{q_t\cdot q_0^{(q_i,q_j,K)}\over
(q_0^{(q_i,q_j,K)})^2} (\a^{(q_i,q_j)}(u)-1)(-\b
q_0^{(q_i,q_j,K)})+Q_t(u=0), \qquad \Label{W-Q-t-1}\eea
and
\bea (q_0)_\mu^{(q_i,q_j,K)} &\equiv& {1 \over K^2} \epsilon_{\mu\nu
\rho \xi} q_i^\nu q_j^\rho K^\xi  = {1 \over K^2}
 \epsilon_{\mu\nu \rho \xi} K_i^\nu K_j^\rho
K^\xi, \qquad \Label{q-0} \eea
\bea \a^{(q_i,q_j)}(u) \equiv {\sqrt{\b^2(1-u)+ {4K^2[
\a_i\a_j(2q_i\cdot q_j)-\a_i^2 q_j^2-\a_j^2 q_i^2]\over (2q_i\cdot
q_j)^2-4 q_i^2 q_j^2}}\over \sqrt{\b^2+ {4K^2[ \a_i\a_j(2q_i\cdot
q_j)-\a_i^2 q_j^2-\a_j^2 q_i^2]\over (2q_i\cdot q_j)^2-4 q_i^2
q_j^2}}}. \qquad \Label{def-alpha-box} \eea

The triangle coefficients are given by
\bea \textrm{Tri}[Q_s,K] & = & { (K^2)^{1+n}\over
2}\frac{1}{(-\b\sqrt{1-u})^{n+1}(\sqrt{-4q_s^2
K^2})^{n+1}}\frac{1}{(n+1)! \vev{P_{s,1}~P_{s,2}}^{n+1}} \nonumber
\\ & & \times \frac{d^{n+1}}{d\tau^{n+1}}\left.\left({\prod_{j=1}^{k+n}
\vev{P_{s,1}-\tau P_{s,2} |R_j(u) Q_s(u)|P_{s,1}-\tau P_{s,2}}\over
\prod_{t=1,t\neq s}^k \vev{P_{s,1}-\tau P_{s,2}|Q_t(u) Q_s(u)
|P_{s,1}-\tau P_{s,2}}} + \{P_{s,1}\leftrightarrow
P_{s,2}\}\right)\right|_{\tau=0}.~~~~~\Label{tri-exp-0}\eea

The bubble coefficients are given by
\bea \textrm{Bub}[K] = (K^2)^{1+n} \sum_{q=0}^n {1\over q!} {d^q
\over ds^q}\left.\left( {\cal
B}_{n,n-q}^{(0)}(s)+\sum_{r=1}^k\sum_{a=q}^n \left({\cal
B}_{n,n-a}^{(r;a-q;1)}(s)-{\cal
B}_{n,n-a}^{(r;a-q;2)}(s)\right)\right)\right|_{s=0},~~~~~\Label{bub-exp--1}
\eea
where
\bea {\cal B}_{n,t}^{(0)}(s)\equiv {d^n\over
d\tau^n}\left.\left(\left. {1 \over n! [\eta|\W \eta K|\eta]^{n}}
{(2\eta\cdot K)^{t+1} \over (t+1) (K^2)^{t+1}}{\prod_{j=1}^{n+k}
\vev{\ell|R_j(u) (K-s\eta)|\ell}\over \vev{\ell~\eta}^{n+1}
\prod_{p=1}^k \vev{\ell|
Q_p(u)(K-s\eta)|\ell}}\right|_{\ket{\ell}\to |K-\tau \W \eta|\eta]
}\right)\right|_{\tau= 0},~~~\eea
\bea & & {\cal B}_{n,t}^{(r;b;1)}(s)  \equiv  {(-1)^{b+1}\over
 b! (-\b\sqrt{1-u})^{b+1}\sqrt{-4q_r^2 K^2}^{b+1} \vev{P_{r,1}~P_{r,2}}^b}{d^b \over d\tau^{b}}
\left({1\over (t+1)} {\gb{P_{r,1}-\tau
P_{r,2}|\eta|P_{r,1}}^{t+1}\over \gb{P_{r,1}-\tau
P_{_r,2}|K|P_{r,1}}^{t+1}}\right. \nonumber \\ & & \times
\left.\left. {\vev{P_{r,1}-\tau P_{r,2}|Q_r(u) \eta|P_{r,1}-\tau
P_{r,2}}^{b} \prod_{j=1}^{n+k} \vev{P_{r,1}-\tau P_{r,2}|R_j(u)
(K-s\eta)|P_{r,1}-\tau P_{r,2}}\over \vev{P_{r,1}-\tau P_{r,2}|\eta
K|P_{r,1}-\tau P_{r,2}}^{n+1} \prod_{p=1,p\neq r}^k
\vev{P_{r,1}-\tau P_{r,2}| Q_p(u)(K-s\eta)|P_{r,1}-\tau
P_{r,2}}}\right)\right|_{\tau=0},~~~\eea
\bea & & {\cal B}_{n,t}^{(r;b;2)}(s)  \equiv  {(-1)^{b+1}\over
 b! (-\b\sqrt{1-u})^{b+1}\sqrt{-4q_r^2 K^2}^{b+1} \vev{P_{r,1}~P_{r,2}}^{b}}{d^{b} \over d\tau^{b}}
\left({1\over (t+1)} {\gb{P_{r,2}-\tau
P_{r,1}|\eta|P_{r,2}}^{t+1}\over \gb{P_{r,2}-\tau
P_{r,1}|K|P_{r,2}}^{t+1}}\right. \nonumber \\ & & \times
\left.\left. {\vev{P_{r,2}-\tau P_{r,1}|Q_r(u) \eta|P_{r,2}-\tau
P_{r,1}}^{b} \prod_{j=1}^{n+k} \vev{P_{r,2}-\tau P_{r,1}|R_j(u)
(K-s\eta)|P_{r,2}-\tau P_{r,1}}\over \vev{P_{r,2}-\tau P_{r,1}|\eta
K|P_{r,2}-\tau P_{r,1}}^{n+1} \prod_{p=1,p\neq r}^k
\vev{P_{r,2}-\tau P_{r,1}| Q_p(u)(K-s\eta)|P_{r,2}-\tau
P_{r,1}}}\right)\right|_{\tau=0}.~~~\eea

All the $u$-dependence is explicitly presented in the formulas. They
are only from $\W R(u), \W Q(u)$, $R(u), Q(u)$ and the $\sqrt{1-u}$
factors.

%%%%%%%%%%%%%%%%%%%%%%%%%%%%%%%%%%%%%%%%%%%%%%%%%%%%%%%%%%%%%%%%%%%%%
\section{\label{compare-pen} Compare the pentagon formulas}
%%%%%%%%%%%%%%%%%%%%%%%%%%%%%%%%%%%%%%%%%%%%%%%%%%%%%%%%%%%%%%%%%%%%%

For the standard form input
\bea {\cal T}(\W\ell) = {\prod_{s=1}^{n+k} (-2\W\ell\cdot P_s) \over
\prod_{t=1}^k (-2\W\ell\cdot K_t + K_t^2 +M_1^2-m_t^2)} , \quad \eea
the pentagon coefficient by using (\ref{Pen}) is
\bea \textrm{Pen}[K_i,K_j,K_r,K] &=& {\prod_{s=1}^{n+k}
(-2\W\ell_{(i,j,r)} \cdot P_s) \over \prod_{t=1,t\neq i,j,r}^k
(-2\W\ell_{(i,j,r)}\cdot K_t + K_t^2 +M_1^2-m_t^2)} \quad
\Label{pen-new} \eea
where $\W\ell_{(i,j,r)}$ is given by (\ref{pen-ell-solution}). We
need to show that the above formula is equivalent to the pentagon
coefficient (\ref{pen-standform})
\bea \textrm{Pen}[Q_i, Q_j, Q_r, K] = (K^2)^{n+3} {\prod_{s=1}^{n+k}
\b_s^{(K_i,K_j,K_r;P_s)} \over \prod_{t=1,t\neq i,j,r}^k
\gamma_t^{(K_i,K_j,K_t,K_r)}}. \qquad \Label{pen-old} \eea
where $\b_s^{(K_i,K_j,K_r;P_s)}, \gamma_t^{(K_i,K_j,K_t,K_r)}$ are
given by (\ref{beta-pen}) and (\ref{gamma-pen}).

To prove the equivalence, we first show that
\bea -2\W\ell_{(i,j,r)} \cdot P_s &=& K^2 \b_s^{(q_i,q_j,q_r;p_s)} .
\qquad \Label{beta-proof} \eea
By using the expansion for $P_s$ as
\bea P_s = b_0 K + b_i K_i + b_j K_j + b_r K_r  \quad
\Label{pen-solution} \eea
where
\bean \begin{pmatrix} b_0 \cr b_i \cr b_j \cr b_r
\end{pmatrix} = \begin{pmatrix}
K^2 & K_i\cdot K & K_j\cdot K & K_r\cdot K \cr
K\cdot K_i & K_i^2 & K_j\cdot K_i & K_r\cdot K_i \cr
K\cdot K_j & K_i\cdot K_j & K_j^2 & K_r\cdot K_j \cr
K\cdot K_r & K_i\cdot K_r & K_j\cdot K_r & K_r^2
\end{pmatrix}^{-1} \cdot
\begin{pmatrix}
K\cdot P_s \cr K_i\cdot P_s \cr K_j\cdot P_s \cr K_r\cdot P_s
\end{pmatrix} = {1\over \eps(K_i,K_j,K,K_r)}\begin{pmatrix} \eps(K_i,K_j,P_s,K_r) \cr
\eps(P_s,K_j,K, K_r) \cr \eps(K_i,P_s,K, K_r) \cr
\eps(K_i,K_j,K,P_s)
\end{pmatrix},  \qquad \eean
and (\ref{l-exp0ijr}), we can find that
\bea -2 (l_0 K\cdot P_s + l_i K_i \cdot P_s + l_j K_j \cdot P_s +
l_r K_r \cdot P_s) &=& (K^2 +M_1^2-M_2^2) b_0 + (K^2 +M_1^2-m_i^2)
b_i \nonumber\\ && + (K^2 +M_1^2-m_j^2) b_j + (K^2 +M_1^2-m_r^2) b_r
. \qquad \eea
This is just the relation (\ref{beta-proof}). Then, with this
relation, we can find directly that
\bea -2\W\ell_{(i,j,r)}\cdot K_t + K_t^2 +M_1^2-m_t^2 &=& K^2
\gamma_t^{(K_i,K_j,K_t,K_r)}. \qquad \Label{gama-proof} \eea
With (\ref{beta-proof}) and (\ref{gama-proof}), it's obvious that
the two pentagon formulas (\ref{pen-new}) and (\ref{pen-old}) are
equivalent.

%%%%%%%%%%%%%%%%%%%%%%%%%%%%%%%%%%%%%%%%%%%%%%%%%%%%%%%%%%%%%%%%%%%%
\section{\label{box-u-sim-II} Another equivalent expression for $\W \ell_{ij}$}
%%%%%%%%%%%%%%%%%%%%%%%%%%%%%%%%%%%%%%%%%%%%%%%%%%%%%%%%%%%%%%%%%%%%

As in section \ref{box-u-sim-I}, we may give another equivalent
expression for $\W\ell_{ij}$ that avoid the appearance of
$q_0^{(q_i,q_j,K)}$. By using the expansion
\bea p_s & = & a_0^{(q_i,q_j,K; p_s)}
q_0^{(q_i,q_j,K)}+a_i^{(q_i,q_j,K; p_s)} q_i+a_j^{(q_i,q_j,K; p_s)}
q_j, \quad \eea
where the coefficients are:
\bea a_0^{(q_i,q_j,K; p_s)} & = & { (P_s\cdot
q_0^{(q_i,q_j,K)})\over
(q_0^{(q_i,q_j,K)})^2}={\eps(P_s,K_i,K_j,K)\over
K^2(q_0^{(q_i,q_j,K)})^2},
\\a_i^{(q_i,q_j,K; p_s)} & = & {  (P_s\cdot q_i) q_j^2- (P_s\cdot
q_j) (q_i\cdot q_j)\over q_i^2 q_j^2- (q_i\cdot q_j)^2} = P_s \cdot
\xi_i^{(q_i,q_j)} , \qquad \xi_i^{(q_i,q_j)} = { q_i q_j^2- q_j
(q_i\cdot q_j)\over q_i^2 q_j^2- (q_i\cdot q_j)^2},
\qquad \Label{xi-i} \\ a_j^{(q_i,q_j,K; p_s)} & = & {  (P_s\cdot
q_j) q_i^2- (P_s\cdot q_i) (q_i\cdot q_j)\over q_i^2 q_j^2-
(q_i\cdot q_j)^2} = P_s \cdot \xi_j^{(q_i,q_j)}, \qquad
\xi_j^{(q_i,q_j)} = { q_j q_i^2- q_i (q_i\cdot q_j)\over q_i^2
q_j^2- (q_i\cdot q_j)^2} . \qquad \Label{xi-j}\eea
we can write
\bea \W R_s(u) & = &  {p_s\cdot q_0^{(q_i,q_j,K)}\over
(q_0^{(q_i,q_j,K)})^2} (\a^{(q_i,q_j)}(u)-1)(-\b q_0^{(q_i,q_j,K)})
+ R_s(u=0) \nonumber\\ & =&
-(\a^{(q_i,q_j)}(u)-1)\left(a_i^{(q_i,q_j,K; p_s)} Q_i
(u=0)+a_j^{(q_i,q_j,K; p_s)} Q_j(u=0)+ \b_s^{(q_i,q_j,K;p_s)} K-
R_s(u=0)\right) \nonumber\\ && +~R_s(u=0)  \nonumber\\ & =&
-(\a^{(q_i,q_j)}(u)-1)\left(a_i^{(q_i,q_j,K; p_s)} Q_i
(u=0)+a_j^{(q_i,q_j,K; p_s)} Q_j(u=0)\right)\nonumber\\ &&
-(\a^{(q_i,q_j)}(u)-1) \b_s^{(q_i,q_j,K;p_s)} K + \a^{(q_i,q_j)}(u)
R_s(u=0) , \qquad \eea
where we have defined
\bea \b_s^{(q_i,q_j,K;p_s)} & \equiv & (\b_s-a_i^{(q_i,q_j,K; p_s)}
\a_i -a_j^{(q_i,q_j,K; p_s)} \a_j) \nonumber\\ & = &
\left(-\a{K\over K^2}- \a_i \xi_i^{(q_i,q_j)} -\a_j
\xi_j^{(q_i,q_j)} \right)\cdot P_s .~~~~\Label{beta-package} \eea
Making use of the properties
\bean
\gb{P_{ji,1}(u=0)|Q_i(u=0)|P_{ji,2}(u=0)}=\gb{P_{ji,1}(u=0)|Q_j(u=0)|P_{ji,2}(u=0)}=0,\eean
we have
\bea \W R_s(u) & = & -(\a^{(q_i,q_j)}(u)-1) \b_s^{(q_i,q_j,K;p_s)} K
+ \a^{(q_i,q_j)}(u) R_s(u=0) . \qquad \eea
Using (\ref{Ru=0}) and (\ref{beta-package}), we can further rewrite
it as
\bean \W R_s(u) & = & (\a^{(q_i,q_j)}(u)-1) \left[\left(\a{K\over
K^2}+ \a_i \xi_i^{(q_i,q_j)} +\a_j \xi_j^{(q_i,q_j)} \right)\cdot
P_s\right] K + \a^{(q_i,q_j)}(u) \left(-\b P_s + (\b-\a){P_s \cdot
K\over K^2}K\right) \nonumber\\ & = & (\a^{(q_i,q_j)}(u) \b-\a)
{K\over K^2}(K\cdot P_s) +(\a^{(q_i,q_j)}(u)-1) \left[\left(\a_i
\xi_i^{(q_i,q_j)} +\a_j \xi_j^{(q_i,q_j)} \right)\cdot P_s\right] K
- \a^{(q_i,q_j)}(u) \b P_s  \qquad \eean
Then we have
\bean \frac{K^{2}}{\left\langle \ell|K|\ell\right]}\gb{\ell|\W
R_s|\ell} & = & -2 \left\{ -{1\over2}\left[ (\a^{(q_i,q_j)}(u)
\b-\a) K +(\a^{(q_i,q_j)}(u)-1) K^2 \left(\a_i \xi_i^{(q_i,q_j)}
+\a_j \xi_j^{(q_i,q_j)} \right) \right] \right. \\ && \left. -~
\b\frac{K^{2}}{\left\langle \ell|K|\ell\right]} \a^{(q_i,q_j)}(u)
P_{\la \W\la} \right\}\cdot P_s \qquad \nonumber \eean
Therefore we find the equivalent rule is that
\bea \W\ell \rightarrow \W\ell_{ij} &=& -{1\over2}\left[
(\a^{(q_i,q_j)}(u) \b-\a) K +(\a^{(q_i,q_j)}(u)-1) K^2 \left(\a_i
\xi_i^{(q_i,q_j)} +\a_j \xi_j^{(q_i,q_j)} \right) \right]
\nonumber\\ && - \b\frac{K^{2}}{\left\langle \ell|K|\ell\right]}
\a^{(q_i,q_j)}(u) P_{\la \W\la} . \qquad \Label{box-ellij-II}\eea
%

%%%%%%%%%%%%%%%%%%%%%%%%%%%%%%%%%%%%%%%%%%%%%%%%%%%%%%%%%%%%%%%%%
\section{\label{mu2-box} Complete Box coefficients of the example}
%%%%%%%%%%%%%%%%%%%%%%%%%%%%%%%%%%%%%%%%%%%%%%%%%%%%%%%%%%%%%%%%%

In this appendix, we give  complete box coefficients.

\bea c_{[23|4|5|1]} & = & \frac{\langle 1|4\rangle \langle
1|5\rangle  [3|2] [5|4]^2}{
 \Delta } (\mu^2)^2 - \frac{s_{23}{}^2 s_{45}{}^2
\langle 1|2\rangle  \langle 1|5\rangle ^2
\langle 3|4\rangle  [5|1] [5|4]}{2 \text{$\Delta $}^3} \nonumber \\
&& \times \Big(\langle 4|1|2|3|1|5\rangle +\langle
4|1|2|3|2|5\rangle -\langle 4|1|4|3|2|5\rangle +\langle
4|3|2|4|1|5\rangle \Big) \mu^2 \qquad \eea
\bea c_{[45|1|2|3]} & = &  \frac{\langle 1|2\rangle \langle
1|3\rangle [3|2]^2 [5|4]}{ \Delta } (\mu^2)^2 + \frac{s_{23}{}^2
s_{45}{}^2 \langle 1|5\rangle  \langle 1|2\rangle ^2
\langle 3|4\rangle  [2|1] [3|2]}{2 \text{$\Delta $}^3} \nonumber \\
&& \times \Big(\langle 2|1|3|5|4|3\rangle +\langle
2|1|4|5|1|3\rangle -\langle 2|5|4|3|1|3\rangle +\langle
2|5|4|5|1|3\rangle \Big) \mu^2 \qquad \eea
\bea c_{[51|2|3|4]} & = &  -\frac{\langle 4|5\rangle  (\langle
1|2|4|3|2|1|5]+\langle 1|3|2|4|3|1|5]+\langle
1|5|4|2|3|4|5])^2}{s_{15} s_{23}{}^2  \langle 2|4\rangle  \langle
3|4\rangle \Delta } (\mu^2)^2 + \frac{\langle 1|2\rangle  \langle
2|3\rangle ^3 [3|2] [4|3]}{2 \text{$\Delta $}^3 \langle 1|5\rangle
\langle 2|4\rangle ^2 \langle 4|5\rangle ^3}\times
\nonumber\\ && \Big\{3 \langle 1|4\rangle ^2 \langle 1|5\rangle
\langle 4|5\rangle ^2 \langle 1|4|1|4\rangle  \langle 4|2|3|4\rangle
(-\langle 5|1|3|2|1]-\langle 5|3|2|5|1])^2 \nonumber\\
&& +\langle 1|4\rangle ^5 \langle 4|5\rangle ^2 (-\langle
5|1|3|2|1]-\langle 5|3|2|5|1])^3-2 \langle 1|5\rangle  \langle
4|5\rangle  \langle 1|4|1|4\rangle \langle 1|4|1|5\rangle  \langle
4|2|3|4\rangle ^2  \nonumber\\
&& (\langle 5|1|2] \langle 4|5\rangle  (-\langle 2|3|1]+\langle
2|5|1])+2 \langle 4|5\rangle
\langle 5|3|2|5|1]) \nonumber\\
&& -2 \langle 1|5\rangle ^2 \langle 1|4|1|4\rangle \langle
4|2|3|4\rangle ^2 \Big(\langle 4|5\rangle  (-\langle 2|5\rangle
\langle 4|3|1]+\langle 2|4\rangle (-3 \langle 5|2|1]-2 \langle
5|3|1])) \langle 5|1|5|1|2] \nonumber\\
&& -\langle 4|5\rangle ^2 \langle 5|3|2|3|2|5|1]-\langle 1|5\rangle
\big(2 \langle 2|4\rangle \langle 3|5\rangle  \langle 4|2|1] (2
\langle 5|2|1]+\langle 5|3|1]) \nonumber\\
&& -\langle 2|5\rangle  \langle 3|4\rangle \langle 4|3|1] (\langle
5|2|1]-\langle 5|3|1])+\langle 2|4\rangle \langle 3|4\rangle (-2
\langle 5|2|1]^2+3 \langle 5|2|1] \langle 5|3|1]+\langle
5|3|1]^2)\big)
[3|2]\Big) \nonumber\\
&& +\langle 1|5\rangle ^3 \langle 2|4\rangle  \langle 4|5\rangle
\langle 4|2|3|4\rangle ^2 \langle 4|K_{51}|4]^2
\langle 4|5\rangle  ([2|1|5|1]-[2|3|2|1]-[2|4|5|1])  \nonumber\\
&& +\langle 1|4\rangle  \langle 1|5\rangle ^2 \langle 4|2|3|4\rangle
^2 \Big(\langle 4|5\rangle  (\langle 2|5\rangle \langle 4|3|1]
+\langle 2|4\rangle  (6 \langle 5|2|1]+5 \langle 5|3|1])) \langle
5|1|5|1|5|1|2] \nonumber\\
&&-\langle 4|5\rangle ^2 \langle 5|3|2|3|2|3|2|5|1] +\langle
1|5|1|5\rangle  \big(-\langle 2|4\rangle \langle 3|5\rangle \langle
4|2|1] (-13 \langle 5|2|1]-10 \langle 5|3|1])\nonumber\\
&& -\langle 2|5\rangle \langle 3|4\rangle  \langle 4|3|1] (2 \langle
5|2|1]-\langle 5|3|1])  -2 \langle 2|4\rangle \langle 3|4\rangle (5
\langle 5|2|1]^2+\langle 5|2|1] \langle
5|3|1]-\langle 5|3|1]^2)\big) [3|2] \nonumber\\
&& -\langle 1|5\rangle \big(-\langle 2|4\rangle  \langle 3|5\rangle
\langle 4|2|1] (-8 \langle 5|2|1]-5 \langle 5|3|1])-\langle
2|5\rangle  \langle 3|4\rangle  \langle 4|3|1] (\langle 5|2|1]-2
\langle 5|3|1]) \nonumber\\
&& +2 \langle 2|4\rangle  \langle 3|4\rangle  (-2 \langle 5|2|1]^2+2
\langle 5|2|1] \langle 5|3|1]+\langle 5|3|1]^2)\big) [3|2|3|2]\Big)
\Big\} \mu^2 \eea
\bea c_{[12|3|4|5]} &=& -\frac{\langle 2|3\rangle (\langle
1|2|3|5|4|3|2]+\langle 1|4|5|3|4|1|2]+\langle
1|5|3|4|5|1|2])^2}{s_{12} s_{45}{}^2 \langle 3|4\rangle  \langle
3|5\rangle \Delta } (\mu^2)^2 - \frac{\langle 1|5\rangle  \langle
4|5\rangle ^3 [4|3] [5|4]}{2 \Delta ^3 \langle 1|2\rangle  \langle
2|3\rangle ^3 \langle 3|5\rangle ^2}\times
\nonumber\\ && \Big\{ 2 \langle 1|3\rangle  \langle 2|3\rangle
\langle 1|3|1|2\rangle ^2 \langle 3|4|5|3\rangle ^2 (2 \langle
2|4|5|2\rangle  \langle 3|2|1]+\langle 1|3\rangle ^5 \langle
2|3\rangle ^2 (\langle 2|1|5|4|1]+\langle 2|5|4|2|1])^3
\nonumber\\ && -3 \langle 1|3\rangle ^3 \langle 2|3\rangle ^2
\langle 1|3|1|2\rangle  \langle 3|4|5|3\rangle  (\langle
2|1|5|4|1]+\langle 2|5|4|2|1])^2+\langle 2|1|5] \langle 2|3\rangle
(\langle 5|4|1]-\langle 5|2|1])
\nonumber\\ && -2 \langle 1|2\rangle  \langle 1|3\rangle  \langle
1|3|1|2\rangle  \langle 3|4|5|3\rangle ^2 \Big(\langle 2|3\rangle
\langle 2|4|5|4|5|2\rangle  \langle 3|2|1]-\langle 2|3\rangle
(\langle 3|5\rangle  (2 \langle 2|4|1]+3 \langle 2|5|1])
\nonumber\\ && +\langle 2|5\rangle \langle 3|4|1]) \langle
2|1|2|1|5]+\langle 1|2\rangle \big(\langle 3|4\rangle  \langle
3|5\rangle  (\langle 2|4|1]^2+3 \langle 2|4|1] \langle 2|5|1]-2
\langle 2|5|1]^2)
\nonumber\\ && -\langle 2|5\rangle \langle 3|4\rangle (-\langle
2|4|1]+\langle 2|5|1]) \langle 3|4|1]-2 \langle 2|4\rangle \langle
3|5\rangle  (-\langle 2|4|1]-2 \langle 2|5|1]) \langle 3|5|1]\big)
[5|4]\Big)
\nonumber\\ && +\langle 1|2\rangle ^3 \langle 2|3\rangle \langle
3|5\rangle \langle 3|4|5|3\rangle ^2 \langle 3|K_{12}|3]^2 \langle
2|3\rangle ([5|4|5|1]+[5|3|2|1] - [5|1|2|1])
\nonumber\\ &&+\langle 1|2\rangle ^2 \langle 1|3\rangle \langle
3|4|5|3\rangle ^2 \Big(\langle 2|3\rangle \langle
2|4|5|4|5|4|5|2\rangle \langle 3|2|1] +\langle 2|3\rangle (\langle
3|5\rangle  (5 \langle 2|4|1]+6 \langle 2|5|1])
\nonumber\\ &&+\langle 2|5\rangle \langle 3|4|1]) \langle
2|1|2|1|2|1|5] -\langle 1|2|1|2\rangle \big(-2 \langle 3|4\rangle
\langle 3|5\rangle (-\langle 2|4|1]^2+\langle 2|4|1] \langle
2|5|1]+5 \langle 2|5|1]^2)
\nonumber\\ && -\langle 2|5\rangle  \langle 3|4\rangle  (-\langle
2|4|1]+2 \langle 2|5|1]) \langle 3|4|1]-\langle 2|4\rangle  \langle
3|5\rangle  (-10 \langle 2|4|1]-13 \langle 2|5|1]) \langle
3|5|1]\big) [5|4]
\nonumber\\ && +\langle 1|2\rangle  \big(2 \langle 3|4\rangle
\langle 3|5\rangle (\langle 2|4|1]^2+2 \langle 2|4|1] \langle
2|5|1]-2 \langle 2|5|1]^2)+\langle 2|5\rangle  \langle 3|4\rangle (2
\langle 2|4|1]-\langle 2|5|1]) \langle 3|4|1]
\nonumber\\ && +\langle 2|4\rangle \langle 3|5\rangle  (5 \langle
2|4|1]+8 \langle 2|5|1]) \langle 3|5|1]\big)
[5|4|5|4]\Big)\Big\}\mu^2 \eea
\bea c_{[34|5|1|2]} &=& \frac{\langle 1|2\rangle  \langle 1|5\rangle
\langle 2|3\rangle \langle 4|5\rangle [4|3] [5|2]^2}{\langle
2|5\rangle \langle 3|4\rangle \Delta } (\mu^2)^2 - \frac{\langle
1|2\rangle ^2 \langle 1|5\rangle ^2 \langle 2|3\rangle ^3 [2|1]
[4|3] [5|1] [5|2]}{2 \text{$\Delta $}^3 \langle 2|5\rangle
^2 \langle 3|4\rangle  \langle 4|5\rangle } \nonumber\\
&& \Big\{ \langle 4|3|2]^2 \Big(2 \langle 4|5\rangle ^2 \langle
2|3|2|5\rangle ^2-\langle 1|5\rangle  \langle 4|5\rangle  \langle
2|3|2|5\rangle (-4 \langle 2|5\rangle  \langle 4|3|1]+\langle
2|4\rangle  (-5 \langle 5|2|1]-\langle 5|3|1]))
\nonumber\\ && +\langle 1|5\rangle ^2 (2 \langle 2|5\rangle ^2
\langle 4|3|1]^2-\langle 2|4\rangle \langle 2|5\rangle \langle
4|3|1] (-5 \langle 5|2|1]-\langle 5|3|1])
\nonumber\\ && +\langle 2|4\rangle ^2 (4 \langle 5|2|1]^2+3 \langle
5|2|1] \langle 5|3|1]+\langle 5|3|1]^2))\Big)
\nonumber\\ && +\langle 1|4\rangle  \langle 4|3|2] \Big(\langle
4|5\rangle \langle 2|3|2|5\rangle  (\langle 3|5\rangle  (\langle
4|2|1]+5 \langle 4|3|1])+4 \langle 3|4\rangle \langle 5|2|1])
\langle 5|2|3]
\nonumber\\ && +\langle 4|5\rangle  \langle 5|1|2|3|1|5\rangle
(-\langle 2|5\rangle \langle 4|3|1]+\langle 2|4\rangle  (-3 \langle
5|2|1]-2 \langle 5|3|1]))+\langle 1|5\rangle \langle 5|2|3]
\nonumber\\ && \big(\langle 2|5\rangle  \langle 3|4\rangle \langle
4|3|1] (3 \langle 5|2|1]+5 \langle 5|3|1])+\langle 2|4\rangle
\langle 3|5\rangle  \langle 4|2|1] (3 \langle 5|2|1]+\langle 5|3|1])
\nonumber\\ && +\langle 2|4\rangle \langle 3|4\rangle (5 \langle
5|2|1]^2+12 \langle 5|2|1] \langle 5|3|1]+3 \langle
5|3|1]^2)\big)\Big)
\nonumber\\ && +\langle 1|4\rangle ^2 \Big(\big(\langle 3|5\rangle
^2 (\langle 4|2|1]^2+3 \langle 4|2|1] \langle 4|3|1]+4 \langle
4|3|1]^2)
\nonumber\\ && -\langle 3|4\rangle \langle 3|5\rangle  (-\langle
4|2|1]-5 \langle 4|3|1]) \langle 5|2|1]+2 \langle 3|4\rangle ^2
\langle 5|2|1]^2 \big) \langle 5|2|3]^2+\langle 4|5\rangle ^2
\langle 5|1|3|2|1]^2
\nonumber\\ && -\langle 4|5\rangle  \langle 5|2|1] (\langle
3|5\rangle (-2 \langle 4|2|1]-3 \langle 4|3|1])-\langle 3|4\rangle
\langle 5|2|1]) \langle 5|1|3|2|3]\Big) \Big\} \mu^2 \eea
%

%%%%%%%%%%%%%%%%%%%%%%%%%%%%%%%%%%%%%%%%%%%%%%%%%%%%%%%%%%%%%%%%%


\begin{thebibliography}{999}

%%%%%%%%%%%%%%%%%%%%%%%%%%%%%%%%%%%%%%%%%%%%%
% {Report of recent progress}

\bibitem{Report08}
  Z.~Bern {\it et al.},
  %``The NLO multileg working group: summary report,''
  arXiv:0803.0494 [hep-ph].
  %%CITATION = ARXIV:0803.0494;%%

%%%%%%%%%%%%%%%%%%%%%%%%%%%%%%%%%%%%%%%%%%%%%%
% {Basis expansion}

%\cite{Passarino:1978jh}
\bibitem{Passarino-Veltman}
  G.~Passarino and M.~J.~G.~Veltman,
  %``One Loop Corrections For E+ E- Annihilation Into Mu+ Mu- In The Weinberg
  %Model,''
  Nucl.\ Phys.\  B {\bf 160}, 151 (1979).
  %%CITATION = NUPHA,B160,151;%%



\bibitem{IntegralRecursion}
Z.~Bern, L.~J.~Dixon and D.~A.~Kosower,
%``Dimensionally Regulated One Loop Integrals,''
Phys.\ Lett.\  B {\bf 302}, 299 (1993)
[Erratum-ibid.\  B {\bf 318}, 649 (1993)]
[hep-ph/9212308];\\
%%CITATION = PHLTA,B302,299;%%
Z.~Bern, L.~J.~Dixon and D.~A.~Kosower,
%``Dimensionally regulated pentagon integrals,''
Nucl.\ Phys.\  B {\bf 412}, 751 (1994)
[hep-ph/9306240].\\

%%%%%%%%%%%%%%%%%%%%%%%%%%%%%%%%%%%%%%%%%%%%%%
% {Baisi of the Basis expansion}

\bibitem{MasterIntegrals}
{
 L.~M.~Brown and R.~P.~Feynman,
  %``Radiative corrections to Compton scattering,''
  Phys.\ Rev.\  {\bf 85}, 231 (1952);\\
D.~B.~Melrose,
  %``Reduction Of Feynman Diagrams,''
  Nuovo Cim.\  {\bf 40}, 181 (1965);\\
%
  G.~'t Hooft and M.~J.~G.~Veltman,
  %``Scalar One Loop Integrals,''
  Nucl.\ Phys.\  B {\bf 153}, 365 (1979); \\
  W.~L.~van Neerven and J.~A.~M.~Vermaseren,
  %``Large Loop Integrals,''
  Phys.\ Lett.\  B {\bf 137}, 241 (1984);\\
R.~G.~Stuart,
  %``ALGEBRAIC REDUCTION OF ONE LOOP FEYNMAN DIAGRAMS TO SCALAR INTEGRALS,''
  Comput.\ Phys.\ Commun.\  {\bf 48}, 367 (1988);\\
  R.~G.~Stuart and A.~Gongora,
  %``ALGEBRAIC REDUCTION OF ONE LOOP FEYNMAN DIAGRAMS TO SCALAR INTEGRALS. 2,''
  Comput.\ Phys.\ Commun.\  {\bf 56}, 337 (1990);\\
G.~J.~van Oldenborgh and J.~A.~M.~Vermaseren,
  %``New Algorithms for One Loop Integrals,''
  Z.\ Phys.\  C {\bf 46}, 425 (1990);\\
 %%CITATION = NUPHA,B153,365;%%
J.~Fleischer, F.~Jegerlehner and O.~V.~Tarasov,
%``Algebraic reduction of one-loop Feynman graph amplitudes,''
Nucl.\ Phys.\  B {\bf 566}, 423 (2000)
[hep-ph/9907327];\\
%%CITATION = NUPHA,B566,423;%%
T.~Binoth, J.~P.~Guillet and G.~Heinrich,
%``Reduction formalism for dimensionally regulated one-loop N-point
%integrals,''
Nucl.\ Phys.\  B {\bf 572}, 361 (2000)
[hep-ph/9911342];\\
%%CITATION = NUPHA,B572,361;%%
  A.~Denner and S.~Dittmaier,
  %``Reduction of one-loop tensor 5-point integrals,''
  Nucl.\ Phys.\  B {\bf 658}, 175 (2003)
  [hep-ph/0212259];\\
  %%CITATION = NUPHA,B658,175;%%
G.~Duplan\v{c}i\'c and B.~Ni\v{z}i\'c,
%``Reduction method for dimensionally regulated one-loop N-point Feynman
%integrals,''
Eur.\ Phys.\ J.\  C {\bf 35}, 105 (2004)
[hep-ph/0303184]; \\
%%CITATION = EPHJA,C35,105;%%
  A.~Denner and S.~Dittmaier,
  %``Reduction schemes for one-loop tensor integrals,''
  Nucl.\ Phys.\  B {\bf 734}, 62 (2006)
  [hep-ph/0509141];\\
  %%CITATION = NUPHA,B734,62;%%
  R.~K.~Ellis and G.~Zanderighi,
  %``Scalar one-loop integrals for QCD,''
  arXiv:0712.1851 [hep-ph].
  %%CITATION = ARXIV:0712.1851;%%
}

%%%%%%%%%%%%%%%%%%%%%%%%%%%%%%%%%%%%%%%%%%%%%%
% {Unitarity Method - BDDK}

%\cite{Bern:1994zx}
\bibitem{BDDK}
  Z.~Bern, L.~J.~Dixon, D.~C.~Dunbar and D.~A.~Kosower,
%``One loop n point gauge theory amplitudes, unitarity and
% collinear limits,''
  Nucl.\ Phys.\ B {\bf 425}, 217 (1994)
  [hep-ph/9403226];\\
  %%CITATION = HEP-PH 9403226;%%
%\cite{Bern:1994cg}
%\bibitem{BDDK2}
  Z.~Bern, L.~J.~Dixon, D.~C.~Dunbar and D.~A.~Kosower,
  %``Fusing gauge theory tree amplitudes into loop amplitudes,''
  Nucl.\ Phys.\  B {\bf 435}, 59 (1995)
  [hep-ph/9409265].
  %%CITATION = NUPHA,B435,59;%%

%\cite{Bern:1997sc}
\bibitem{BDK-ee4p}
  Z.~Bern, L.~J.~Dixon and D.~A.~Kosower,
  %``One-loop amplitudes for e+ e- to four partons,''
  Nucl.\ Phys.\  B {\bf 513}, 3 (1998)
  [hep-ph/9708239].
  %%CITATION = NUPHA,B513,3;%%

%%%%%%%%%%%%%%%%%%%%%%%%%%%%%%%%%%%%%%%%%%%%%%%%%%%%%%

%\cite{Witten:2003nn}
\bibitem{Witten03}
  E.~Witten,
  %``Perturbative gauge theory as a string theory in twistor space,''
  Commun.\ Math.\ Phys.\  {\bf 252}, 189 (2004)
  [hep-th/0312171].
  %%CITATION = CMPHA,252,189;%%



%%%%%%%%%%%%%%%%%%%%%%%%%%%%%%%%%%%%%%%%%%%%%%%%%%%%%%%%
% {One-loop Amplitude calculations}


%\cite{Cachazo:2004by}
\bibitem{Cachazo:2004by}
  F.~Cachazo, P.~Svrcek and E.~Witten,
  %``Gauge theory amplitudes in twistor space and holomorphic anomaly,''
  JHEP {\bf 0410}, 077 (2004)
  [hep-th/0409245].
  %%CITATION = HEP-TH 0409245;%%

%\cite{Bena:2004xu}
\bibitem{Bena:2004xu}
  I.~Bena, Z.~Bern, D.~A.~Kosower and R.~Roiban,
  %``Loops in twistor space,''
  Phys.\ Rev.\ D {\bf 71}, 106010 (2005)
  [hep-th/0410054].
  %%CITATION = HEP-TH 0410054;%%

%\cite{Cachazo:2004dr}
\bibitem{Cachazo:2004dr}
  F.~Cachazo,
%   ``Holomorphic anomaly of unitarity cuts and one-loop gauge theory
  %amplitudes,''
  [hep-th/0410077].
  %%CITATION = HEP-TH 0410077;%%

%\cite{Britto:2004nj}
\bibitem{Britto:2004nj}
  R.~Britto, F.~Cachazo and B.~Feng,
  %``Computing one-loop amplitudes from the holomorphic anomaly of unitarity
  %cuts,''
  Phys.\ Rev.\ D {\bf 71}, 025012 (2005)
  [hep-th/0410179].
  %%CITATION = HEP-TH 0410179;%%

%\cite{Bern:2004ky}
\bibitem{Bern:2004ky}
  Z.~Bern, V.~Del Duca, L.~J.~Dixon and D.~A.~Kosower,
  %``All non-maximally-helicity-violating one-loop seven-gluon amplitudes in  N
  %= 4 super-Yang-Mills theory,''
  Phys.\ Rev.\ D {\bf 71}, 045006 (2005)
  [hep-th/0410224].
  %%CITATION = HEP-TH 0410224;%%

%\cite{Quigley:2004pw}
\bibitem{Quigley:2004pw}
  C.~Quigley and M.~Rozali,
  %``One-loop MHV amplitudes in supersymmetric gauge theories,''
  JHEP {\bf 0501}, 053 (2005)
  [hep-th/0410278].
  %%CITATION = JHEPA,0501,053;%%

%\cite{Bidder:2004tx}
\bibitem{Bidder:2004tx}
  S.~J.~Bidder, N.~E.~J.~Bjerrum-Bohr, L.~J.~Dixon and D.~C.~Dunbar,
  %``N = 1 supersymmetric one-loop amplitudes and the holomorphic anomaly of
  %unitarity cuts,''
  Phys.\ Lett.\  B {\bf 606}, 189 (2005)
  [hep-th/0410296].
  %%CITATION = PHLTA,B606,189;%%


\bibitem{oneloop-MHVvertices}
%\cite{Brandhuber:2004yw}
%\bibitem{Brandhuber:2004yw}
  A.~Brandhuber, B.~J.~Spence and G.~Travaglini,
  %``One-loop gauge theory amplitudes in N = 4 super Yang-Mills from MHV
  %vertices,''
  Nucl.\ Phys.\  B {\bf 706}, 150 (2005)
  [hep-th/0407214];\\
  %%CITATION = NUPHA,B706,150;%%
%\cite{Bedford:2004py}
%\bibitem{Bedford:2004py}
  J.~Bedford, A.~Brandhuber, B.~J.~Spence and G.~Travaglini,
  %``A twistor approach to one-loop amplitudes in N = 1 supersymmetric
  %Yang-Mills theory,''
  Nucl.\ Phys.\  B {\bf 706}, 100 (2005)
  [hep-th/0410280];\\
  %%CITATION = NUPHA,B706,100;%%
%\cite{Bedford:2004nh}
%\bibitem{Bedford:2004nh}
  J.~Bedford, A.~Brandhuber, B.~J.~Spence and G.~Travaglini,
  %``Non-supersymmetric loop amplitudes and MHV vertices,''
  Nucl.\ Phys.\  B {\bf 712}, 59 (2005)
  [hep-th/0412108].
  %%CITATION = NUPHA,B712,59;%%


%\cite{Britto:2004nc}
\bibitem{BCF04-N=4}
  R.~Britto, F.~Cachazo and B.~Feng,
  %``Generalized unitarity and one-loop amplitudes in N = 4  super-Yang-Mills,''
  Nucl.\ Phys.\ B {\bf 725}, 275 (2005)
  [hep-th/0412103].
  %%CITATION = HEP-TH 0412103;%%

%\cite{Britto:2005ha}
\bibitem{BBCF05-SQCD}
  R.~Britto, E.~Buchbinder, F.~Cachazo and B.~Feng,
  %``One-loop amplitudes of gluons in SQCD,''
  Phys.\ Rev.\ D {\bf 72}, 065012 (2005)
  [hep-ph/0503132].
  %%CITATION = HEP-PH 0503132;%%

%\cite{Britto:2006sj}
\bibitem{BFM06-QCD}
  R.~Britto, B.~Feng and P.~Mastrolia,
  %``The cut-constructible part of QCD amplitudes,''
  Phys.\ Rev.\ D {\bf 73}, 105004 (2006)
  [hep-ph/0602178].
  %%CITATION = HEP-PH 0602178;%%




%\cite{Mastrolia:2006ki}
\bibitem{MastroliaTriCut}
  P.~Mastrolia,
  %``On triple-cut of scattering amplitudes,''
  Phys.\ Lett.\  B {\bf 644}, 272 (2007)
  [hep-th/0611091].
  %%CITATION = PHLTA,B644,272;%%


%\cite{BjerrumBohr:2007vu}
\bibitem{BjerrumBohr:2007vu}
  N.~E.~J.~Bjerrum-Bohr, D.~C.~Dunbar and W.~B.~Perkins,
  %``Analytic Structure of Three-Mass Triangle Coefficients,''
  arXiv:0709.2086 [hep-ph].
  %%CITATION = ARXIV:0709.2086;%%





%%%%%%%%%%%%%%%%%%%%%%%%%%%%%%%%%%%%%%%%%%%%%%%%%%%%%%%%%%%%%%%
%%%%%%%%%%%%%%%%%%%%%%%%%%%%%%%%%%%%%%%%%%%%%%%%%%%%%%%%%%%%%%%
% {Rational Part}

%%%%%%%%%%%%%%%%%%%%%%%%%%%%%%%%%%%%%%%%%%%%%%%%%%%%%%%%%%%%%%
% {XYZ}

%\cite{Xiao:2006vr}
\bibitem{XYZ}
  Z.~Xiao, G.~Yang and C.~J.~Zhu,
  %``The rational part of QCD amplitude. I: The general formalism,''
  Nucl.\ Phys.\ B {\bf 758}, 1 (2006)
  [hep-ph/0607015];\\
  %%CITATION = HEP-PH 0607015;%%
%\cite{Su:2006vs}
  X.~Su, Z.~Xiao, G.~Yang and C.~J.~Zhu,
  %``The rational part of QCD amplitude. II: The five-gluon,''
  Nucl.\ Phys.\ B {\bf 758}, 35 (2006)
  [hep-ph/0607016];\\
  %%CITATION = HEP-PH 0607016;%%
%\cite{Xiao:2006vt}
  Z.~Xiao, G.~Yang and C.~J.~Zhu,
  %``The rational part of QCD amplitude. III: The six-gluon,''
  Nucl.\ Phys.\ B {\bf 758}, 53 (2006)
  [hep-ph/0607017].
  %%CITATION = HEP-PH 0607017;%%

%%%%%%%%%%%%%%%%%%%%%%%%%%%%%%%%%%%%%%%%%%%%%%%%%%%%%%%%%%%%%%%%
% {BGH06}

%\cite{Binoth:2006hk}
\bibitem{BGH06}
  T.~Binoth, J.~P.~Guillet and G.~Heinrich,
  %``Algebraic evaluation of rational polynomials in one-loop amplitudes,''
  JHEP {\bf 0702}, 013 (2007)
  [hep-ph/0609054].
  %%CITATION = JHEPA,0702,013;%%

%%%%%%%%%%%%%%%%%%%%%%%%%%%%%%%%%%%%%%%%%%%%%%%%%%%%%%%%%%%%%%%%
% {On-shell, Bootstrapping method}

\bibitem{BDK-Bootstrap}
%\cite{Bern:2005hs}
%\bibitem{Bern:2005hs}
  Z.~Bern, L.~J.~Dixon and D.~A.~Kosower,
  %``On-shell recurrence relations for one-loop QCD amplitudes,''
  Phys.\ Rev.\ D {\bf 71}, 105013 (2005)
  [hep-th/0501240];\\
  %%CITATION = HEP-TH 0501240;%%
%
%\cite{Bern:2005ji}
%\bibitem{Bern:2005ji}
  Z.~Bern, L.~J.~Dixon and D.~A.~Kosower,
  %``The last of the finite loop amplitudes in QCD,''
  Phys.\ Rev.\ D {\bf 72}, 125003 (2005)
  [hep-ph/0505055];\\
  %%CITATION = HEP-PH 0505055;%%
%
%\cite{Bern:2005cq}
%\bibitem{Bern:2005cq}
  Z.~Bern, L.~J.~Dixon and D.~A.~Kosower,
  %``Bootstrapping multi-parton loop amplitudes in QCD,''
  Phys.\ Rev.\ D {\bf 73}, 065013 (2006)
  [hep-ph/0507005];\\
  %%CITATION = HEP-PH 0507005;%%
%
%\cite{Forde:2005hh}
%\bibitem{Forde:2005hh}
  D.~Forde and D.~A.~Kosower,
  %``All-multiplicity one-loop corrections to MHV amplitudes in QCD,''
  Phys.\ Rev.\  D {\bf 73}, 061701 (2006)
  [hep-ph/0509358].
  %%CITATION = PHRVA,D73,061701;%%


\bibitem{BBDFK-Bootstrap}
%\cite{Berger:2006ci}
%\bibitem{Berger:2006ci}
  C.~F.~Berger, Z.~Bern, L.~J.~Dixon, D.~Forde and D.~A.~Kosower,
  %``Bootstrapping one-loop QCD amplitudes with general helicities,''
  Phys.\ Rev.\ D {\bf 74}, 036009 (2006)
  [hep-ph/0604195];\\
  %%CITATION = HEP-PH 0604195;%%
%
%\cite{Berger:2006vq}
%\bibitem{Berger:2006vq}
  C.~F.~Berger, Z.~Bern, L.~J.~Dixon, D.~Forde and D.~A.~Kosower,
  %``All one-loop maximally helicity violating gluonic amplitudes in QCD,''
  Phys.\ Rev.\  D {\bf 75}, 016006 (2007)
  [hep-ph/0607014].
  %%CITATION = PHRVA,D75,016006;%%

\bibitem{Bootstrap-PhiMHV}
%\cite{Badger:2007si}
%\bibitem{Badger:2007si}
  S.~D.~Badger, E.~W.~N.~Glover and K.~Risager,
  %``One-loop phi-MHV amplitudes using the unitarity bootstrap,''
  JHEP {\bf 0707}, 066 (2007)
  arXiv:0704.3914 [hep-ph];\\
  %%CITATION = JHEPA,0707,066;%%
%\cite{Glover:2008ff}
%\bibitem{Glover:2008ff}
  E.~W.~N.~Glover, P.~Mastrolia and C.~Williams,
  %``One-loop phi-MHV amplitudes using the unitarity bootstrap: the general
  %helicity case,''
  arXiv:0804.4149 [hep-ph].
  %%CITATION = ARXIV:0804.4149;%%


%%%%%%%%%%%%%%%%%%%%%%%%%%%%%%%%%%%%%%%%%%%%%%%%%%%%%%%%

%\cite{Berger:2008sj}
\bibitem{Berger:2008sj}
  C.~F.~Berger {\it et al.},
  %``An Automated Implementation of On-Shell Methods for One-Loop Amplitudes,''
  arXiv:0803.4180 [hep-ph].
  %%CITATION = ARXIV:0803.4180;%%


%%%%%%%%%%%%%%%%%%%%%%%%%%%%%%%%%%%%%%%%%%%%%%%%%%%%%%
% {BDK-OnShell-Review}

%\cite{Bern:2007dw}
\bibitem{BDK-OnShell-Review}
  Z.~Bern, L.~J.~Dixon and D.~A.~Kosower,
  %``On-Shell Methods in Perturbative QCD,''
  Annals Phys.\  {\bf 322}, 1587 (2007)
  arXiv:0704.2798 [hep-ph].
  %%CITATION = APNYA,322,1587;%%

%\cite{Risager:2008yz}
\bibitem{Risager-PhDThesis}
  K.~Risager,
  %``Unitarity and On-Shell Recursion Methods for Scattering Amplitudes,''
  U. Copenhagen Ph.D. thesis (2008)
  arXiv:0804.3310 [hep-th].
  %%CITATION = ARXIV:0804.3310;%%



%%%%%%%%%%%%%%%%%%%%%%%%%%%%%%%%%%%%%%%%%%%%%%%%%
%%%%%%%%%%%%%%%%%%%%%%%%%%%%%%%%%%%%%%%%%%%%%%%%%
% {OPP-Reduction}

%\cite{Ossola:2006us}
\bibitem{OPP}
  G.~Ossola, C.~G.~Papadopoulos and R.~Pittau,
  %``Reducing full one-loop amplitudes to scalar integrals at the integrand
  %level,''
  Nucl.\ Phys.\  B {\bf 763}, 147 (2007)
  [hep-ph/0609007];\\
  %%CITATION = NUPHA,B763,147;%%
%
%\cite{Ossola:2007bb}
%\bibitem{OPP-photon}
  G.~Ossola, C.~G.~Papadopoulos and R.~Pittau,
  %``Numerical Evaluation of Six-Photon Amplitudes,''
  JHEP {\bf 0707}, 085 (2007)
  arXiv:0704.1271 [hep-ph].
  %%CITATION = JHEPA,0707,085;%%

%\cite{delAguila:2004nf}
\bibitem{delAguila-Pittau}
  F.~del Aguila and R.~Pittau,
  %``Recursive numerical calculus of one-loop tensor integrals,''
  JHEP {\bf 0407}, 017 (2004)
  [hep-ph/0404120].
  %%CITATION = HEP-PH 0404120;%%


%%%%%%%%%%%%%%%%%%%%%%%%%%%%%%%%%%%%%%%%%%%%%%%%%
% {Numerical development based on OPP-Reduction}

%\cite{Ellis:2007br}
\bibitem{Ellis:2007br}
  R.~K.~Ellis, W.~T.~Giele and Z.~Kunszt,
  %``A Numerical Unitarity Formalism for Evaluating One-Loop Amplitudes,''
  JHEP {\bf 0803}, 003 (2008)
  arXiv:0708.2398 [hep-ph].
  %%CITATION = JHEPA,0803,003;%%

%\cite{Ossola:2007ax}
\bibitem{Ossola:2007ax}
  G.~Ossola, C.~G.~Papadopoulos and R.~Pittau,
  %``CutTools: a program implementing the OPP reduction method to compute
  %one-loop amplitudes,''
  JHEP {\bf 0803}, 042 (2008)
  arXiv:0711.3596 [hep-ph].
  %%CITATION = JHEPA,0803,042;%%

%\cite{Giele:2008ve}
\bibitem{Giele:2008ve}
  W.~T.~Giele, Z.~Kunszt and K.~Melnikov,
  %``Full one-loop amplitudes from tree amplitudes,''
  arXiv:0801.2237 [hep-ph].
  %%CITATION = ARXIV:0801.2237;%%

%\cite{Ossola:2008xq}
\bibitem{Ossola:2008xq}
  G.~Ossola, C.~G.~Papadopoulos and R.~Pittau,
  %``On the Rational Terms of the one-loop amplitudes,''
  arXiv:0802.1876 [hep-ph].
  %%CITATION = ARXIV:0802.1876;%%

%\cite{Mastrolia:2008jb}
\bibitem{Mastrolia:2008jb}
  P.~Mastrolia, G.~Ossola, C.~G.~Papadopoulos and R.~Pittau,
  %``Optimizing the Reduction of One-Loop Amplitudes,''
  arXiv:0803.3964 [hep-ph].
  %%CITATION = ARXIV:0803.3964;%%

%\cite{Ellis:2008ir}
\bibitem{Ellis:2008ir}
  R.~K.~Ellis, W.~T.~Giele, Z.~Kunszt and K.~Melnikov,
  %``Masses, fermions and generalized D-dimensional unitarity,''
  arXiv:0806.3467 [hep-ph].
  %%CITATION = ARXIV:0806.3467;%%

%\cite{Giele:2008bc}
\bibitem{Giele:2008bc}
  W.~T.~Giele and G.~Zanderighi,
  %``On the Numerical Evaluation of One-Loop Amplitudes: the Gluonic Case,''
  arXiv:0805.2152 [hep-ph].
  %%CITATION = ARXIV:0805.2152;%%



%%%%%%%%%%%%%%%%%%%%%%%%%%%%%%%%%%%%%%%%%%%%%%%%%%%%%%%%
% {Analytic developments of OPP-reduction}

%\cite{Forde:2007mi}
\bibitem{Forde07}
  D.~Forde,
  %``Direct extraction of one-loop integral coefficients,''
  Phys.\ Rev.\  D {\bf 75}, 125019 (2007)
  arXiv:0704.1835 [hep-ph].
  %%CITATION = PHRVA,D75,125019;%%

%\cite{Kilgore:2007qr}
\bibitem{Kilgore07}
  W.~B.~Kilgore,
  %``One-loop Integral Coefficients from Generalized Unitarity,''
  arXiv:0711.5015 [hep-ph].
  %%CITATION = ARXIV:0711.5015;%%


%\cite{Badger:2008cm}
\bibitem{Badger:2008cm}
  S.~D.~Badger,
  %``Direct Extraction Of One Loop Rational Terms,''
  arXiv:0806.4600 [hep-ph].
  %%CITATION = ARXIV:0806.4600;%%



%%%%%%%%%%%%%%%%%%%%%%%%%%%%%%%%%%%%%%%%%%%%%%%%
% {D-dimensional Unitarity}

%\cite{vanNeerven:1985xr}
\bibitem{vanNeerven85}
  W.~L.~van Neerven,
%   ``Dimensional Regularization Of Mass And Infrared Singularities In Two Loop
  %On-Shell Vertex Functions,''
  Nucl.\ Phys.\ B {\bf 268}, 453 (1986).
  %%CITATION = NUPHA,B268,453;%%

%\cite{Bern:1995db}
\bibitem{Bern-Morgan}
  Z.~Bern and A.~G.~Morgan,
  %``Massive Loop Amplitudes from Unitarity,''
  Nucl.\ Phys.\ B {\bf 467}, 479 (1996)
  [hep-ph/9511336].
  %%CITATION = HEP-PH 9511336;%%

%\cite{Bern:1996ja}
\bibitem{BDDK-N=4}
   Z.~Bern, L.~J.~Dixon, D.~C.~Dunbar and D.~A.~Kosower,
  %``One-loop self-dual and N = 4 superYang-Mills,''
  Phys.\ Lett.\ B {\bf 394}, 105 (1997)
  [hep-th/9611127].
  %%CITATION = HEP-TH 9611127;%%

%\cite{Brandhuber:2005jw}
\bibitem{Brandhuber05}
  A.~Brandhuber, S.~McNamara, B.~J.~Spence and G.~Travaglini,
  %``Loop amplitudes in pure Yang-Mills from generalised unitarity,''
  JHEP {\bf 0510}, 011 (2005)
  [hep-th/0506068].
  %%CITATION = HEP-TH 0506068;%%

%\cite{Anastasiou:2006jv}
\bibitem{ABFKM06}
  C.~Anastasiou, R.~Britto, B.~Feng, Z.~Kunszt and P.~Mastrolia,
  %``D-dimensional unitarity cut method,''
  Phys.\ Lett.\  B {\bf 645}, 213 (2007)
  [hep-ph/0609191];\\
  %%CITATION = PHLTA,B645,213;%%
%\cite{Anastasiou:2006gt}
%\bibitem{Anastasiou:2006gt}
  C.~Anastasiou, R.~Britto, B.~Feng, Z.~Kunszt and P.~Mastrolia,
  %``Unitarity cuts and reduction to master integrals in d dimensions for
  %one-loop amplitudes,''
  JHEP {\bf 0703}, 111 (2007)
  [hep-ph/0612277].
  %%CITATION = JHEPA,0703,111;%%



%%%%%%%%%%%%%%%%%%%%%%%%%%%%%%%%%%%%%%%%%%%%%%%%
% {Formulas of coefficients}

%\cite{Britto:2006fc}
\bibitem{BF06}
  R.~Britto and B.~Feng,
  %``Unitarity cuts with massive propagators and algebraic expressions for
  %coefficients,''
  Phys.\ Rev.\  D {\bf 75}, 105006 (2007)
  [hep-ph/0612089].
  %%CITATION = PHRVA,D75,105006;%%

%\cite{Britto:2007tt}
\bibitem{BF07}
  R.~Britto and B.~Feng,
  %``Integral Coefficients for One-Loop Amplitudes,''
  JHEP {\bf 0802}, 095 (2008)
  arXiv:0711.4284 [hep-ph].
  %%CITATION = JHEPA,0802,095;%%

%\cite{Britto:2008vq}
\bibitem{BFM08}
  R.~Britto, B.~Feng and P.~Mastrolia,
  %``Closed-Form Decomposition of One-Loop Massive Amplitudes,''
  arXiv:0803.1989 [hep-ph].
  %%CITATION = ARXIV:0803.1989;%%

%\cite{Britto:2008sw}
\bibitem{BFG08}
  R.~Britto, B.~Feng and G.~Yang,
  %``Complete One-Loop Amplitudes With Massless Propagators,''
  arXiv:0803.3147 [hep-ph].
  %%CITATION = ARXIV:0803.3147;%%


%%%%%%%%%%%%%%%%%%%%%%%%%%%%%%%%%%%%%%%%%%%%%%%


%\cite{Bern:1995ix}
\bibitem{Bern-Chalmers}
  Z.~Bern and G.~Chalmers,
  %``Factorization in one loop gauge theory,''
  Nucl.\ Phys.\  B {\bf 447}, 465 (1995)
  [hep-ph/9503236].
  %%CITATION = NUPHA,B447,465;%%




%%%%%%%%%%%%%%%%%%%%%%%%%%%%%%%%%%%%%%%%%%%%%%%%%%%%%%%%
% {BCFW recursion relation}

%\cite{Britto:2004ap}
\bibitem{BCFW}
  R.~Britto, F.~Cachazo and B.~Feng,
  %``New recursion relations for tree amplitudes of gluons,''
  Nucl.\ Phys.\  B {\bf 715}, 499 (2005)
  [hep-th/0412308];\\
  %%CITATION = NUPHA,B715,499;%%
%
%\cite{Britto:2005fq}
%\bibitem{BCFW05}
  R.~Britto, F.~Cachazo, B.~Feng and E.~Witten,
  %``Direct proof of tree-level recursion relation in Yang-Mills theory,''
  Phys.\ Rev.\ Lett.\  {\bf 94}, 181602 (2005)
  [hep-th/0501052].
  %%CITATION = PRLTA,94,181602;%%


%%%%%%%%%%%%%%%%%%%%%%%%%%%%%%%%%%%%%%%%%%%%%%%%%%%%
% {Off-shell tree}
%\cite{BG}
\bibitem{BG} F.A.~Berends and W.T.~Giele, Nucl.Phys.B306(1988), 759.





%%%%%%%%%%%%%%%%%%%%%%%%%%%%%%%%%%%%%%%%%%%%%%%%%%%%%
% {SpinorFormalism}

\bibitem{ChineseMagic}
{
%\cite{Berends:1981rb}
%\bibitem{Berends:1981rb}
  F.~A.~Berends, R.~Kleiss, P.~De Causmaecker, R.~Gastmans and T.~T.~Wu,
  %``Single Bremsstrahlung Processes In Gauge Theories,''
  Phys.\ Lett.\  B {\bf 103}, 124 (1981).
  %%CITATION = PHLTA,B103,124;%%

%\cite{De Causmaecker:1981bg}
%\bibitem{De Causmaecker:1981bg}
  P.~De Causmaecker, R.~Gastmans, W.~Troost and T.~T.~Wu,
  %``Multiple Bremsstrahlung In Gauge Theories At High-Energies. 1. General
  %Formalism For Quantum Electrodynamics,''
  Nucl.\ Phys.\  B {\bf 206}, 53 (1982).
  %%CITATION = NUPHA,B206,53;%%

%\cite{Kleiss:1985yh}
%\bibitem{Kleiss:1985yh}
  R.~Kleiss and W.~J.~Stirling,
  %``Spinor Techniques For Calculating P Anti-P $\to$ W+- / Z0 + Jets,''
  Nucl.\ Phys.\  B {\bf 262}, 235 (1985).
  %%CITATION = NUPHA,B262,235;%%

%\cite{Gastmans:1990xh}
%\bibitem{Gastmans:1990xh}
  R.~Gastmans and T.~T.~Wu,
  %``The Ubiquitous Photon: Helicity Method For QED And QCD,''
%\href{http://www.slac.stanford.edu/spires/find/hep/www?irn=2297604}{SPIRES entry}
{\it  Oxford, UK: Clarendon (1990) 648 p. (International series of
  monographs on physics, 80)}

%\cite{Xu:1986xb}
%\bibitem{Xu:1986xb}
  Z.~Xu, D.~H.~Zhang and L.~Chang,
  %``Helicity Amplitudes for Multiple Bremsstrahlung in Massless Nonabelian
  %Gauge Theories,''
  Nucl.\ Phys.\  B {\bf 291}, 392 (1987).
  %%CITATION = NUPHA,B291,392;%%

%\cite{Gunion:1985vca}
%\bibitem{Gunion:1985vca}
  J.~F.~Gunion and Z.~Kunszt,
  %``Improved Analytic Techniques For Tree Graph Calculations And The G G Q
  %Anti-Q Lepton Anti-Lepton Subprocess,''
  Phys.\ Lett.\  B {\bf 161}, 333 (1985).
  %%CITATION = PHLTA,B161,333;%%
}


%%%%%%%%%%%%%%%%%%%%%%%%%%%%%%%%%%%%%%%%%%%%%%%%%%%%%
% {Color Decomposition}

\bibitem{ColorDecomposition}
%\cite{Berends:1987cv}
%\bibitem{Berends:1987cv}
  F.~A.~Berends and W.~Giele,
  %``The Six Gluon Process As An Example Of Weyl-Van Der Waerden Spinor
  %Calculus,''
  Nucl.\ Phys.\  B {\bf 294}, 700 (1987);\\
  %%CITATION = NUPHA,B294,700;%%
%
%\cite{Mangano:1987xk}
%\bibitem{Mangano:1987xk}
  M.~L.~Mangano, S.~J.~Parke and Z.~Xu,
  %``Duality and Multi - Gluon Scattering,''
  Nucl.\ Phys.\  B {\bf 298}, 653 (1988);\\
  %%CITATION = NUPHA,B298,653;%%
%
%\cite{Mangano:1988kk}
%\bibitem{Mangano:1988kk}
  M.~L.~Mangano,
  %``The Color Structure Of Gluon Emission,''
  Nucl.\ Phys.\  B {\bf 309}, 461 (1988);\\
  %%CITATION = NUPHA,B309,461;%%
%
%\cite{Bern:1990ux}
%\bibitem{Bern:1990ux}
  Z.~Bern and D.~A.~Kosower,
  %``Color Decomposition Of One Loop Amplitudes In Gauge Theories,''
  Nucl.\ Phys.\  B {\bf 362}, 389 (1991).
  %%CITATION = NUPHA,B362,389;%%


%%%%%%%%%%%%%%%%%%%%%%%%%%%%%%%%%%%%%%%%%%%%%%%%%%%
% {Reviews for spinor formalism and color decomposition.}

%\cite{Mangano:1990by}
\bibitem{Mangano-Parke}
  M.~L.~Mangano and S.~J.~Parke,
  %``Multiparton amplitudes in gauge theories,''
  Phys.\ Rept.\  {\bf 200}, 301 (1991)
  [hep-th/0509223].
  %%CITATION = PRPLC,200,301;%%

%\cite{Dixon:1996wi}
\bibitem{DixonTASI}
L.~J.~Dixon,
%``Calculating scattering amplitudes efficiently,''
in {\it QCD \& Beyond: Proceedings of TASI '95}, ed. D.~E.~Soper
(World Scientific, 1996) [hep-ph/9601359].
%%CITATION = HEP-PH 9601359;%%


%%%%%%%%%%%%%%%%%%%%%%%%%%%%%%%%%%%%%%%%%%%%%%%%%%%%
%\cite{Cachazo:2004kj}
\bibitem{CSW04-MHV}
  F.~Cachazo, P.~Svrcek and E.~Witten,
  %``MHV vertices and tree amplitudes in gauge theory,''
  JHEP {\bf 0409}, 006 (2004)
  [hep-th/0403047].
  %%CITATION = JHEPA,0409,006;%%


%%%%%%%%%%%%%%%%%%%%%%%%%%%%%%%%%%%%%%%%%%%%%%%%%%%%
% {Mathematica Package S@M}

%\cite{Maitre:2007jq}
\bibitem{S@M}
  D.~Maitre and P.~Mastrolia,
  %``S@M, a Mathematica Implementation of the Spinor-Helicity Formalism,''
  arXiv:0710.5559 [hep-ph].
  %%CITATION = ARXIV:0710.5559;%%


%%%%%%%%%%%%%%%%%%%%%%%%%%%%%%%%%%%%%%%%%%%%%%%%%%%
% {Supersymmetric identities}

%\cite{Grisaru:1977px}
\bibitem{Grisaru1977}
  M.~T.~Grisaru and H.~N.~Pendleton,
  %``Some Properties Of Scattering Amplitudes In Supersymmetric Theories,''
  Nucl.\ Phys.\  B {\bf 124}, 81 (1977).
  %%CITATION = NUPHA,B124,81;%%


%%%%%%%%%%%%%%%%%%%%%%%%%%%%%%%%%%%%%%%%%%%%%%%%%%%
% {BDK-FiveGluon}
%\cite{Bern:1993mq}
\bibitem{BDK-FiveGluon}
  Z.~Bern, L.~J.~Dixon and D.~A.~Kosower,
  %``One loop corrections to five gluon amplitudes,''
  Phys.\ Rev.\ Lett.\  {\bf 70}, 2677 (1993)
  [hep-ph/9302280].
  %%CITATION = PRLTA,70,2677;%%


%%%%%%%%%%%%%%%%%%%%%%%%%%%%%%%%%%%%%%%%%%%%%%%%%%%%
% {O(eps) in two loops}
%\cite{Bern:1998sc}
\bibitem{Bern:1998sc}
  Z.~Bern, V.~Del Duca and C.~R.~Schmidt,
  %``The infrared behavior of one-loop gluon amplitudes at
  %next-to-next-to-leading order,''
  Phys.\ Lett.\  B {\bf 445}, 168 (1998)
  [hep-ph/9810409];\\
  %%CITATION = PHLTA,B445,168;%%
%\cite{Bern:1999ry}
%\bibitem{Bern:1999ry}
  Z.~Bern, V.~Del Duca, W.~B.~Kilgore and C.~R.~Schmidt,
  %``The infrared behavior of one-loop {QCD} amplitudes at
  %next-to-next-to-leading order,''
  Phys.\ Rev.\  D {\bf 60}, 116001 (1999)
  [hep-ph/9903516].
  %%CITATION = PHRVA,D60,116001;%%


%%%%%%%%%%%%%%%%%%%%%%%%%%%%%%%%%%%%%%%%%%%%%%%%%%%%%
% {Massive on-shell recursive relation}

%\cite{Badger:2005zh}
\bibitem{BGKS-MassiveOnShell}
  S.~D.~Badger, E.~W.~N.~Glover, V.~V.~Khoze and P.~Svrcek,
  %``Recursion relations for gauge theory amplitudes with massive particles,''
  JHEP {\bf 0507}, 025 (2005)
  [hep-th/0504159].
  %%CITATION = JHEPA,0507,025;%%


\end{thebibliography}
\end{document}